\documentclass[twocolumn]{aastex63}

\newcommand{\teff}{T_{\rm eff}}

\newcommand{\ebv}{E(B\, -\, V)}
\newcommand{\dmn}{(m\, -\, M)_0}

\shorttitle{A Blueprint for the Milky Way's Stellar Populations.\ V}
\shortauthors{An et al.}

\begin{document}

\title{A Blueprint for the Milky Way's Stellar Populations.\ V. 3D Local Dust Extinction}

\submitjournal{The Astrophysical Journal Supplement Series}
\received{November 23, 2023}
\accepted{March 15, 2024}

\author{Deokkeun An}
\affiliation{Department of Science Education, Ewha Womans University, 52 Ewhayeodae-gil, Seodaemun-gu, Seoul 03760, Republic of Korea; deokkeun@ewha.ac.kr}

\author{Timothy C.\ Beers}
\affiliation{Department of Physics and Astronomy and JINA Center for the Evolution of the Elements, University of Notre Dame, Notre Dame, IN 46556, USA}

\author{Anirudh Chiti}
\affiliation{Department of Astronomy \& Astrophysics, University of Chicago, 5640 S. Ellis Avenue, Chicago, IL 60637, USA}
\affiliation{Kavli Institute for Cosmological Physics, University of Chicago, Chicago, IL 60637, USA}

\begin{abstract}

Using a grid of empirically calibrated synthetic spectra developed in our previous study, we construct an all-sky 3D extinction map from the large collection of low-resolution XP spectra in Gaia DR3. Along each line of sight, with an area ranging from $0.2$ to $13.4$ deg$^2$, we determine both the reddening and metallicity of main-sequence stars and model the foreground extinction up to approximately $3$~kpc from the Sun. Furthermore, we explore variations in the total-to-selective extinction ratio in our parameter search and identify its mean systematic change across diverse cloud environments in both hemispheres. In regions outside the densest parts of the clouds, our reddening estimates are validated through comparisons with previous reddening maps. However, a notable discrepancy arises when compared to other independent work based on XP spectra, although our metallicity scale shows reasonable agreement with the high-resolution spectroscopic abundance scale. We also assess the accuracy of the XP spectra by applying our calibrated models, and we confirm an increasing trend of flux overestimation at shorter wavelengths below $400$~nm.

\end{abstract}


\section{Introduction}

The availability of a vast amount of observational data for Galactic stars has significantly advanced our understanding of the Milky Way's structure and evolution \citep[e.g.,][]{ivezic:12,blandhawthorn:16,helmi:20}. Photometric surveys are particularly noteworthy in this regard, as they can offer the largest and most representative sample of stars through observing a much greater number of stars than spectroscopic surveys. As demonstrated in this series of papers \citep[][hereafter Paper~I through Paper~IV, respectively]{paper1,paper2,paper3,paper4}, photometric data can serve as a valuable resource for constraining the fundamental properties of stars, including effective temperatures ($\teff$) and metallicities ([Fe/H]), thereby facilitating the construction of comprehensive phase-space distributions of Galactic stars on a global scale.

In this regard, the set of low-resolution spectra for $220$ million objects obtained using the Blue and Red Photometer (BP/RP; hereafter XP) on board the Gaia spacecraft is a new exciting addition to its data release 3 \citep[DR3;][]{gaia:dr3}. As described in \citet{deangeli:23}, the XP spectra were taken using the BP/RP photometers, which cover the wavelength ranges [330, 680]~nm and [640, 1050]~nm, respectively, resulting in a composite spectrum that spans from the near-ultraviolet (UV) to near-infrared wavelengths. The XP data product contains information on almost all sources brighter than $G=17.65$, making it an invaluable resource across the entire sky, including low-latitude regions near the Galactic plane. Unlike the Gaia mission's medium-resolution Radial Velocity Spectrometer \citep[RVS;][]{katz:23}, however, the spectral resolution of XP spectra varies between $30 < \lambda/\Delta\lambda < 100$ in BP and $70 < \lambda/\Delta\lambda < 100$ in RP, depending on the brightness of the source, the position on the detector, and the wavelength \citep{carrasco:21}, making it challenging to derive accurate stellar parameters using traditional absorption-line analysis methods. Nonetheless, XP spectra essentially provide approximately ten times higher resolution than multiwavelength broadband photometry, allowing for the precise estimation of fundamental stellar parameters.

In this study, we utilize the XP data to investigate the foreground reddening of stars, by taking advantage of its finely sampled flux measurements across a broad wavelength range. To accomplish this, it is essential to establish a well-calibrated relationship that links stellar spectral energy distributions with stellar parameters. To address this need, we employ a suite of empirically calibrated isochrones in this study, as developed in our previous paper of this series. Specifically, in Paper~IV, we utilized a comprehensive dataset consisting of precise observations of main-sequence (MS) stars in Galactic clusters and those with spectroscopic measurements, spanning a broad range of $\teff$ and [Fe/H], which facilitated the development of an empirical calibration technique for theoretical synthetic spectra. Notably, our calibration incorporated data from numerous photometric passbands obtained from various photometric surveys, which was a crucial step toward creating accurate spectral templates over a wide range of wavelengths.

Over the past year, a number of studies have also attempted to determine stellar parameters from XP spectra. \citet{andrae:23a} estimated the metallicity and foreground extinction for around 500 million sources with XP spectra using the General Stellar Parameterizer from Photometry (GSP-Phot). However, the authors acknowledged biases in their metallicity estimates due to the systematic differences between the purely theoretical models they have adopted and the observed data. This situation has been improved in a subsequent work \citep{andrae:23b}, by using a machine-learning algorithm to train empirical models based on spectroscopic data from the Apache Point Observatory Galactic Evolution Experiment \citep[APOGEE;][]{apogee} in the Sloan Digital Sky Survey \citep[SDSS;][]{sdss:dr17}, augmented by spectra for a list of metal-poor stars \citep{li:22}.

In parallel to \citet{andrae:23b}, \citet{zhang:23b} developed an empirical forward model connecting the observed XP spectra to stellar parameters, determined through medium-resolution spectroscopy from the Large Sky Area Multi-Object Fiber Spectroscopic Telescope \citep[LAMOST;][]{cui:12}. By incorporating near-infrared photometry from the Two Micron All Sky Survey \citep[2MASS;][]{2mass} and the Wide-field Infrared Survey Explorer \citep[WISE;][]{wise} and employing Bayesian statistics, they constrained astrophysical parameters ($\teff$, [Fe/H], distance, and extinction) for the majority of stars with XP spectra, achieving uncertainties of $\sim0.15$~dex in [Fe/H] and $\sim0.03$~mag in $\ebv$. Unlike the approach adopted in \citet{andrae:23a}, who derived astrophysical parameters based on purely theoretical models, both \citet{andrae:23b} and \citet{zhang:23b} relied solely on empirical modeling to relate these parameters to observables (XP spectra). This methodology is similar to earlier studies that have utilized deterministic models \citep[e.g.,][]{ivezic:08} or machine-learning techniques \citep[e.g.,][]{whitten:21}.

In this work, we aim to improve the accuracy of extinction estimates by using our empirically calibrated set of theoretical models. Our hybrid approach distinguishes itself from the aforementioned studies in that it does not exclusively rely on either empirical or theoretical relationships to establish connections between observables and physical parameters. In Paper~IV, we made use of precisely known astrophysical parameters obtained from star clusters and individual stars with spectroscopic measurements, similar to the approach employed by \citet{andrae:23b} or \citet{zhang:23b}. However, in contrast to these approaches, we employed them to empirically calibrate theoretical models, thereby allowing us to account for the inherent systematic errors in these models while still benefiting from their predictive powers. Given that the relationship linking observables to physical parameters serves as the primary source of systematic errors in parameter estimation, our calibrated models should help to mitigate such errors by combining the strengths of both empirical and theoretical modeling.

The structure of this paper is organized as follows. In Section~\ref{sec:sample}, we provide a concise overview of XP spectra and present our selection of the XP MS sample. Section~\ref{sec:method} describes our methodology for estimating stellar parameters using a set of four alternative models, allowing us to assess systematic errors in our approach. Additionally, we offer a quality assessment of the XP spectra through the application of best-fit models to individual samples. In Section~\ref{sec:cube}, we construct 3D models of foreground extinction while assuming a constant total-to-selective extinction ratio ($R_V\equiv A_V/\ebv$), and compare these models to previously published work in the literature, as discussed in Section~\ref{sec:comp}. In Section~\ref{sec:rv}, we explore variations in the line-of-sight $R_V$ values across the sky by allowing $R_V$ to vary during our parameter search. Finally, we summarize our findings and discuss the limitations of the current work in Section~\ref{sec:summary}. Additionally, the appendices include a validation of our technique and instructions for accessing our data products.

\section{Gaia XP Spectra}\label{sec:sample}

We obtained time-averaged mean XP spectra from the Gaia DR3 archive, which are provided in the form of a continuous representation in basis functions\footnote{\tt http://cdn.gea.esac.esa.int/Gaia/gdr3/Spectroscopy/}. To calibrate the raw (internally calibrated, continuously represented) mean spectra, we employed a Python library suite, GaiaXPy \citep{deangeli:23}\footnote{\tt https://gaia-dpci.github.io/GaiaXPy-website/}. See \citet{montegriffo:23} for more information on the external calibration of the XP spectra. We binned the spectra on a grid of [336, 1011]~nm with a sampling interval of $9$~nm, which is wider than the default case ($2$~nm) to ensure efficient processing while retaining sufficient information for our parameter estimates.

We matched the XP sources to the main Gaia DR3 catalog using the Gaia source identifier ({\tt source\_id}). Our analysis focuses specifically on MS stars, as our empirically calibrated isochrones are currently only applicable to the parameter space covered by these stars. To ensure that our parameter estimates are not sensitive to our age assumptions in the underlying models, we restricted our analysis to objects with $3.5 < M_G < 7.5$
and $0.4 < G_{\rm BP}\, -\, G_{\rm RP} < 2.0$, where $M_G$ is computed using Gaia parallaxes. In this initial selection of the sample, we adopted foreground extinction from our 3D extinction map in an iterative manner, with extinction coefficients in Gaia passbands provided by \citet{casagrande:18}. Only those having parallax uncertainties less than $20\%$ were kept, but the majority of the objects in the XP catalog are sufficiently bright to have reliable parallax measurements. The above selection reduced the total number of XP sources in our sample from $220$ million to $\sim80$ million.

\section{Stellar Parameter Estimation}\label{sec:method}

\subsection{Strategy}

\begin{deluxetable*}{lll}
\tablecaption{A Family of Solutions on Stellar Parameters\label{tab:tab1}}
%\tabletypesize{\scriptsize}
\tablehead{
   \colhead{Designation} &
   \colhead{Distance Prior} &
   \colhead{Empirical Model Calibration in Paper~IV}
}
\startdata
Case A & Gaia DR3 & Stellar sequences + individual spectroscopic samples \\
Case B & Gaia DR3 & Stellar sequences \\
Case C & \nodata & Stellar sequences + individual spectroscopic samples \\
Case D & \nodata & Stellar sequences \\
\enddata
\end{deluxetable*}

As outlined in Table~\ref{tab:tab1} and further explained below, we have generated four distinct sets of parameter estimates in this study. They were derived from two sets of empirical calibrations, taking into account whether they were computed with or without the constraint from Gaia parallaxes. We refer to each of these sets as Cases~A--D, respectively. While Case~A and Case~B are preferred, due to the additional constraints from Gaia parallaxes, we have included Case~C and Case~D to explore systematic uncertainties in our parameter estimates by analyzing the differences among these solutions. We also note that both Case~A and Case~B are considered equally valid; therefore, we take the average of these two solutions to derive our reddening map, while considering their differences as a realistic measure of systematic errors in our models.

\subsection{Recap: Empirically Calibrated Isochrones}

In this series of papers, we derived stellar parameters, including mass ($M_*$) and [Fe/H], of a star by utilizing stellar isochrones with empirical corrections (see below). Given our primary focus on the lower MS, $M_*$ is directly correlated with $\teff$. Stellar isochrones play a crucial role in our methodology, as they not only establish a fundamental $M_*$-luminosity relation for a given age and chemical abundances but also provide essential information on stellar radii necessary for placing synthetic spectra onto the absolute flux scale. In instances where individual spectroscopic measurements of $\teff$ and surface gravity ($\log{g}$) for cluster stars in our calibration sample were unavailable, we relied on absolute magnitudes (luminosities) of the isochrones to infer these parameters. These isochrones were also employed to construct calibration samples from diverse spectroscopic surveys, all tied to the same $\teff$ scale.

We continued to employ YREC evolutionary models \citep{sills:00}, each intricately linked to a synthetic spectrum generated by the MARCS atmosphere models \citep{gustafsson:08}. Our isochrones cover the range $-3 \leq {\rm [Fe/H]} \leq +0.4$, and for this range we assumed a monotonic relationship between [Fe/H] and $\alpha$-element abundance ([$\alpha$/Fe]): ([Fe/H], [$\alpha$/Fe])$=(-3.0, 0.4)$, $(-2.0, 0.3)$, $(-1.0, 0.3)$, $(-0.75, 0.25)$, $(-0.5, 0.2)$, $(-0.3, 0.0)$, $(0.4, 0.0)$. Furthermore, our models incorporate a linear relationship between age and metallicity, as described by ([Fe/H], age)$=(-3.0, 13\ {\rm Gyr})$, $(-1.2, 13\ {\rm Gyr})$, $(-0.3, 4\ {\rm Gyr})$, $(0.4, 4\ {\rm Gyr})$.

We adopted the linear relationship between [$\alpha$/Fe] and [Fe/H] owing to the predominant influence of $\teff$ and overall metallicity (represented by [Fe/H]) on stellar continuum flux, with the impact of surface gravity being relatively modest. In contrast, the contribution of [$\alpha$/Fe] is minimal. From an observational standpoint, although modern spectroscopic surveys offer valuable [$\alpha$/Fe] information for a large number of field stars, the challenge arises from the limited number of benchmark clusters in our calibration sample with known [$\alpha$/Fe], making it challenging to precisely determine the dependence of stellar flux on [$\alpha$/Fe]. As described above, we have also opted for a simple linear relationship of stellar age with [Fe/H] because of the relative absence of age information for field stars, both in the calibration and when applying the models to observational databases. To mitigate the influence of our assumed age, especially near the MS turnoff, we focused our analysis on the lower MS by restricting the sample to $M_G > 3.5$~mag. Some stars near the MS turnoff could be excluded owing to poor model fits resulting from inaccurate ages, leading to a reduced number of stars in our final sample.

Our empirical corrections were formulated based on our working hypothesis that YREC interior models predict accurate relations among luminosity, $\teff$, and $M_*$, while deviations observed from the models are entirely attributed to how fluxes are computed in the stellar atmosphere. In our earlier calibration exercise \citep[e.g.,][]{an:13}, the empirical correction procedure involved aligning our theoretical isochrones with the observed sequences of both globular and open clusters in broadband photometry. This was achieved by defining a table of magnitude corrections in each filter as a function of $\teff$ and [Fe/H]. The resulting set of empirically corrected isochrones has been extensively utilized in Papers~I--III. In Paper~IV, we extended the calibration by directly adjusting synthetic spectra, rather than making corrections on an individual filter basis. Additionally, we expanded our calibration sample by incorporating extensive spectroscopic observations of individual stars from the Sloan Extension for Galactic Understanding and Exploration \citep[SEGUE;][]{yanny:09,rockosi:22}, APOGEE, and the Galactic Archaeology with HERMES \citep[GALAH;][]{buder:21}.

In this study, we employed two sets of isochrones, each featuring distinct empirical correction schemes. The first set, introduced in Paper~IV, is based on calibrations using both cluster sequences and individual stars with spectroscopic measurements. The second set relies exclusively on the cluster sequences, newly constructed as part of the current work, following the correction procedure outlined in Paper~IV. Corrections for both sets are defined within the ranges $-3 \leq {\rm [Fe/H]} \leq +0.4$ and $4000$~K $\leq \teff \leq 7000$~K. However, the incorporation of spectroscopic samples into the correction procedure depends on our assumed age (see above), leading to systematic differences in the derived [Fe/H] for hot stars (see Figure~4 in Paper~IV). Inconsistent metallicity scales among diverse spectroscopic surveys and the cluster data pose additional challenges. Nevertheless, we regard the spectroscopic samples as valuable resources for extending our calibration, and we incorporated both model sets in the subsequent analysis to address potential systematic uncertainties in our methodology.

\subsection{Parameter Estimation}

We obtained the optimal parameter set, including $\{$[Fe/H], $M_*$, $\ebv$, and/or distance modulus $\dmn$, and/or $R_V \}$, for each XP spectrum utilizing empirically corrected isochrones. To match the resolution of the input XP spectra (see \S~\ref{sec:sample}), we adjusted synthetic spectra accordingly. Given [Fe/H] and stellar mass, we calculated and minimized the total $\chi^2$ of a fit employing the MPFIT package \citep{markwardt:09}, for a robust nonlinear least-squares curve fitting. This process yielded the best-fitting foreground reddening, and/or $R_V$, and/or $\dmn$. In our base case (\S~\ref{sec:cube}), we employed the average Galactic extinction curve from \citet{fitzpatrick:99}, corresponding to the standard value of $R_V \equiv A_V / \ebv = 3.1$, where $A_V$ represents the foreground extinction in the $V$ band. Conversely, in scenarios allowing for the variation of $R_V$ in our parameter estimation (\S~\ref{sec:rv}), we adopted the $R_V$-dependent extinction curve model from \citet{fitzpatrick:19}.

Besides the two different methods for calibrating isochrones, there are two distinct choices regarding whether Gaia parallaxes should be incorporated into the fitting process as a prior. When our sample includes nearby stars with reliable parallaxes, incorporating the Gaia prior results in a more stringent constraint on parameter estimates compared to the other case. On the other hand, the alternative approach, independently determining a stellar distance through the fitting process, can produce a more internally consistent set of parameters, notwithstanding potential systematic errors in the models. As outlined in Table~\ref{tab:tab1}, we explore systematic uncertainties in our derived parameters by examining all four families of solutions.

For each XP spectrum, we calculated the $\chi^2$ distribution within a model grid, with grid intervals set at $\Delta {\rm [Fe/H]} = 0.05$~dex and $\Delta M_* = 0.02\ M_\odot$. The process involved fitting a paraboloid to the $\chi^2$ distribution to locate the minimum $\chi^2$, thereby determining the best-fitting set of parameters for a star. To gauge parameter uncertainties, we considered $\Delta \chi^2 = 3.53$ and $4.72$ from the minimum $\chi^2$ for $3$ and $4$ degrees of freedom, respectively. When incorporating Gaia parallaxes, we propagated parallax uncertainties and combined them with the estimated uncertainties.

\begin{figure*}
\center
\epsscale{1.1}
\plottwo{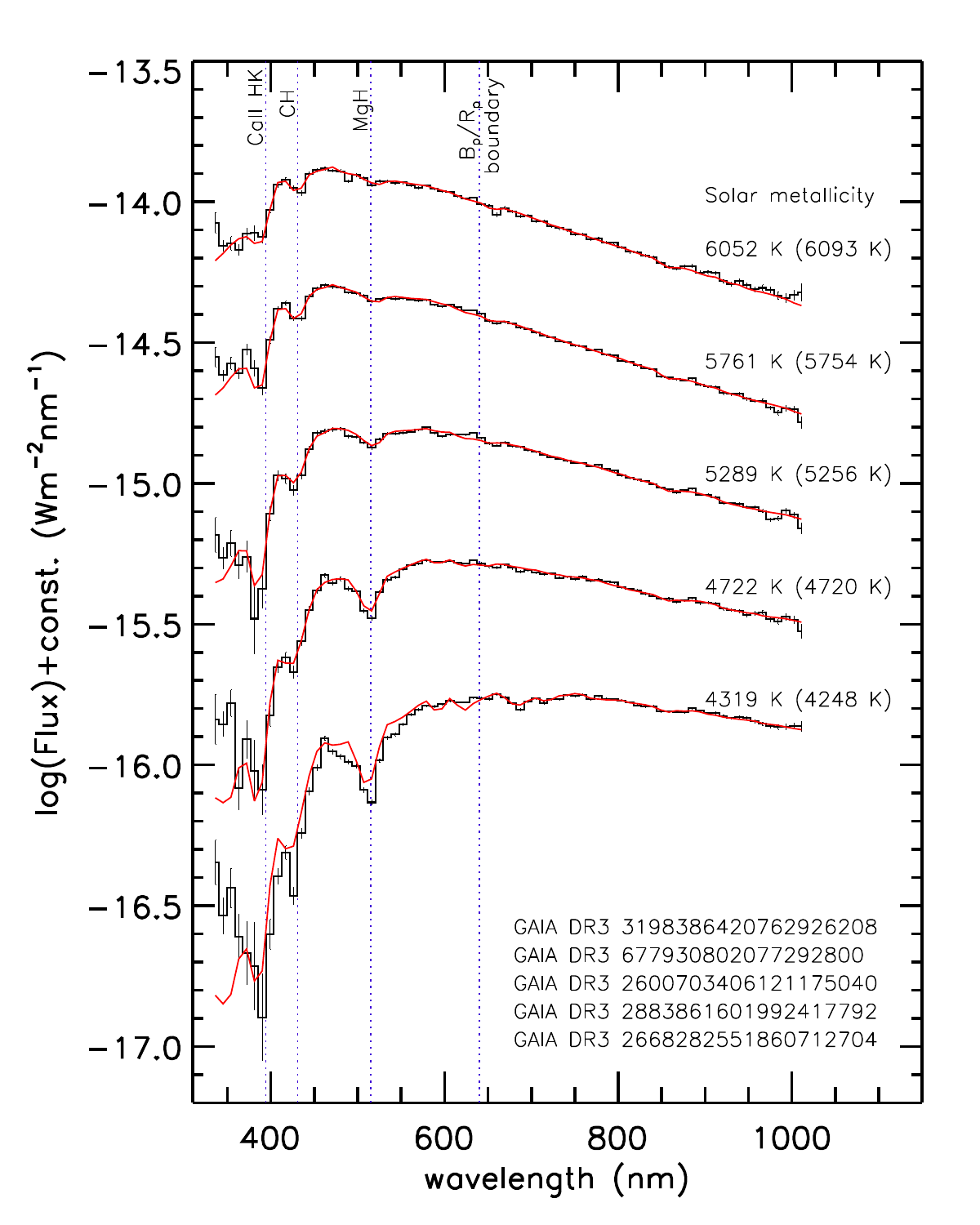}{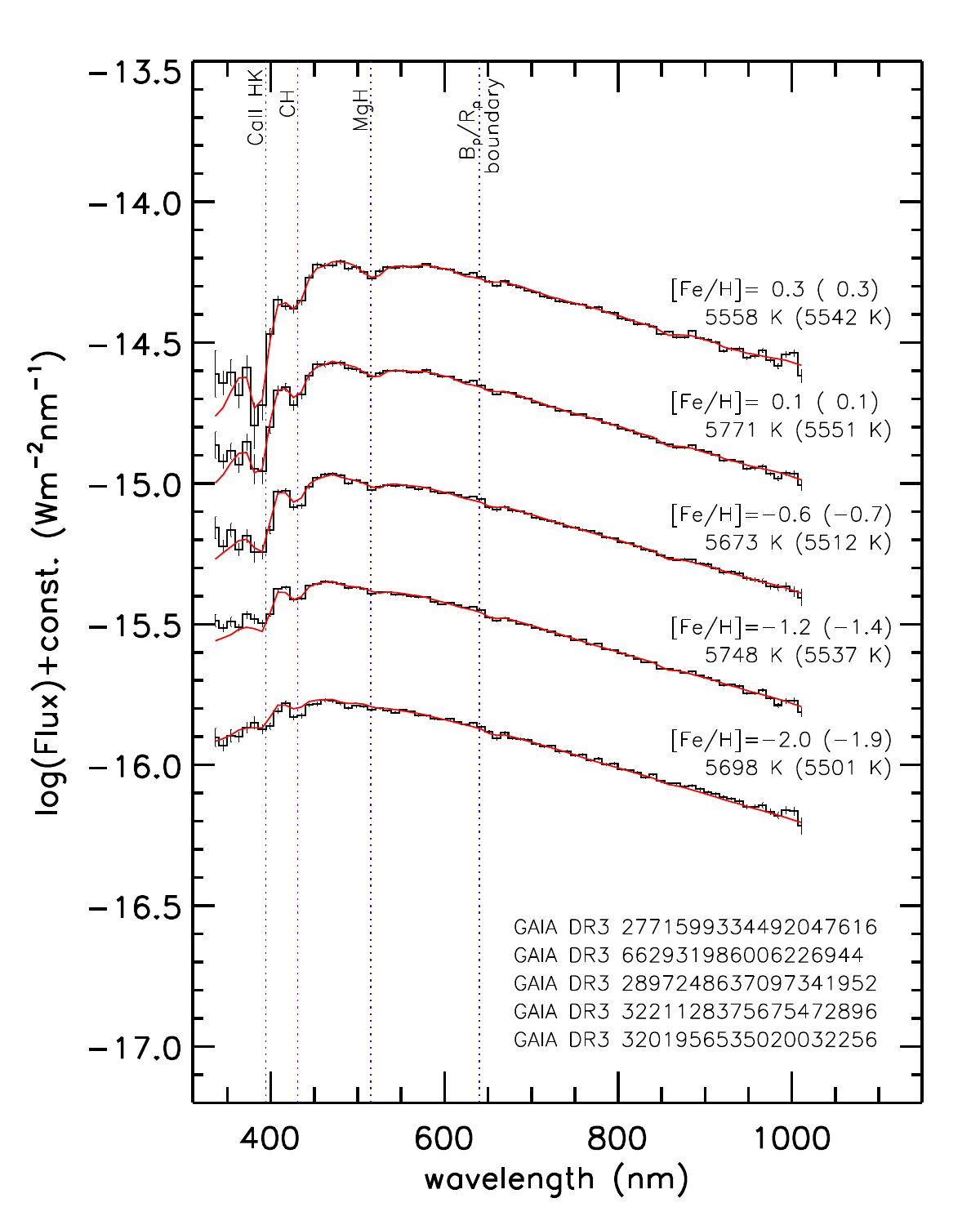}
\caption{Model fits illustrating comparisons for various effective temperatures ($\teff$, left) and metallicities ([Fe/H]; right). Best-fitting estimates for $\teff$ and [Fe/H] are shown, with high-resolution spectroscopic estimates from GALAH indicated in parentheses. The left panel depicts all cases at solar metallicity.}
\label{fig:fit}
\end{figure*}

Figure~\ref{fig:fit} illustrates our model fits to XP spectra for a subset of stars characterized by minimal foreground reddening, $\ebv < 0.05$~mag. The left panel illustrates model comparisons (based on Case~A) as a function of $\teff$, while the right panel shows the case as a function of [Fe/H]. Although individual absorption lines are not distinctly visible in the XP spectra, molecular absorption bands from CH ($430$~nm) and MgH ($515$~nm) become evident in low-$\teff$ or high-[Fe/H] stars. Notably, strong absorption is clearly observed at wavelengths shorter than $400$~nm, primarily driven by the \ion{Ca}{2} H and K lines ($395$~nm). These features, along with the overall spectral shape, exhibit variations with $\teff$ and [Fe/H], providing valuable constraints in our model fitting. Nonetheless, although our models exhibit a satisfactory overall fit above $360$~nm, systematic deviations emerge at shorter wavelengths. In the following section, we delve into a comprehensive discussion of these deviations, attributing them to inherent characteristics of the XP spectra. It is important to note, however, that these discrepancies do not markedly impact the overall fit, given the large uncertainties associated with the affected pixels.

As our study primarily aims to derive the {\it average} reddening structure along each line of sight, we fully utilized our estimated uncertainties to effectively assign weights to each $\ebv$ measurement. Hence, we applied a set of moderate-quality criteria to exclude only instances of stochastic failures, requiring that $\chi^2_{\rm min}/\nu < 5$, $\sigma(\ebv) < 0.3$~mag and $\sigma(\rm [Fe/H]) < 1.5$~dex, where $\nu$ represents the number of degrees of freedom. We retained stars with metallicities falling within our model grid ($-3 < {\rm [Fe/H]} < 0.4$). Furthermore, we required that the distance moduli obtained from our pure spectrophotometric approach (Case~C or Case~D) are within $1$~mag compared to those derived from Gaia parallaxes. This criterion was applied to exclude objects that are less likely to belong to the MS. Alongside the initial selection of the XP MS sample, as detailed in \S~\ref{sec:sample}, these requirements served as the basis for our sample selection, resulting in $2.6\times10^7$, $2.3\times10^7$, $2.8\times10^7$, and $2.5\times10^7$ stars in our final sample from Case~A through Case~D, respectively.

Our XP sample may also include a significant number of unresolved binaries and/or blends, which exhibit systematically higher luminosities and/or redder colors than single stars. For details on the impact of such binaries on photometric metallicity estimates, we refer interested readers to \citet{an:13}. In the current study, we examined the potential bias by selecting stars from the XP sample associated with two open clusters, M67 and NGC~2516, both known to contain a noticeable number of unresolved binaries in the $G_{\rm BP}\, -\, G_{\rm RP}$ color-magnitude diagram. Our analysis revealed that their $\ebv$ values are consistently higher than those of resolved stars. When Gaia parallaxes are employed in the parameter estimation (Case~A and Case~B), these binaries exhibit systematically larger $\ebv$ values, with differences of up to $0.15$~mag, while the purely spectrophotometric solutions (Case~C and Case~D) show differences of less than $\Delta\ebv\sim0.05$~mag. The greater bias found in the former cases arises from the inherent discrepancy between the actual (observed) and expected (modeled) luminosities of the binaries. Nonetheless, the collective influence of binary populations in the sample is anticipated to be modest: When considering the ensemble of objects (both single and binary populations) in these clusters, the overall bias ranges from $0.005$ to $0.04$~mag in Case~A and Case~B and remains below $0.01$~mag in Case~C and Case~D.

\subsection{Systematic Deviations and Uncertainties in XP Spectra}

\begin{figure}
\centering
\epsscale{1.15}
\plotone{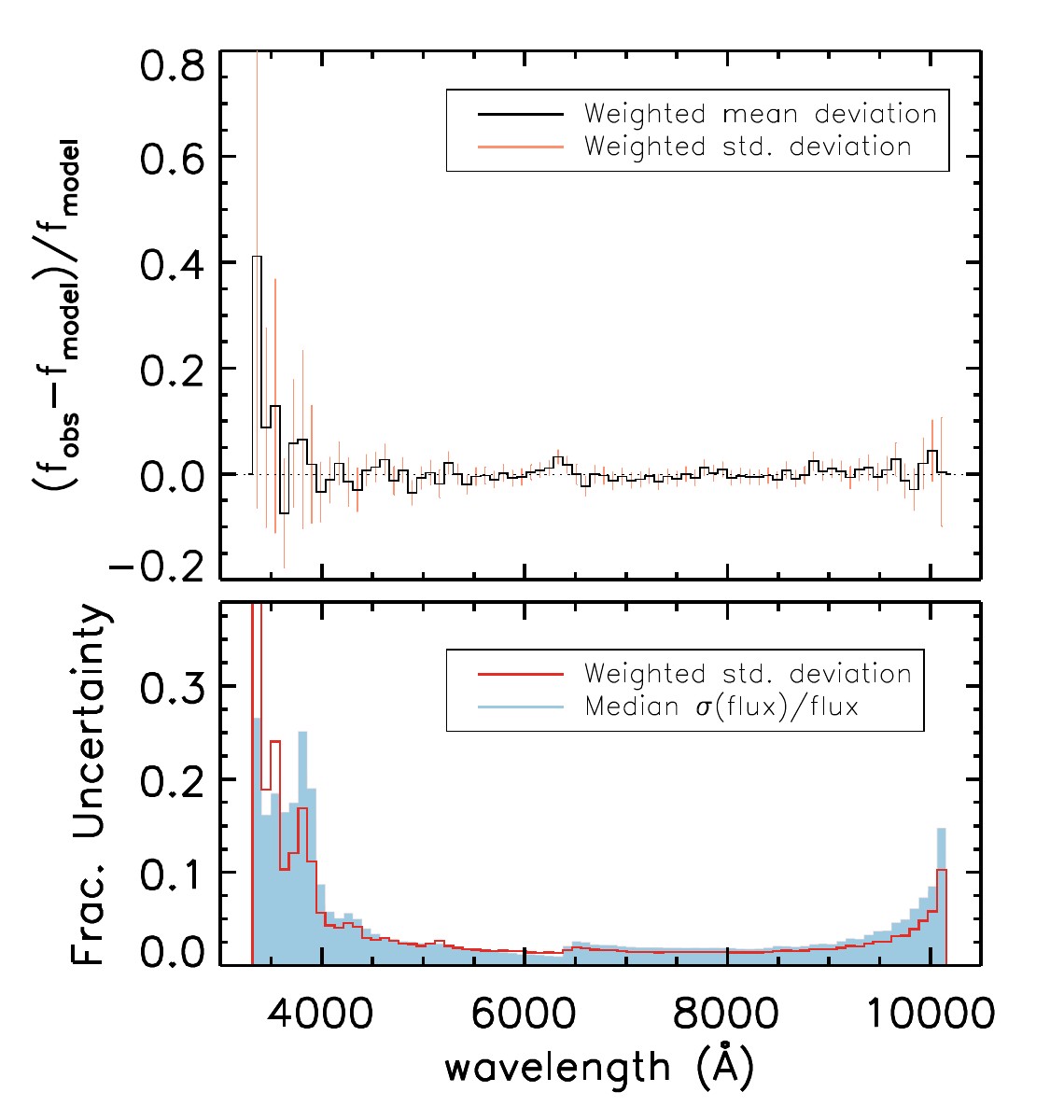}
\caption{Model deviations and uncertainties in XP MS Sample. Top panel: mean model deviations of the XP MS sample as a function of wavelength. Weighted mean differences from all stars with $\ebv < 0.1$ are shown by a histogram, along with weighted standard deviations displayed by error bars. Bottom panel: weighted standard deviations (error bars in the top panel), overlaid on top of median fractional uncertainties in the XP spectra.}
\label{fig:error}
\end{figure}

In the top panel of Figure~\ref{fig:error}, the mean model deviations of the XP MS sample as a function of wavelength are shown by a histogram. It illustrates the weighted mean differences of the best-match models from Case~A for all stars with $\ebv < 0.1$~mag. Error bars indicate the weighted standard deviations of the differences, but uncertainties in the mean differences are too small to display, due to the large number of stars in the sample. Our best-match models are within $3\%$ of the XP spectra at $400$~nm $< \lambda < 1000$~nm, with an rms difference of only $1.4\%$. The mild bump at $\sim640$~nm is associated with the boundary between the BP and RP spectra. However, the mean model deviation gradually increases toward shorter wavelengths, reaching $40\%$ at the blue edge of the spectra ($336$~nm). The large systematic deviation is accompanied by a large scatter of the differences, as shown by the error bars.

The systematic errors of the XP spectra evident in the top panel of Figure~\ref{fig:error} align with earlier validation work conducted by \citet{montegriffo:23}. Their validation process incorporated a set of high-precision spectra from $111$ stars in the Gaia spectrophotometric standard stars survey \citep{pancino:12}, with an additional $60$ stars from the `passband validation library' \citep{pancino:21}. Furthermore, they made use of a dataset consisting of $348$ stars from the Hubble Space Telescope's Next Generation Spectral Library \citep[NGSL;][]{heap:16}. For wavelengths above $370$~nm, the first two datasets displayed excellent agreement with the XP data, with their smoothed median differences falling within $3\%$ (see their Figure~23). Importantly, their work unveiled an intriguing comparison using the NGSL, which was not part of the external calibration for the XP spectra, unlike the other two datasets. In the range [400, 1000]~nm, the smoothed median difference from the NGSL was similar to the other two. However, at wavelengths below $400$~nm, the deviation from the NGSL data exhibited a sharp increase, reaching $\sim10\%$ at $350$~nm---a value that escalates with higher $G_{\rm BP}\, -\, G_{\rm RP}$ colors. The average deviation observed in the NGSL sample closely mirrors our findings obtained through the calibrated models. Therefore, it becomes evident that the substantial systematic deviation at shorter wavelengths primarily arises from errors inherent in the XP spectra themselves, rather than being a consequence of inaccurate isochrone models.

In the bottom panel of Figure~\ref{fig:error}, we compare the weighted standard deviations derived in our study (red solid histogram) with the median flux uncertainties extracted from the XP spectra (blue shaded histogram). It is noteworthy that, within the range [400, 1000]~nm, the above two uncertainty estimates closely align with each other, suggesting that our parameter search is reasonably accurate, given the flux uncertainties in the XP spectra. Below $360$~nm, however, the observed scatter surpasses the estimated uncertainties, which may indicate an underestimation of flux uncertainties in the XP spectra.

\begin{deluxetable}{cccc}
\tablecaption{Mean Model Deviations\label{tab:tab2}}
\tabletypesize{\footnotesize}
\tablewidth{0pt}
\tablehead{
   \colhead{Wavelength} &
   \multicolumn{2}{c}{Fractional Differences\tablenotemark{\scriptsize a}} &
   \colhead{Median Frac.} \\
   \cline{2-3}
   \colhead{(nm)} & 
   \colhead{Weighted Mean} &
   \colhead{Weighted Std.} &
   \colhead{Uncertainty\tablenotemark{\scriptsize b}}
   }
\startdata
     336 &  0.412 & 0.476 & 0.266 \\
     345 &  0.087 & 0.189 & 0.162 \\
     354 &  0.128 & 0.240 & 0.185 \\
     363 & -0.075 & 0.103 & 0.165 \\
     372 &  0.058 & 0.121 & 0.175 \\
     381 &  0.065 & 0.169 & 0.251 \\
     390 &  0.018 & 0.112 & 0.190 \\
     399 & -0.034 & 0.056 & 0.087 \\
     408 & -0.011 & 0.043 & 0.058 \\
\enddata
\tablecomments{Only a portion of this table is shown here to demonstrate its form and content. A machine-readable version of the full table is available.}
\tablenotetext{a}{Fractional differences of the flux, $(f_{\rm obs} - f_{\rm model})/f_{\rm model}$, as shown in the top panel of Figure~\ref{fig:error}, where $f_{\rm obs}$ and $f_{\rm model}$ represent the monochromatic flux in the XP spectra and the best-fitting model, respectively.}
\tablenotetext{b}{Median fractional uncertainties calculated using the cataloged values of flux uncertainties in the XP spectra.}
\end{deluxetable}

A machine-readable version of the data presented in Figure~\ref{fig:error} is available in Table~\ref{tab:tab2}. In the following analysis, we employed XP data spanning the wavelength range of [336, 1011]~nm. Despite the presence of significant systematic errors in the blue and red edges of the XP spectra, alongside underestimated flux uncertainties at these wavelengths, the substantial flux uncertainties still underweight data below $400$~nm and above $1000$~nm, respectively. We provide further evidence of this insensitivity in Appendix~\ref{sec:comp2}, where we demonstrate that including UV data ($\lambda<400$~nm) has minimal impact on our parameter search, based on comparisons to spectroscopic parameter estimates from GALAH and APOGEE.

\section{3D Reddening Cube}\label{sec:cube}

\subsection{Multiresolution HEALPix Scheme}

\begin{figure}
\center
  \gridline{\fig{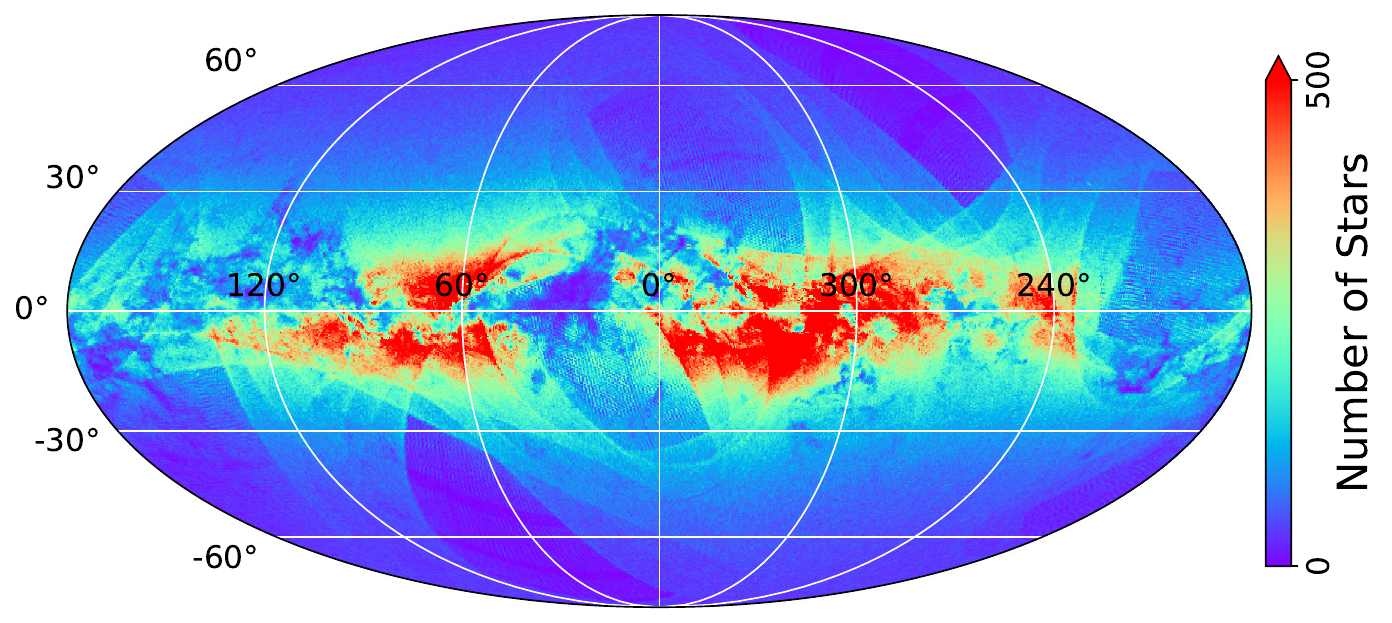}{0.48\textwidth}{\textbf{(a) Number density of the XP MS sample }}}
  \gridline{\fig{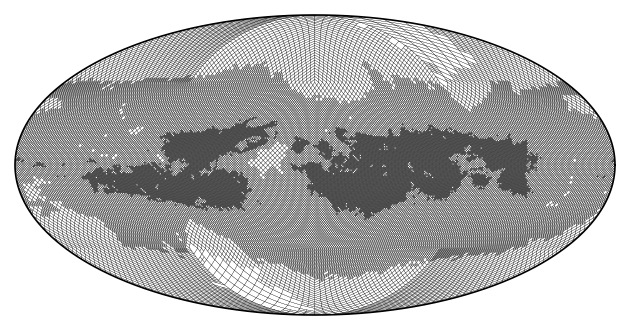}{0.44\textwidth}{\textbf{(b) Multiresolution HEALPix grid }}}
  \gridline{\fig{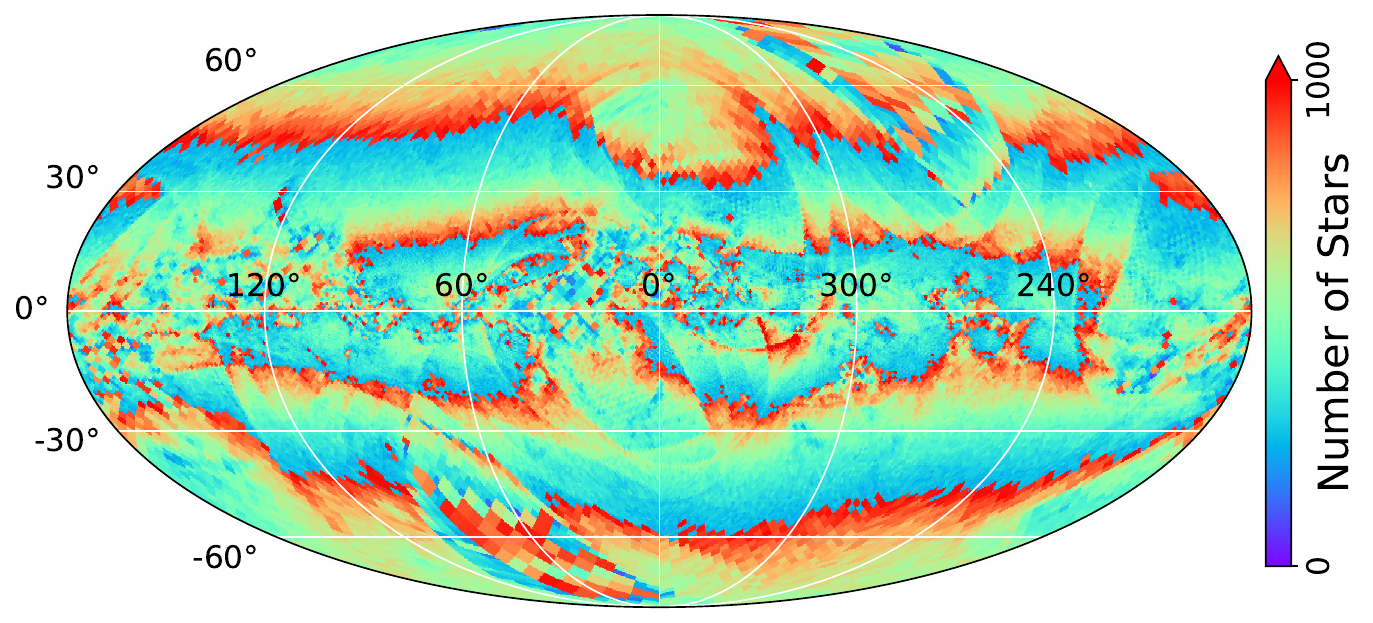}{0.48\textwidth}{\textbf{(c) Number of stars in the multiresolution grid }}}
\caption{Multiresolution HEALPix scheme. (a) The number density of the XP MS sample in the Galactic coordinate system (Case~A). The number of stars per $0.2$~deg$^2$ area is shown. (b) A multiresolution HEALPix map created using the data shown in panel~(a), in the same coordinate system. The map is composed of four distinct sets of HEALPix cells, each with varying surface areas: $0.2$~deg$^2$ ($28,152$ cells), $0.8$~deg$^2$ ($24,802$), $3.4$~deg$^2$ ($3872$), and $13.4$~deg$^2$ ($114$). (c) Distribution of the number of stars per cell, derived from the multiresolution grid scheme shown in panel~(b).}
\label{fig:num}
\end{figure}

Panel~(a) of Figure~\ref{fig:num} presents the number density distribution of the XP MS sample in the Galactic coordinate system. For this visualization, we employed the Hierarchical Equal Area isoLatitude Pixelization (HEALPix) scheme \citep{gorski:05}, which utilizes diamond-shaped cells, ensuring equal surface area coverage on the celestial sphere. The number of cells and their respective areas are determined by a single parameter, denoted as $N_{\rm side}$. In this illustration, we selected $N_{\rm side}=128$, resulting in a pixel area equivalent to $0.2$~deg$^2$. Given that our analysis is confined to MS dwarfs, our parameter estimation is limited to approximately $3$~kpc from the Sun (see below), encompassing a substantial number of stars near the Galactic plane.

\begin{figure*}
\centering
\includegraphics[scale=0.58]{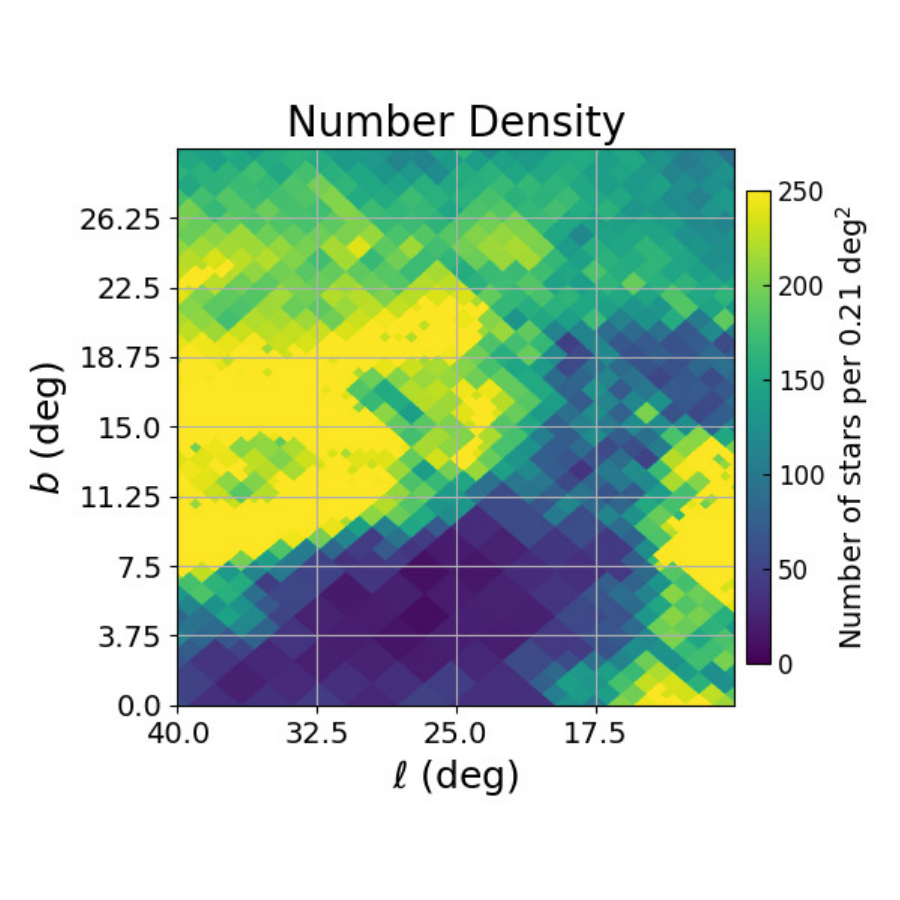}
\includegraphics[scale=0.58]{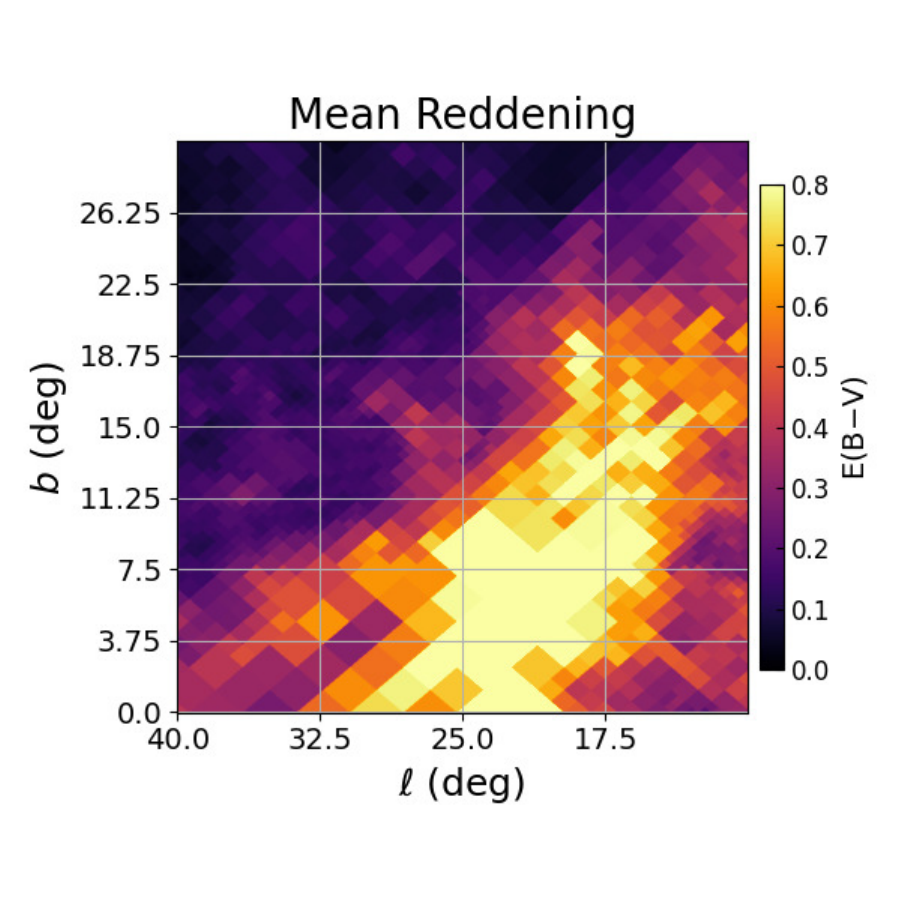}
\caption{Example of the multiresolution HEALPix scheme. Left panel: number density distribution of the XP MS sample in the Galactic coordinate system in a patch of the sky. The number of stars per unit HEALPix cell with an area of $0.2$~deg$^2$ ($N_{\rm side}=128$) is depicted. Right panel: mean reddening for the same region of the sky as in the left panel. We note that the region with high foreground extinction in the right panel (e.g., $(l,b)\sim(20\arcdeg,5\arcdeg$)) is underrepresented by the XP MS sample. Larger multiresolution HEALPix cells are adopted in this area to maintain a sufficient number of stars for modeling the foreground extinction distribution as a function of distance.}
\label{fig:mhealpy}
\end{figure*}

Panel~(a) of Figure~\ref{fig:num} reveals a nonuniform number density distribution along the Galactic plane, primarily due to substantial, patchy foreground extinction in the magnitude-limited star catalog. A more detailed illustration of this issue is provided in Figure~\ref{fig:mhealpy}. It displays the number density (left panel) and foreground reddening (right panel) in a $30\arcdeg\times30\arcdeg$ patch centered around $(l,b)\sim(25\arcdeg,15\arcdeg$), which includes a high-contrast strip of dense clouds near the Aquila Rift. As evidenced in the lower right area, regions with high extinction (right panel) correspond to a lower number density of stars from the XP MS sample (left panel), resulting in a limited number of stars within each HEALPix cell for precise modeling of the distance vs.\ reddening distribution.

To address this challenge, we employed multiresolution HEALPix maps developed by \citet{mhealpy}, to effectively redistribute and adjust the sizes of HEALPix cells, thereby ensuring a comparable number of stars within each merged HEALPix cell. We initially divided the sky into HEALPix cells with $N_{\rm side}=128$, but we adjusted the level of pixelization to ultimately arrive at HEALPix cells as large as having $N_{\rm side}=16$, by ensuring that each merged cell can accommodate a maximum of $1000$ stars. As depicted in the right panel of Figure~\ref{fig:mhealpy}, the adoption of the multiresolution HEALPix approach resulted in larger HEALPix cells in regions characterized by high foreground extinction, while still maintaining a sufficiently large number of stars.

The multiresolution mapping grid constructed in this manner is shown in panel~(b) of Figure~\ref{fig:num}. It consists of $56,940$ merged cells, of which $N_{\rm side}$ and the number of cells correspond to $N_{\rm side}$=$16$ ($114$ cells), $32$ ($3872$), $64$ ($24,802$), and $128$ ($28,152$). The surface area ranges from $0.2$ to $13.4$~deg$^2$. As illustrated in panel~(c), each multiresolution cell contains between $\sim200$ and $\sim1000$ stars, with a median count of $\sim400$ stars, providing a sufficient number for modeling the reddening structure, as demonstrated below. This multiresolution grid is employed throughout the paper unless stated otherwise.

\subsection{Modeling the Cumulative Reddening}\label{sec:model}

\begin{figure*}
\center
\epsscale{0.65}
  \gridline{\fig{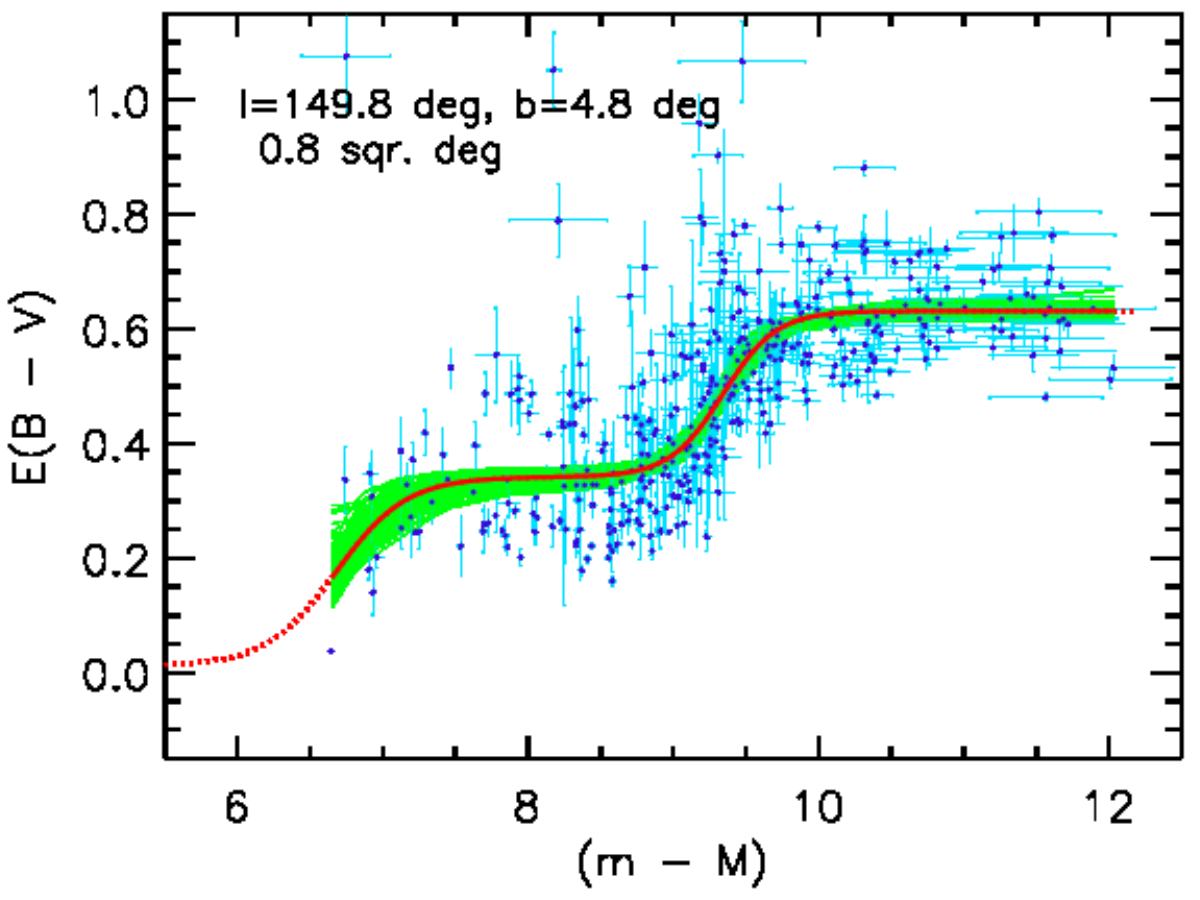}{0.42\textwidth}{\textbf{(a) Case~A }}
                \fig{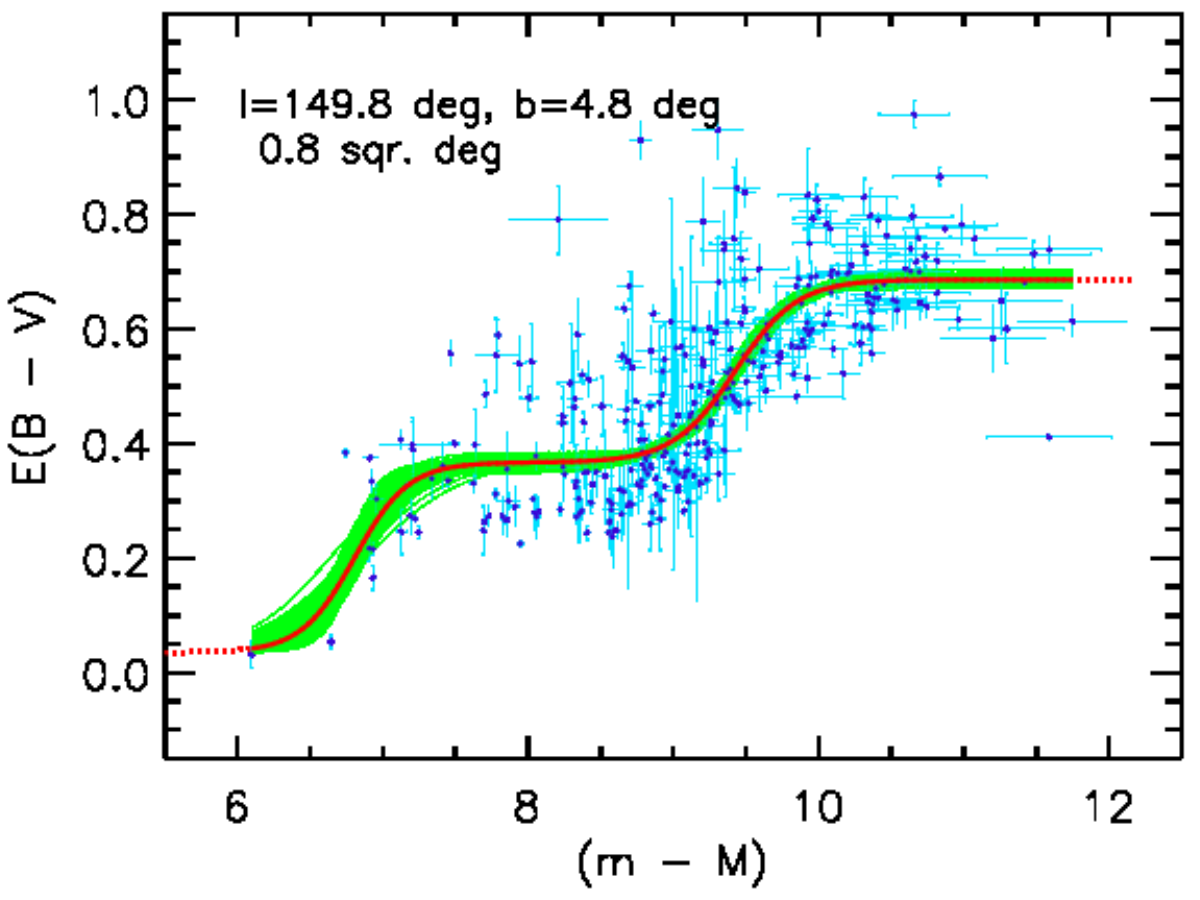}{0.42\textwidth}{\textbf{(b) Case~B }}}
  \gridline{\fig{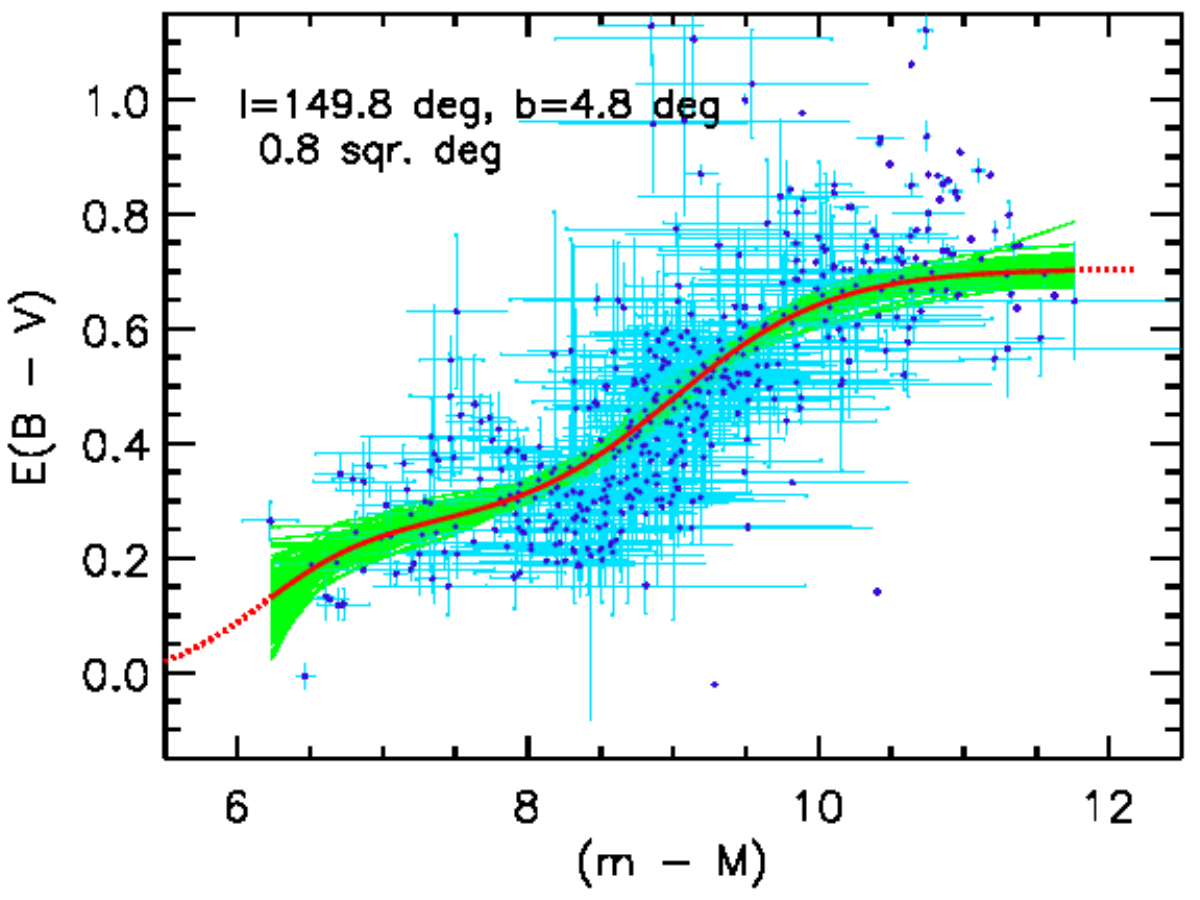}{0.42\textwidth}{\textbf{(c) Case~C}}
                \fig{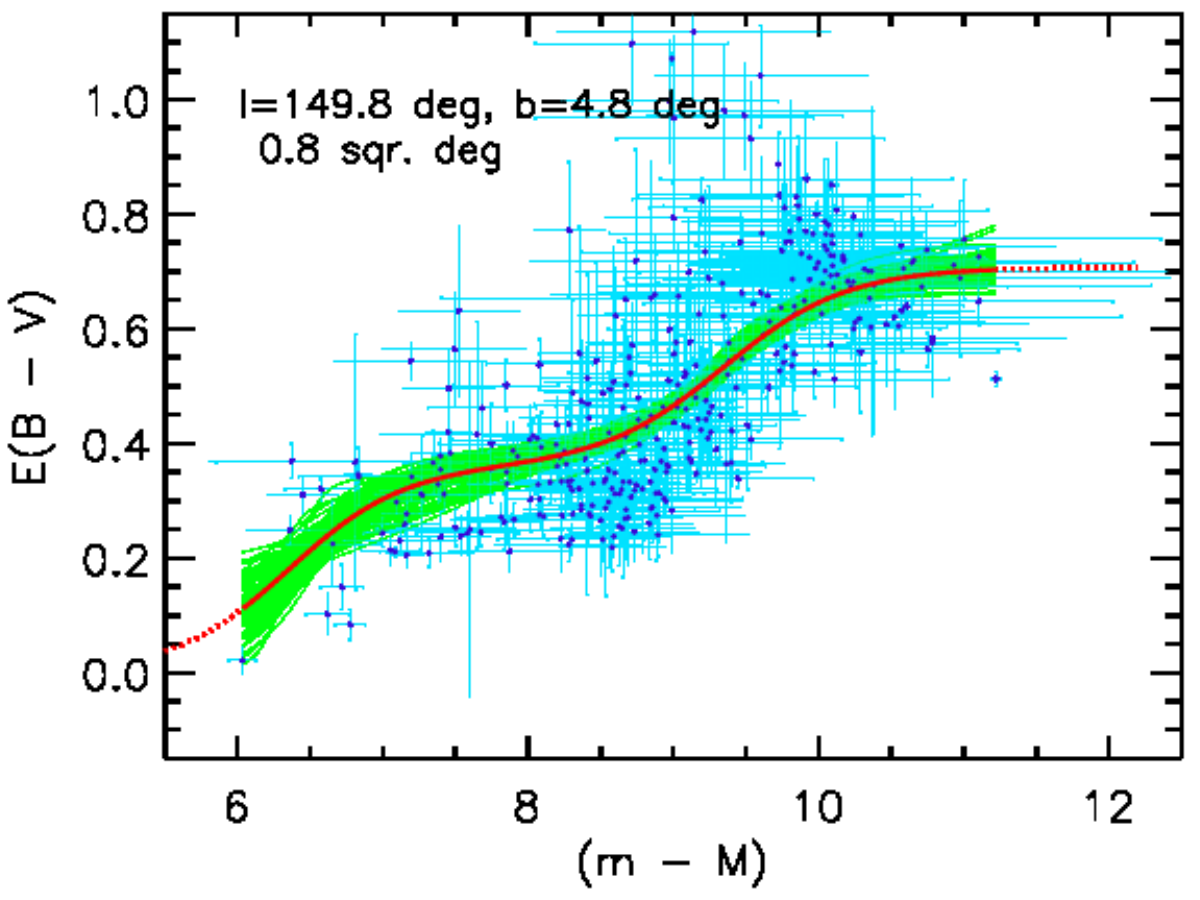}{0.42\textwidth}{\textbf{(d) Case~D }}}
\caption{Reddening of stars against distance modulus in a single, multiresolution HEALPix cell located along $l=149.8\arcdeg$ and $b=4.8\arcdeg$. The four panels display different sets of $\ebv$ measurements, Case~A through Case~D (see Table~\ref{tab:tab1}). Individual fits to the simulated data using a double logistic function in equation~(\ref{eq:1}) are illustrated by green lines, and the best-fitting dust model is depicted by a red solid line.}
\label{fig:raw}
\end{figure*}

Figure~\ref{fig:raw} illustrates the line-of-sight distribution of reddening toward $l = 149.8\arcdeg$ and $b = 4.8\arcdeg$, displaying the distribution of $\ebv$ of individual stars as a function of distance modulus. Four alternative solutions for $\ebv$ measurements are shown in separate panels (Cases~A--D). In this low-latitude example, the observed distribution in all four cases exhibits characteristic jumps in the reddening at distances of approximately $250$ and $800$~pc from the Sun ($\dmn\approx7$ and $9.5$, respectively), representing nearby dust clouds and those possibly associated with large-scale spiral arm(s), respectively.

Figure~\ref{fig:raw} also showcases systematic differences among the four alternative solutions. First, it is apparent that the uncertainties in $\ebv$ are significantly smaller in Case~A and Case~B compared to Case~C and Case~D, mainly because of the additional constraint applied in the parameter estimation. The smaller uncertainties in both axes lead to a tighter correlation in Case~A and Case~B, while in Case~C and Case~D, the distribution appears more dispersed. Despite this limitation, we note that, since metallicity and reddening are two primary factors that influence the shape of stellar spectra, the purely spectrophotometric solutions (Case~C and Case~D) enable us to obtain an essentially more internally consistent result, provided that our calibrated synthetic spectra accurately reflect the shape of stellar spectra. There are also a larger number of nearby stars in Case~C and Case~D, as opposed to the other cases. This can be attributed to the tight parallax constraints imposed on the nearest stars in Case~A and Case~B, which, in turn, could lead to the exclusion of such stars with substantial $\chi^2$ values owing to a small systematic error in the models and/or the presence of unresolved star pairs.

Secondly, Case~A and Case~C generally contain more stars than the cases using isochrones with sequence-based calibration (Case~B and Case~D) by approximately $10$--$20\%$. Furthermore, it is noteworthy in Figure~\ref{fig:raw} that the $\ebv$ estimates derived from the sequence-based solutions (Case~B and Case~D) exhibit values approximately $\Delta \ebv=0.02$~mag higher than those obtained in the other cases, highlighting the distinction between the two calibration methods. This discrepancy indicates that the difference between the two calibration approaches extends beyond merely having additional stars in the calibration, and reflects the unique information on stellar properties provided by each calibration sample. Specifically, our assumptions regarding stellar age, which are made for cluster sequences (with direct information from an MS turnoff) and individual field stars (without reliable age estimates), could have a profound impact on our calibration. This would leave bluer stars near the MS turnoff unconstrained in Case~B and Case~D, since isochrones with sequence-based calibration have a redder MS turnoff than the other cases. The two approaches also exhibit a significant discrepancy in the derived metallicities near the MS turnoff, as demonstrated in Paper~IV. Although a systematic difference between the two calibrations can be reduced by selecting stars fainter than $M_G \approx 4.5$~mag, we kept stars with $M_G > 3.5$~mag as this severely limits the available stars in the XP MS sample.

In order to construct a 3D reddening cube, we employed a logistic function that effectively captures the observed distribution of $\ebv$ on a purely empirical basis:
\begin{equation}
   \ebv=\epsilon_0+\sum_{i=1}^2 A_i (1 + e^{-k_i (\mu - \mu_i)} )^{-1}, \label{eq:1}
\end{equation}
where $\mu$ indicates the distance modulus. Here we employed a double logistic function with $i=\{1, 2\}$ to represent the observed two-step features, such as shown in Figure~\ref{fig:raw}. Both $k_i$ and $\mu_i$ jointly determine the shape of the cumulative $\ebv$ distribution, whereas $A_i$ serves as a normalization factor.

In equation~(\ref{eq:1}), $\epsilon_0$ signifies the zero-point offset in our $\ebv$ measurements. To determine the systematic offsets for each dataset, we employed $\ebv$ of stars at high Galactic latitudes ($|b|>60\arcdeg$) with minimal foreground extinction. The $\ebv$ distribution in these high-latitude regions shows systematic variations across different datasets: we computed the mode of the distribution using a bin size of $0.002$~mag and found $\ebv=\{0.011$, $0.035$, $-0.021$, and $0.001\}$~mag for Case~A through Case~D, respectively. We took these values as $\epsilon_0$ for each case and kept them constant throughout the subsequent modeling.

We fit equation~(\ref{eq:1}) to the observed distribution of $\ebv$ against $\dmn$ along each line of sight utilizing the MPFIT package. To enhance the constraints on models involving a limited number of stars, we used $\ebv$ measurements from $10^4$ stars in the surrounding area to obtain shape parameters $\{k_1, \mu_1, k_2, \mu_2 \}$ in equation~(\ref{eq:1}), assuming that the structural properties of dust clouds remain constant over a moderate angular scale (typically with a radius of $1\arcdeg$--$3\arcdeg$). For these stars, we binned the $\dmn$ vs.\ $\ebv$ data using bin sizes of $0.10$ and $0.05$~mag for $\dmn$ and $\ebv$, respectively, and then used equation~(\ref{eq:1}) to model the mode of the $\ebv$ distribution for a given $\dmn$ with equal weights. This approach was particularly useful for tracing the ridgeline at small $\dmn$, given the relatively small number of available stars in that range.

In the subsequent modeling in the central, multiresolution HEALPix cell, we sought the best-fitting normalization factors $\{A_1, A_2 \}$ in equation~(\ref{eq:1}), after a couple of iterations with a $3\sigma$ outlier rejection. To take into account heteroscedastic uncertainties in $\ebv$ and $\dmn$ in both axes, we repeated all of the above fitting processes using a simple Monte Carlo simulation, in which we assumed a normal distribution of measurement uncertainties in these parameters based on our estimated uncertainties and those from Gaia. The green lines in Figure~\ref{fig:raw} show individual models obtained from $100$ realizations in such simulations, while the red solid line represents their mean trend. To characterize this mean trend, we consistently applied equation~(\ref{eq:1}) to fit the median $\ebv$ from individual models as a function of $\dmn$.

A couple of caveats in our modeling should be noted, including the neglect of correlations between $\ebv$ and $\dmn$ and the internal scatter originating from the finite size of our adopted HEALPix map scheme. However, as evident in Figure~\ref{fig:raw}, the double logistic function effectively captures the two-layered dust structure in all families of solutions. Notably, the mean models obtained from Case~C and Case~D exhibit a more gradual increase in $\ebv$ at $\dmn\sim9.5$, due to the degraded precision in both distance and reddening estimation in the pure spectrophotometric solutions. While the range of $\ebv$ and $\dmn$ remains comparable across all cases, this comparison leads us to the conclusion that both Case~A and Case~B, relying on Gaia parallaxes, outperform the other cases.

\begin{figure*}
\center
\epsscale{0.65}
  \gridline{\fig{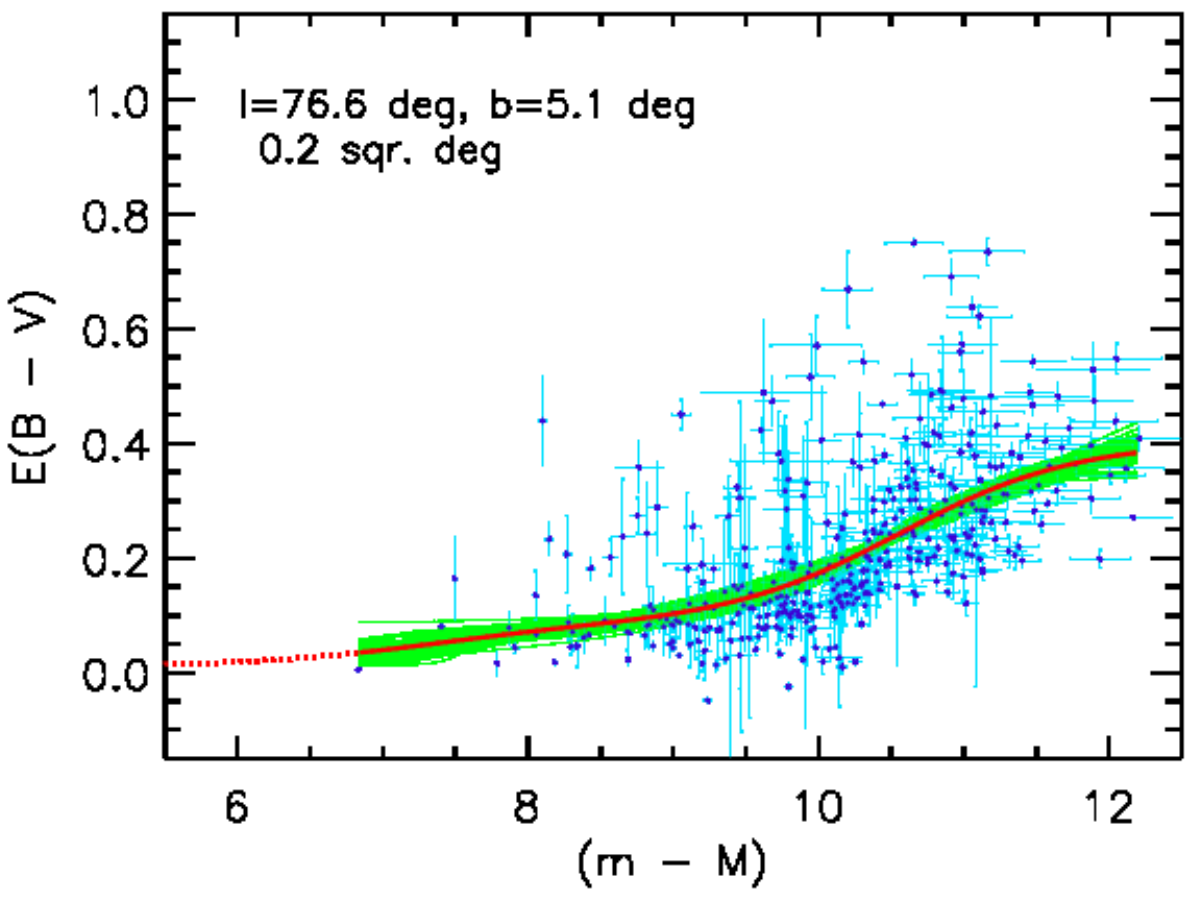}{0.32\textwidth}{\textbf{ }}
                \fig{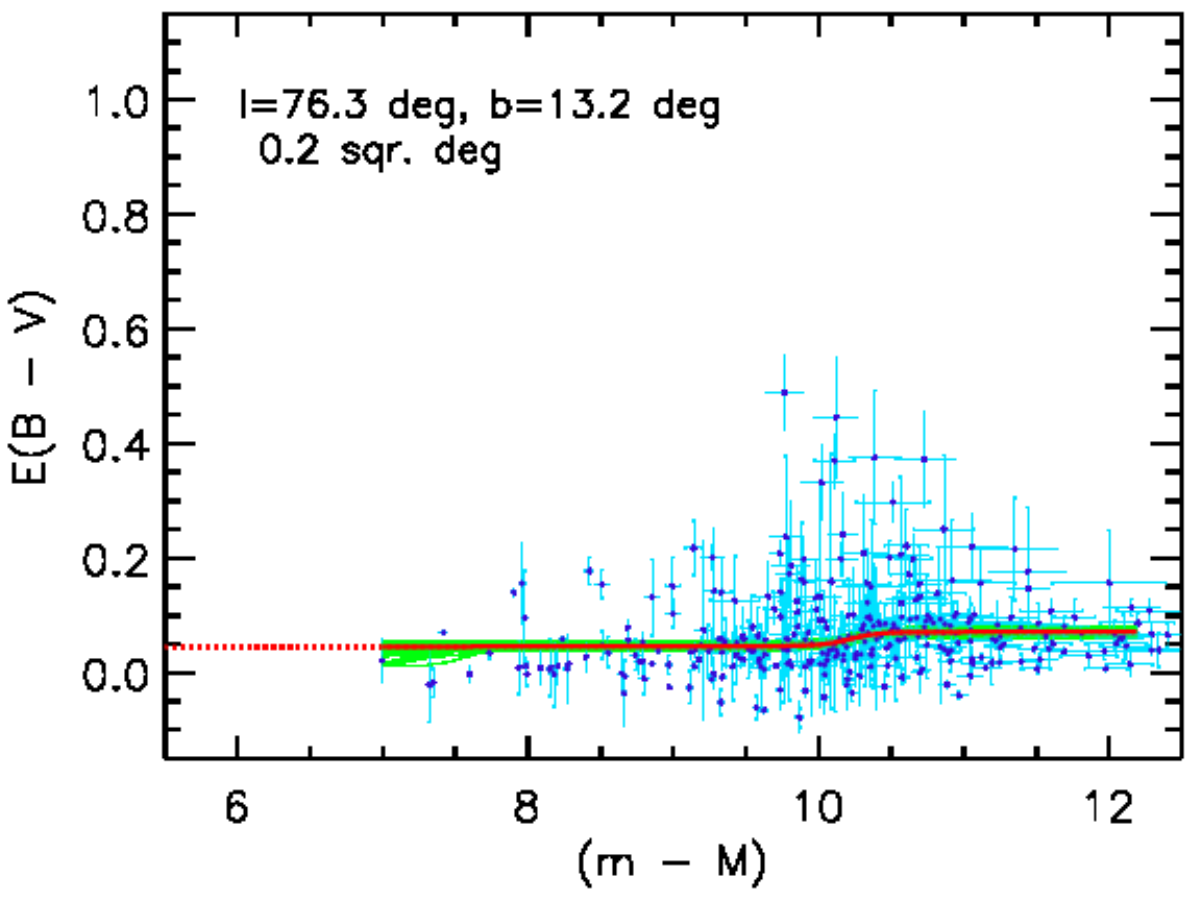}{0.32\textwidth}{\textbf{ }}
                \fig{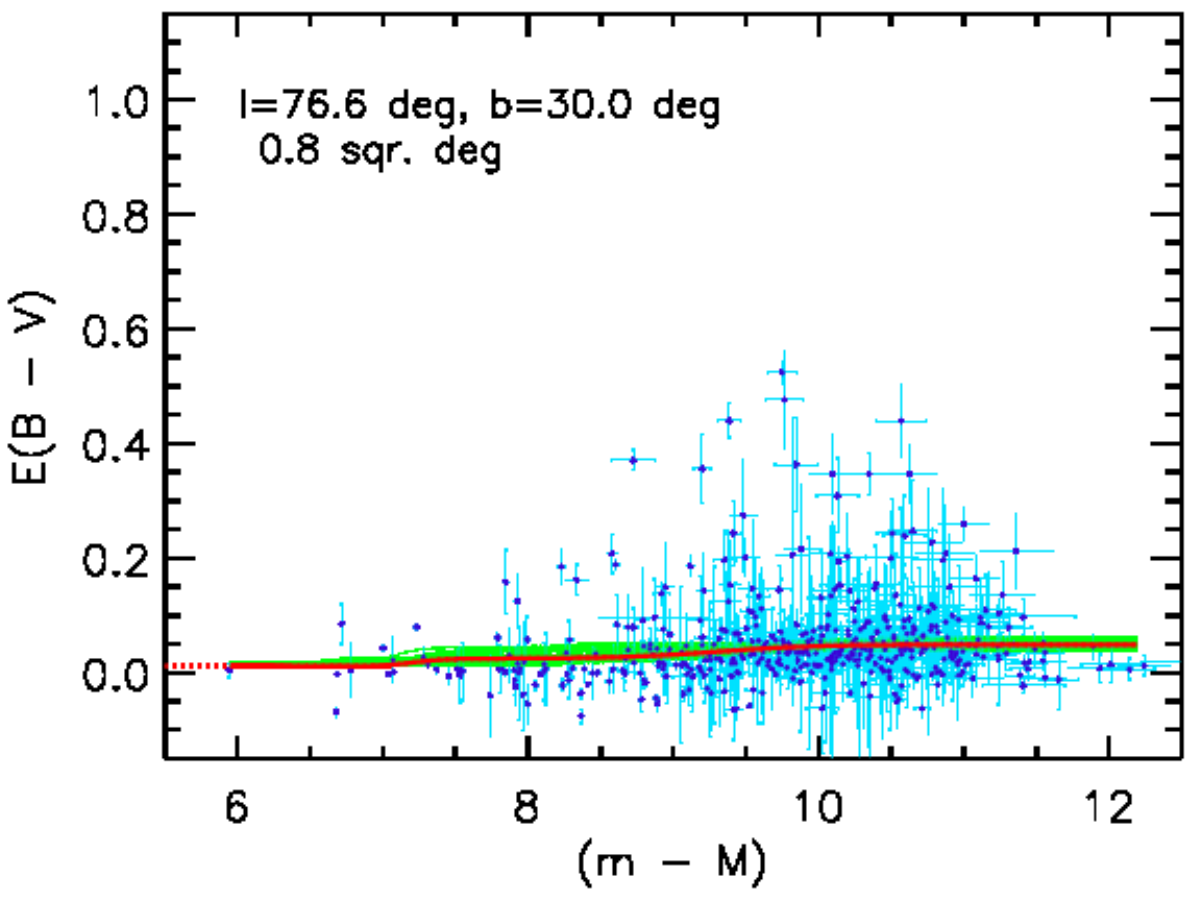}{0.32\textwidth}{\textbf{ }}}
  \gridline{\fig{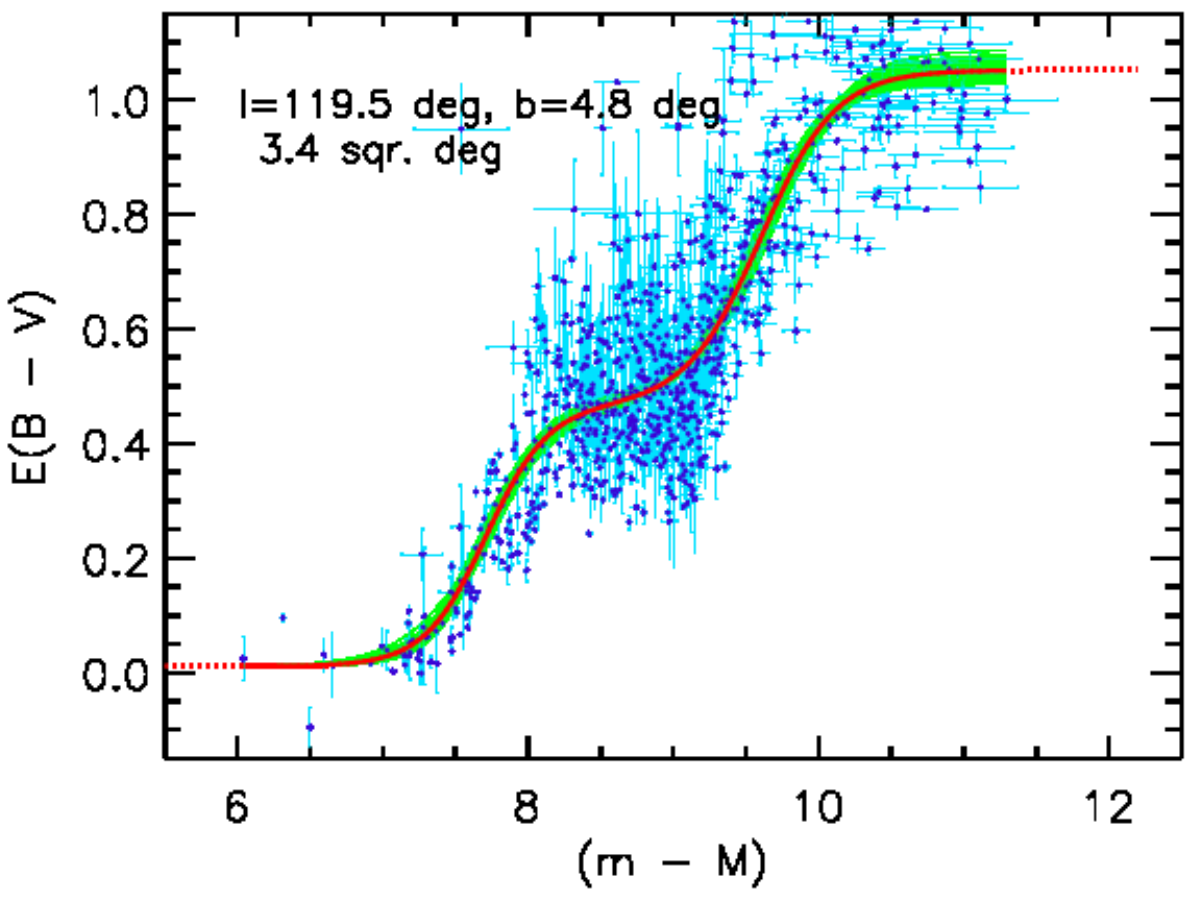}{0.32\textwidth}{\textbf{ }}
                \fig{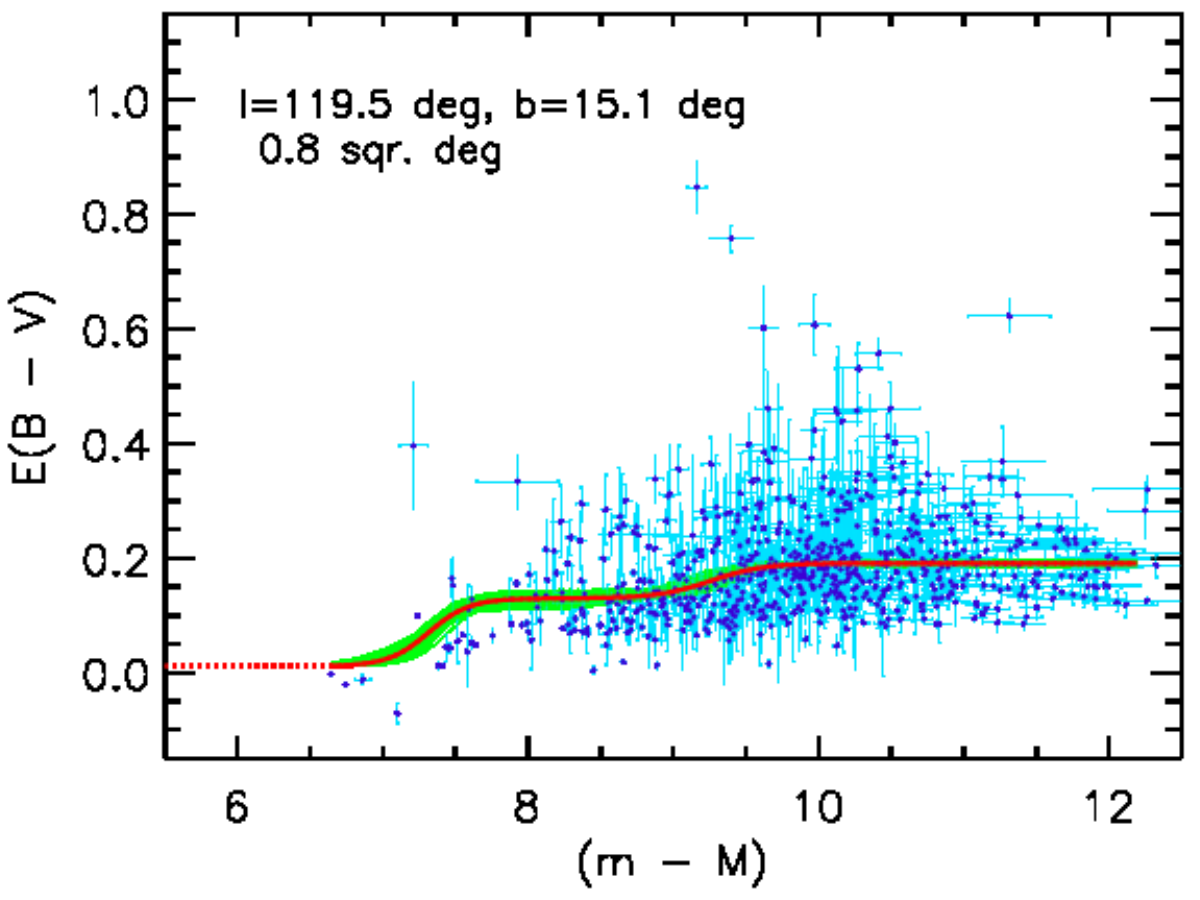}{0.32\textwidth}{\textbf{ }}
                \fig{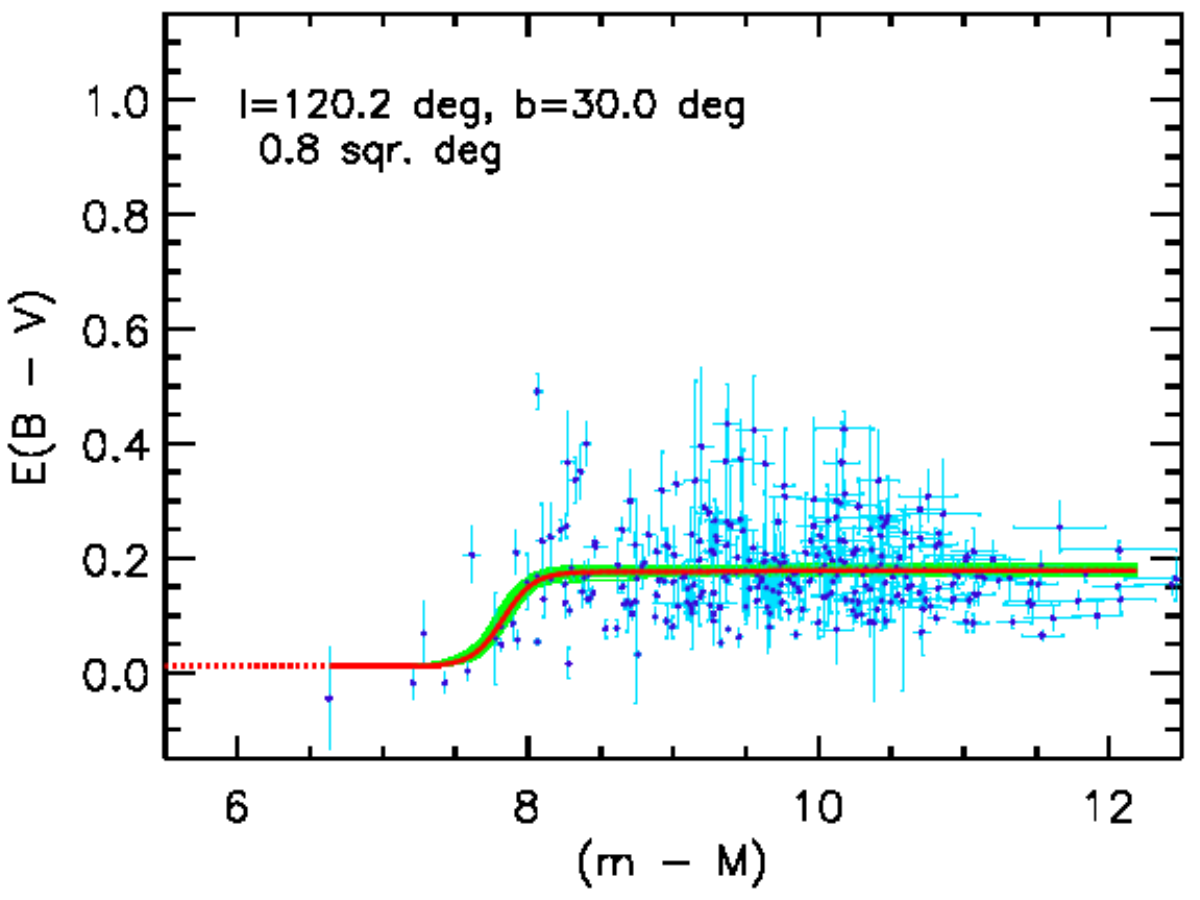}{0.32\textwidth}{\textbf{ }}}
  \gridline{\fig{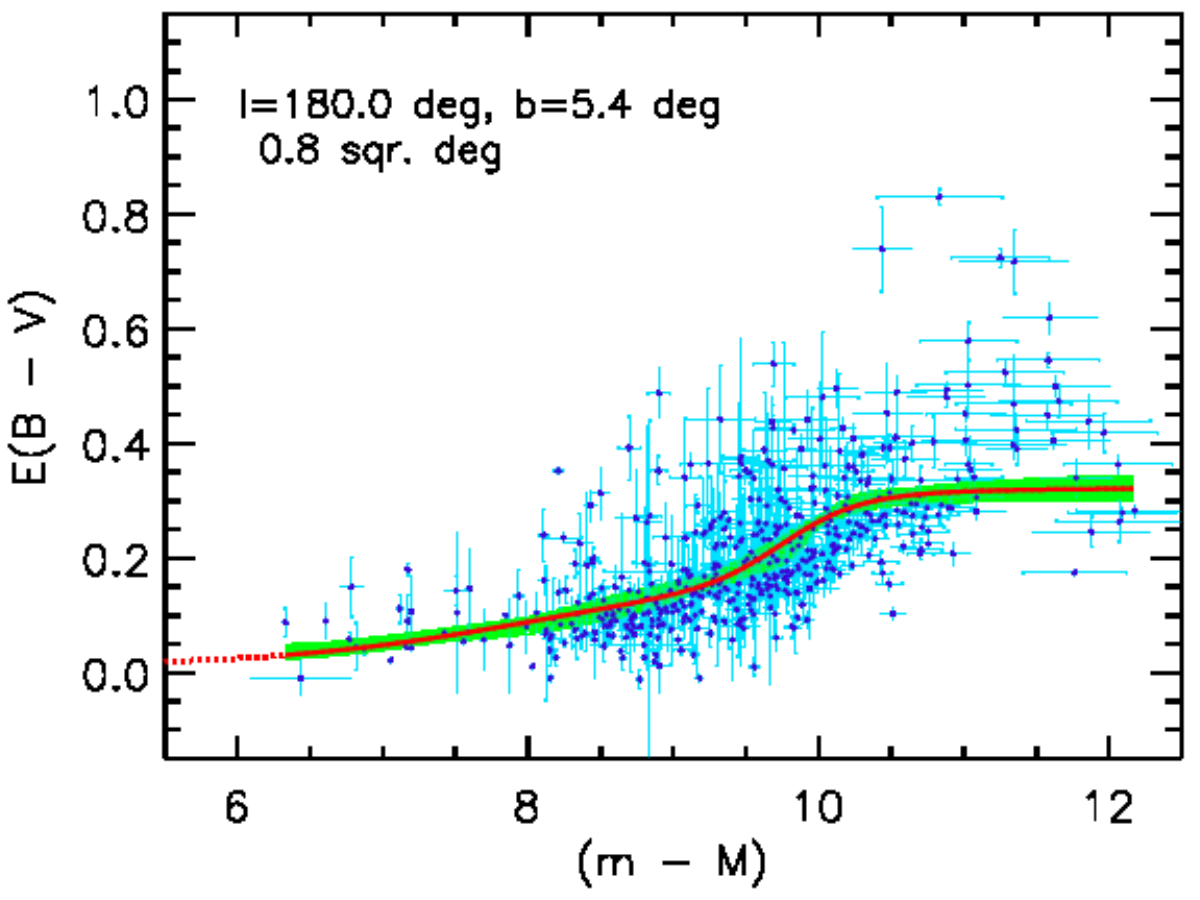}{0.32\textwidth}{\textbf{ }}
                \fig{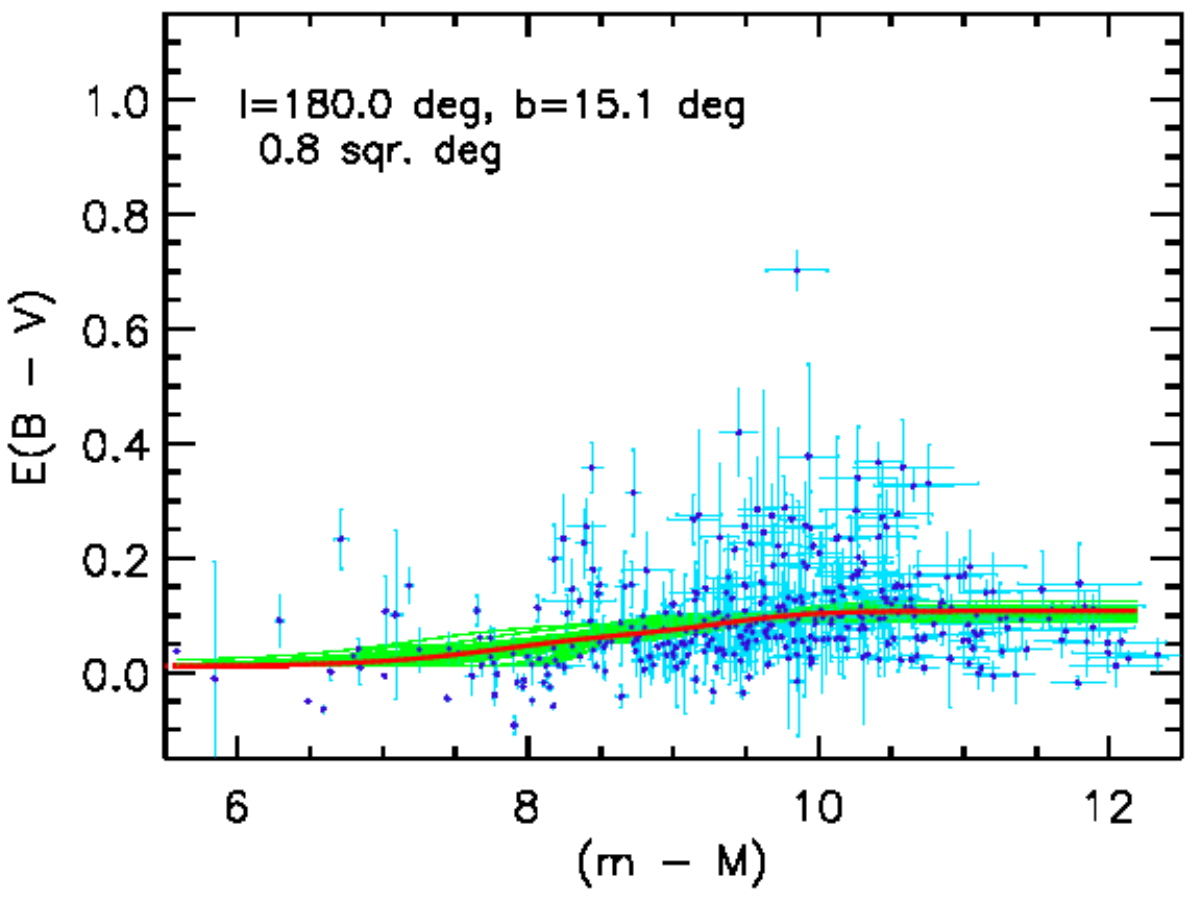}{0.32\textwidth}{\textbf{ }}
                \fig{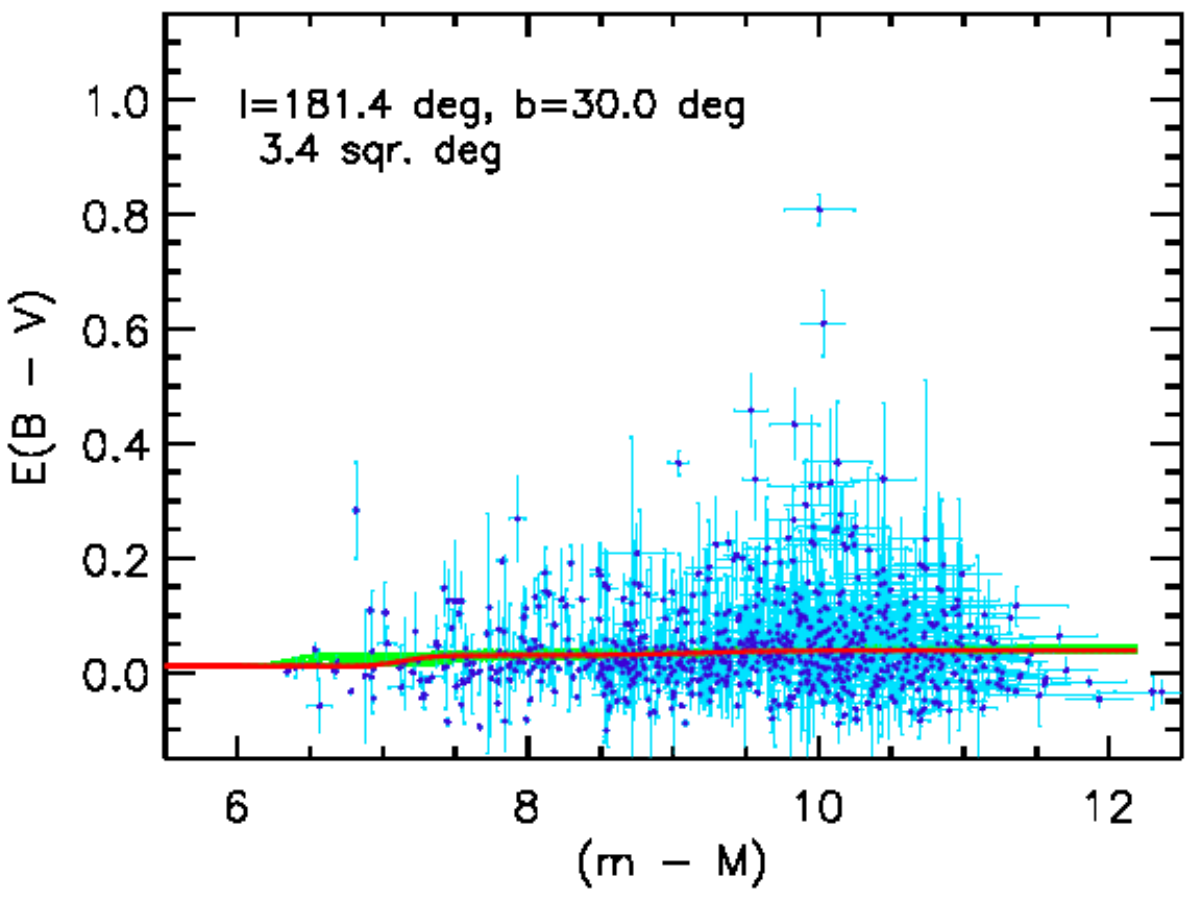}{0.32\textwidth}{\textbf{ }}}
  \gridline{\fig{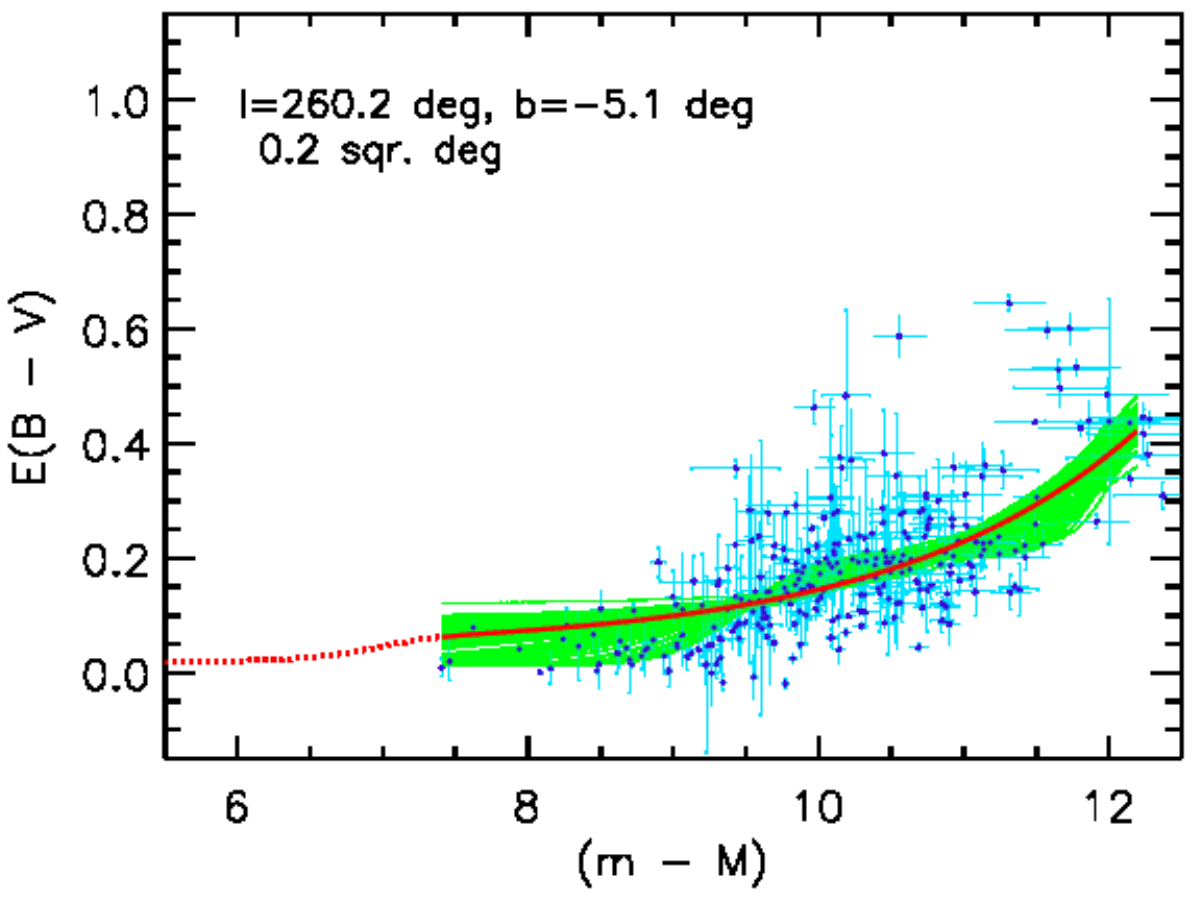}{0.32\textwidth}{\textbf{ }}
                \fig{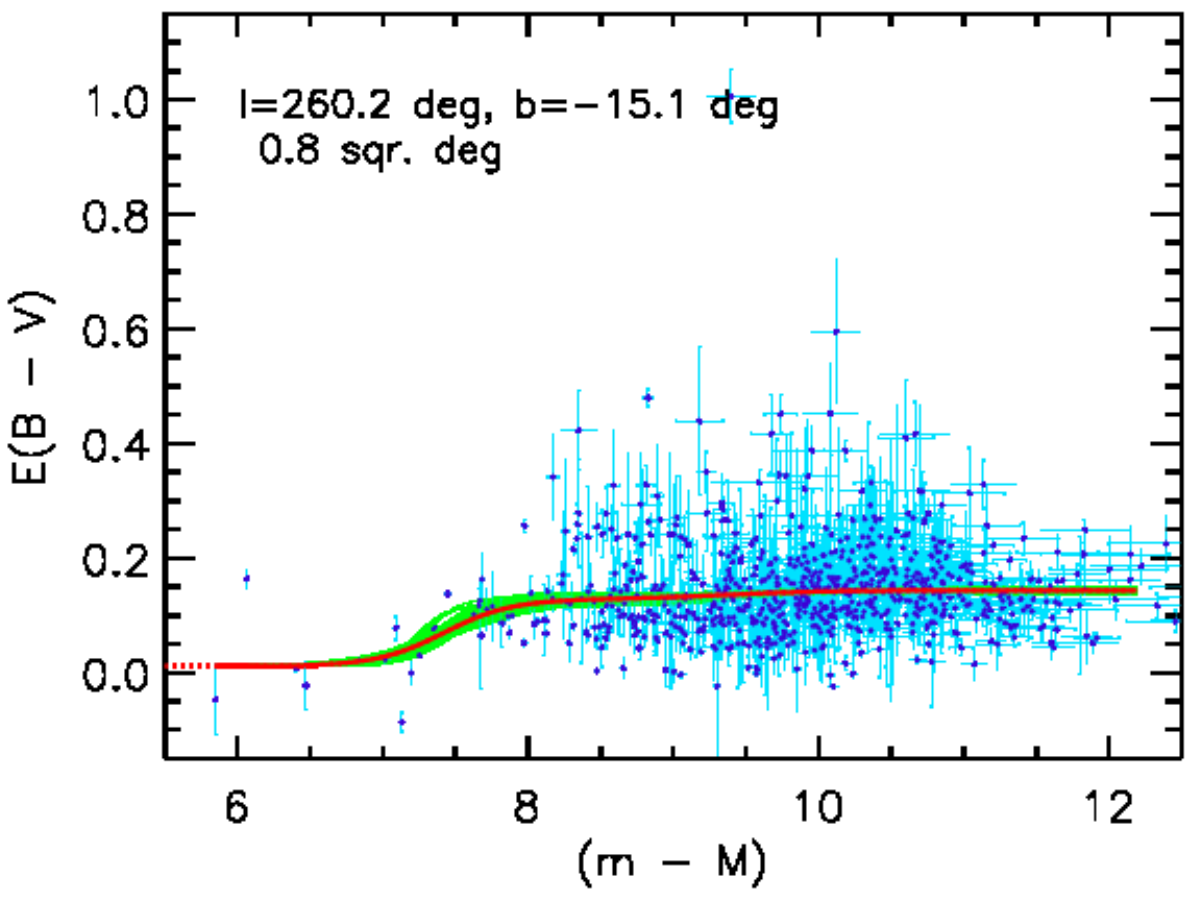}{0.32\textwidth}{\textbf{ }}
                \fig{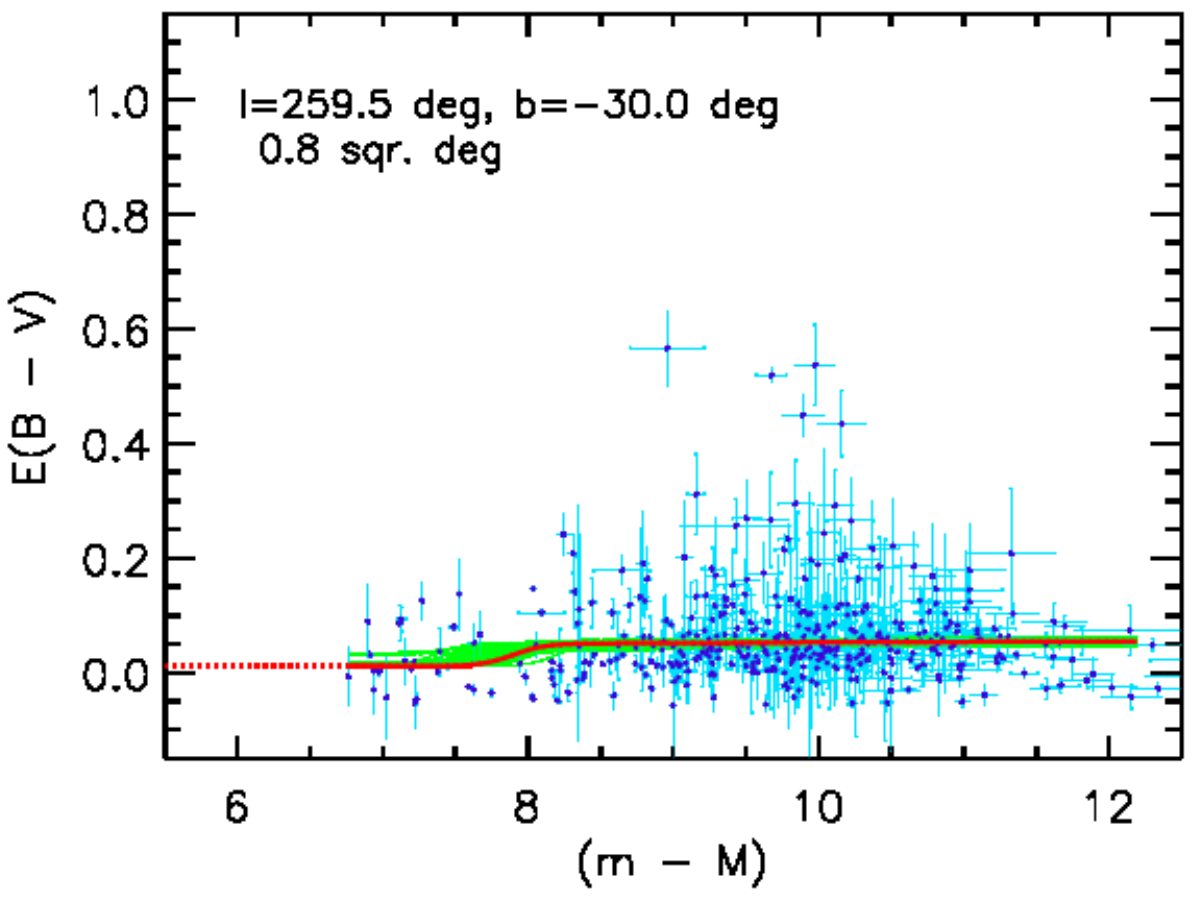}{0.32\textwidth}{\textbf{ }}}
\caption{Same as Figure~\ref{fig:raw}, but displaying Case~A measurements for selected lines of sight. Sight lines directed toward the Galactic plane are displayed on the left, while additional sight lines situated at various Galactic latitudes within the same longitude range are displayed on the right. Note that distinct coverages are employed, with each one representing a single multiresolution HEALPix cell.}
\label{fig:raw2}
\end{figure*}

Figure~\ref{fig:raw2} displays the same $\ebv$ vs.\ $\dmn$ distribution as Figure~\ref{fig:raw} but for some lines of sight, providing a glimpse of the systematic change in Galactic dust structure and the quality of our model fits. Only $\ebv$ derived from Case~A is shown. The left panels display data taken near the Galactic plane, while the right panels show observations at $|b| \approx 30\arcdeg$, where reddening is significantly less pronounced. Our models generally exhibit good agreement with the observed structure in areas with moderate cumulative extinction ($\ebv \la 0.5$). Some regions near the Galactic plane exhibit a two-step functional form similar to that observed in Figure~\ref{fig:raw}. In contrast, for most high-latitude samples, a single logistic function suffices, which is unsurprising given that these lines of sight do not traverse dense clouds in the local spiral arms.

\begin{figure}
\center
  \gridline{\fig{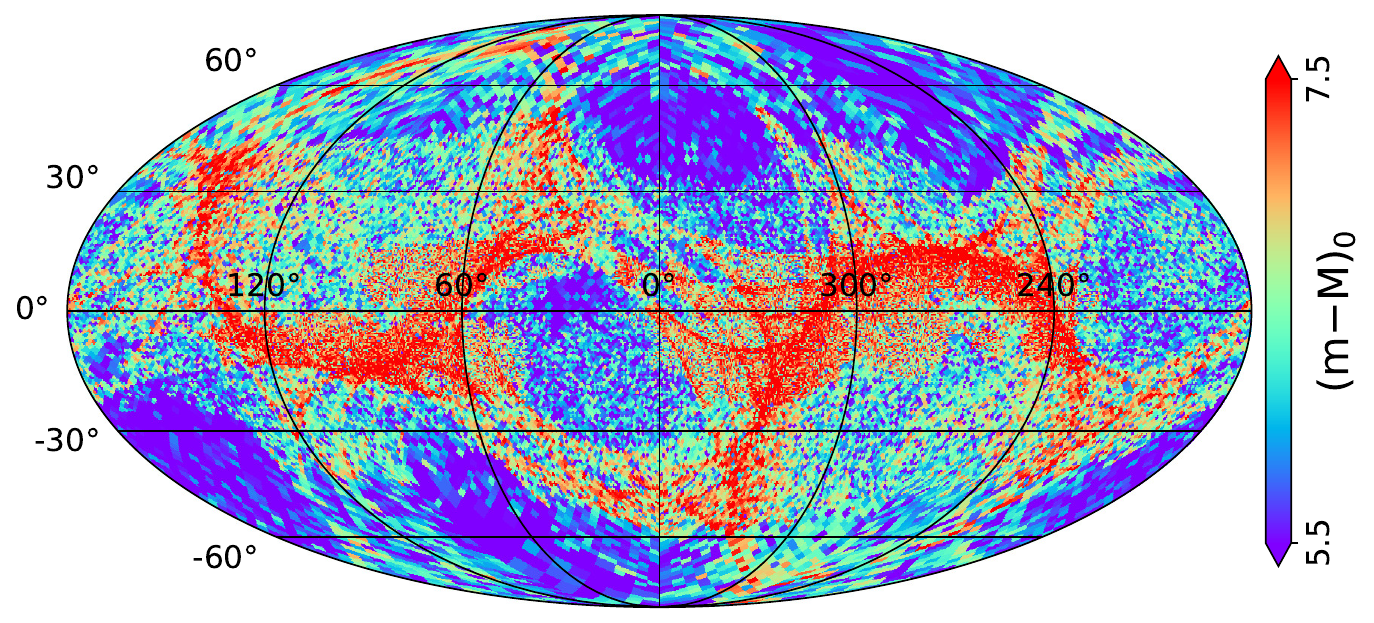}{0.48\textwidth}{\textbf{(a) Minimum distance modulus }}}
  \gridline{\fig{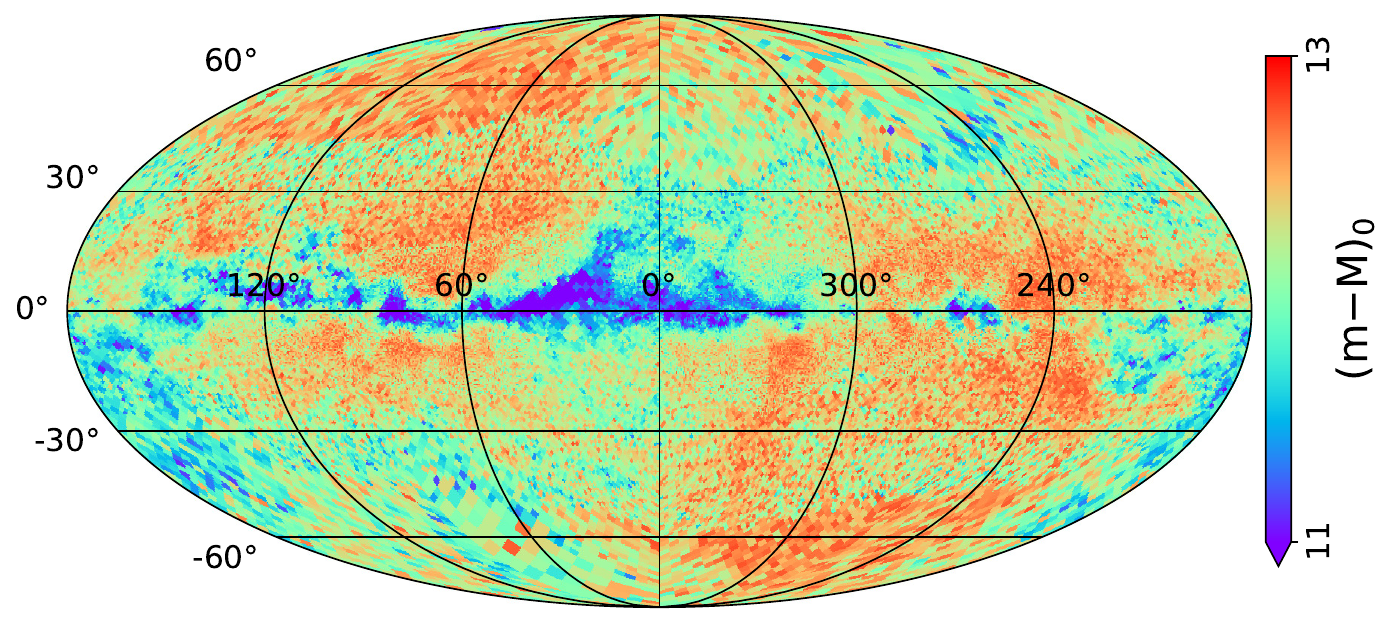}{0.48\textwidth}{\textbf{(b) Maximum distance modulus }}}
\caption{Range of minimum (top panel) and maximum (bottom panel) distance modulus determined by the number of stars accessible in Case~A.}
\label{fig:dmn}
\end{figure}

Figure~\ref{fig:dmn} shows the range of the valid distance modulus of the sample employed in Case~A. Within the most densely obscured regions along the Galactic plane, distances are confined to $d\la1$~kpc, reflecting the limited number of stars within the magnitude-limited XP MS sample. Beyond $|b| > 10\arcdeg$, our sample encompasses volumes extending, on average, up to $\sim3$~kpc, with certain areas exhibiting even greater depth.

\subsection{3D Reddening Map}\label{sec:map}

\begin{figure*}
\center
\epsscale{0.65}
  \gridline{\fig{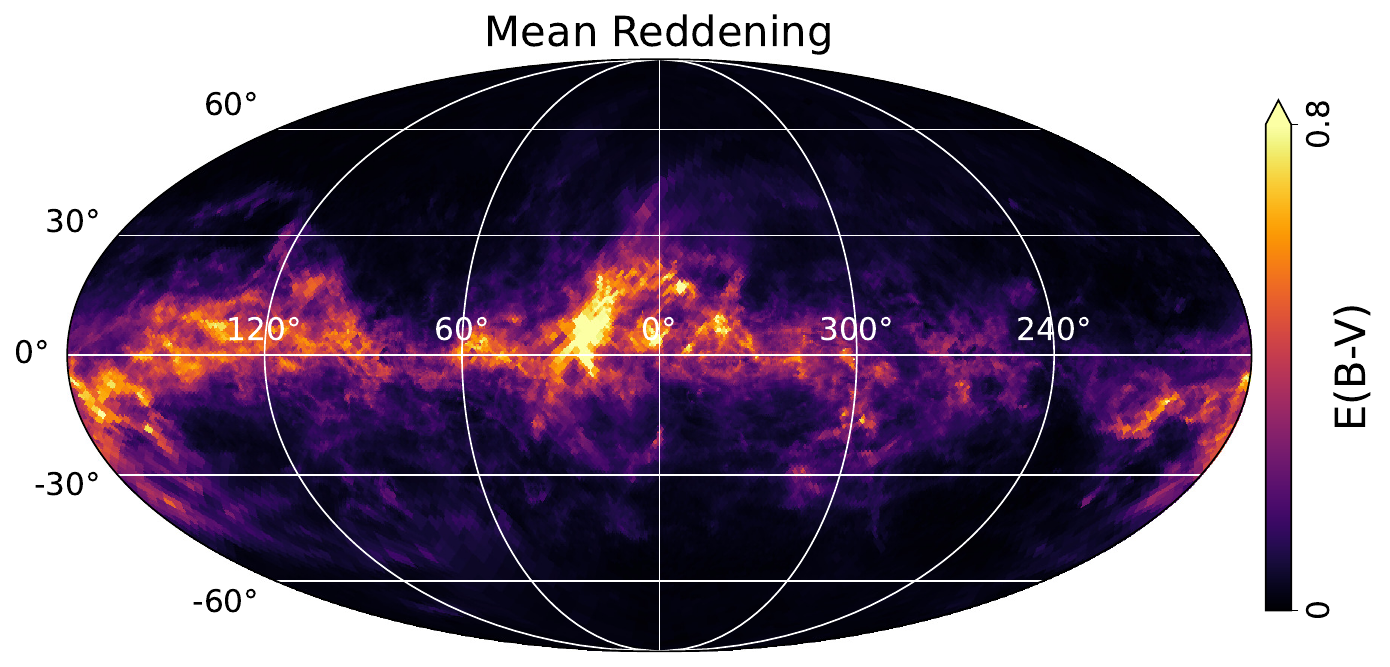}{0.42\textwidth}{\textbf{(a) Case A }}
                \fig{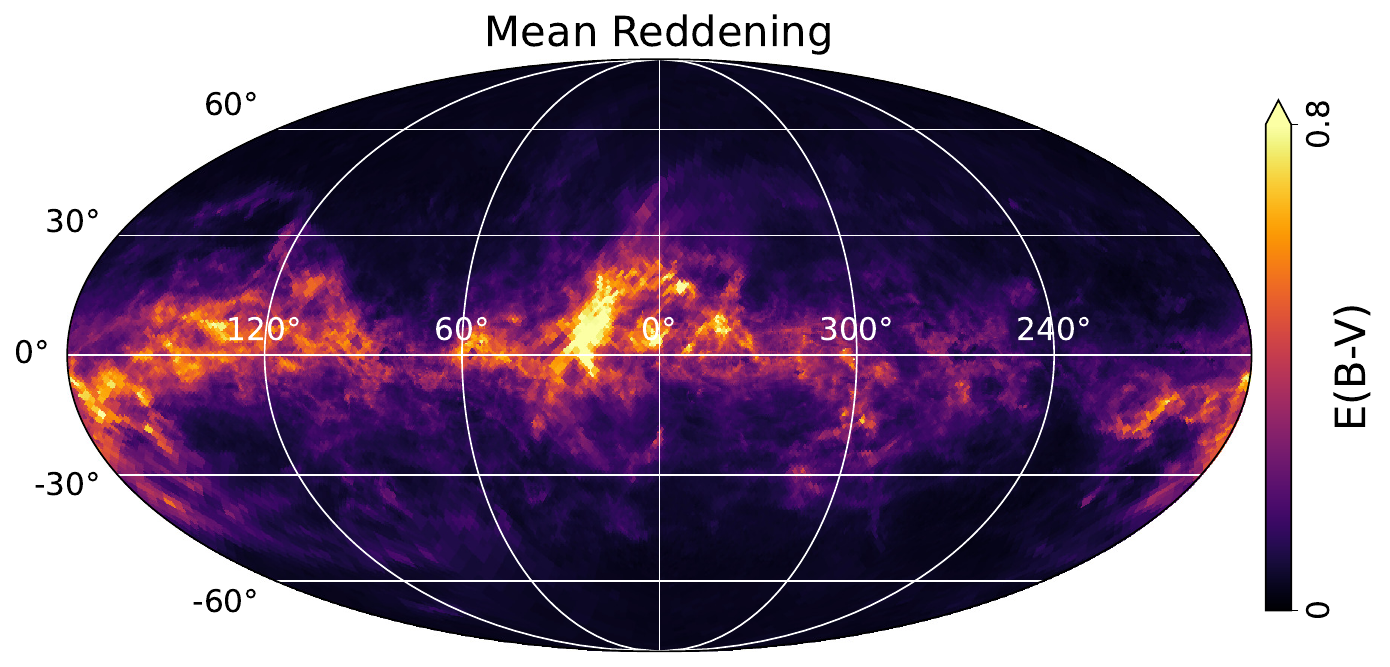}{0.42\textwidth}{\textbf{(b) Case B }}}
  \gridline{\fig{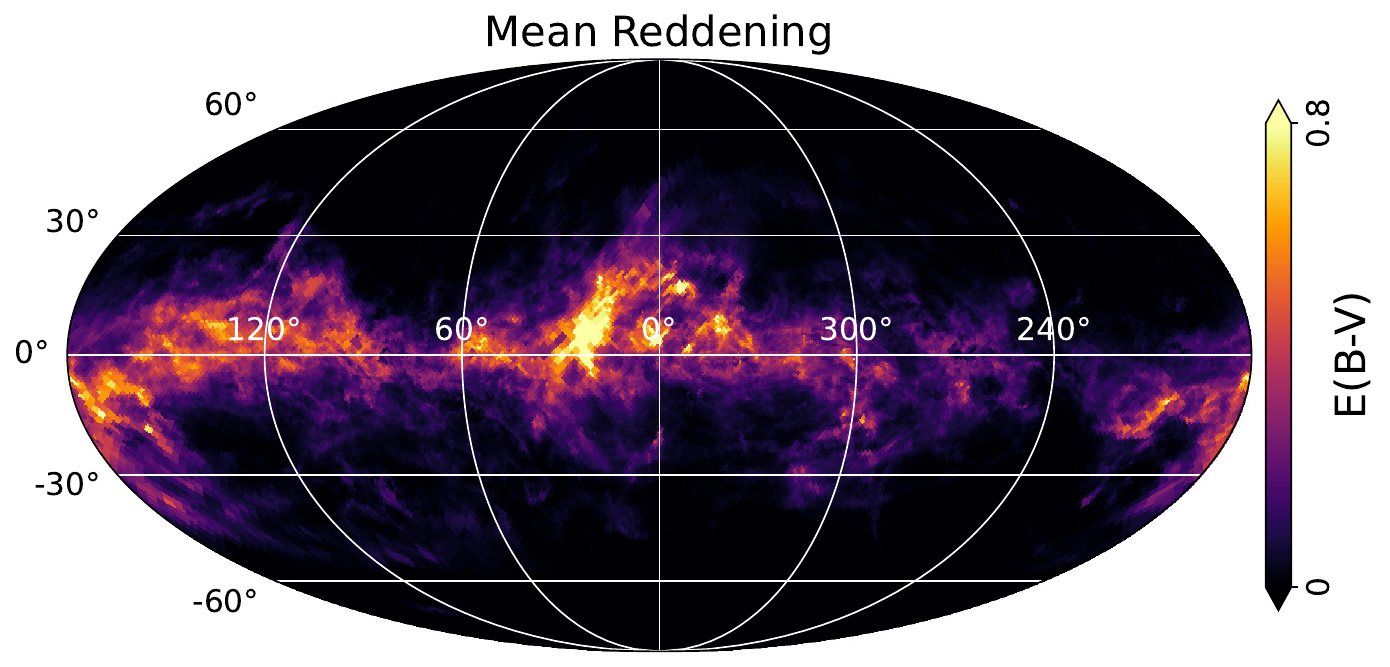}{0.42\textwidth}{\textbf{(c) Case C }}
                \fig{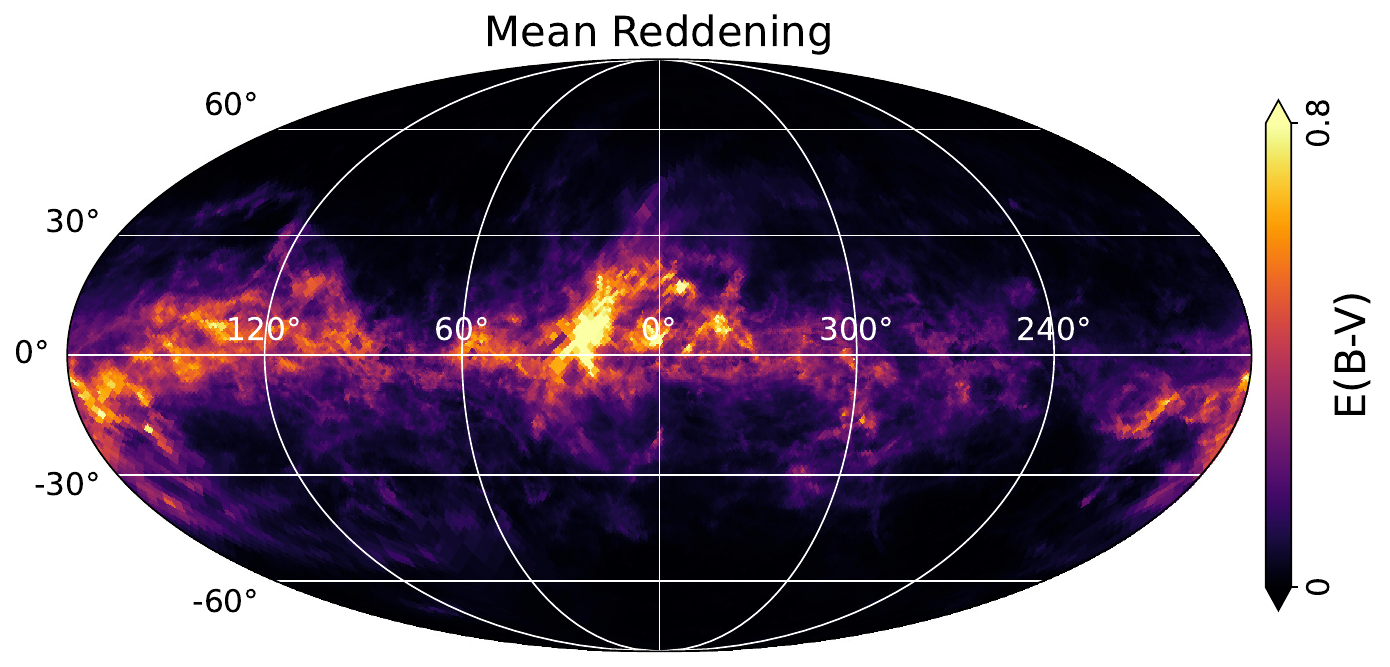}{0.42\textwidth}{\textbf{(d) Case D }}}
\caption{Average reddening of the XP MS sample shown in a Mollweide projection of the Galactic coordinate system, represented by the multiresolution HEALPix. Each panel corresponds to one of the four different solutions obtained in this study (Table~\ref{tab:tab1}).}
\label{fig:avgmap}
\end{figure*}

\begin{figure*}
\center
\epsscale{0.65}
  \gridline{\fig{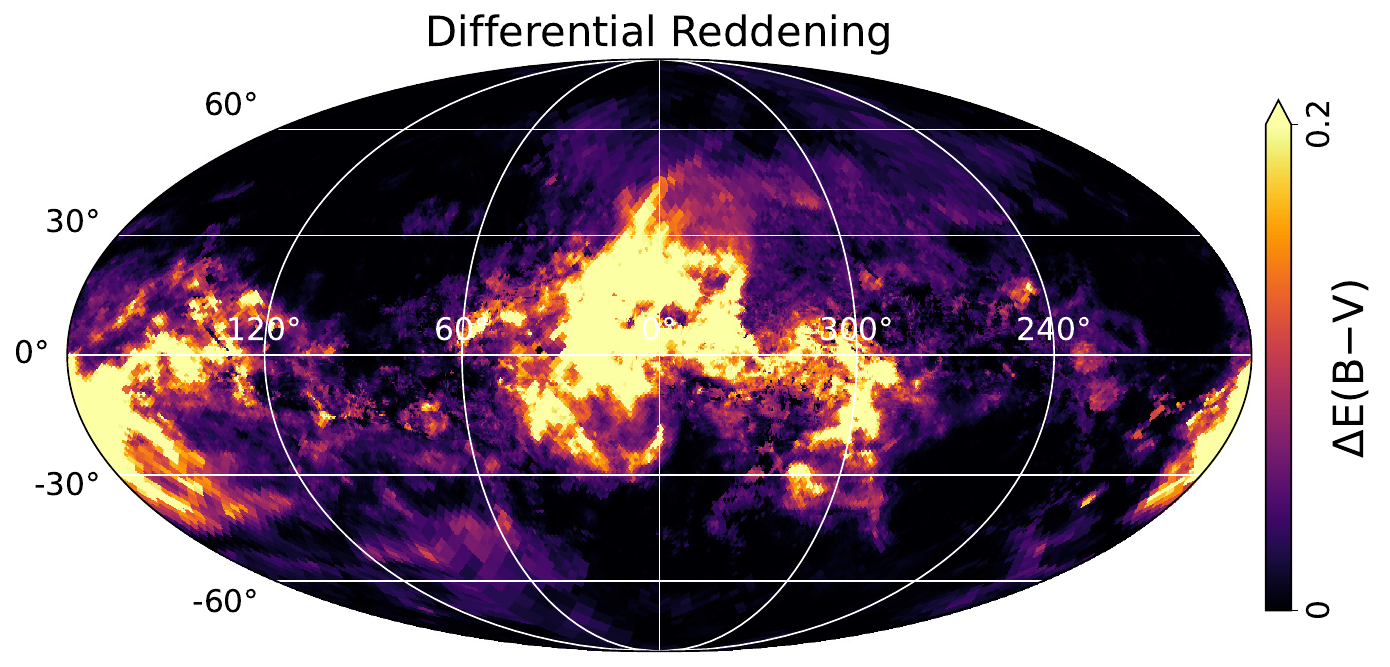}{0.42\textwidth}{\textbf{(a) $d \leq251$~pc }}
                \fig{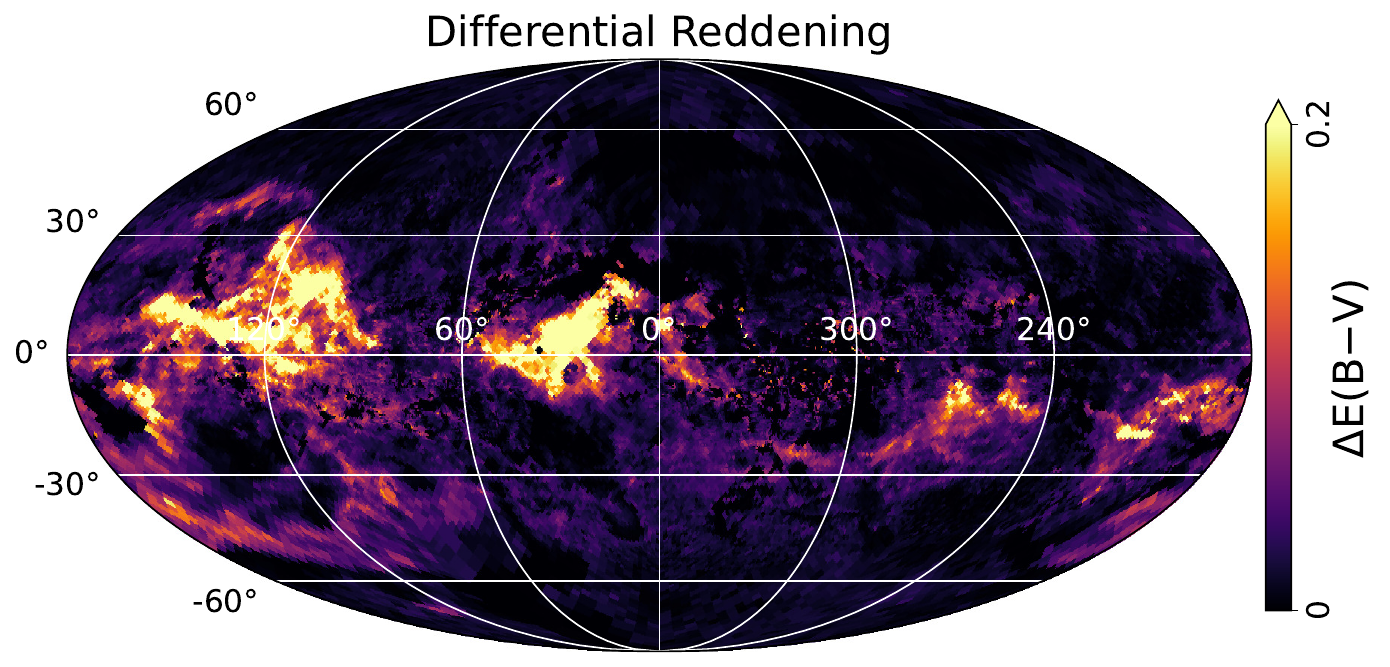}{0.42\textwidth}{\textbf{(b) $251\leq d \leq 398$~pc }}}
  \gridline{\fig{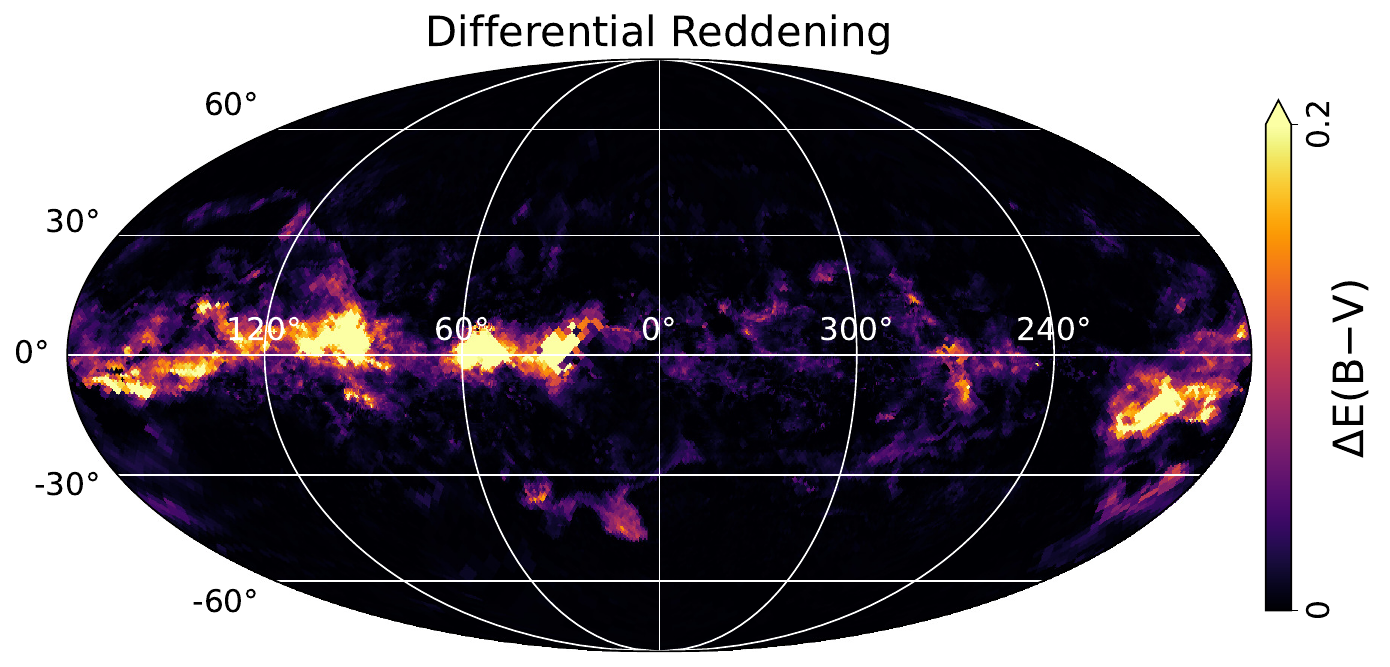}{0.42\textwidth}{\textbf{(c) $398\leq d \leq630$~pc }}
                \fig{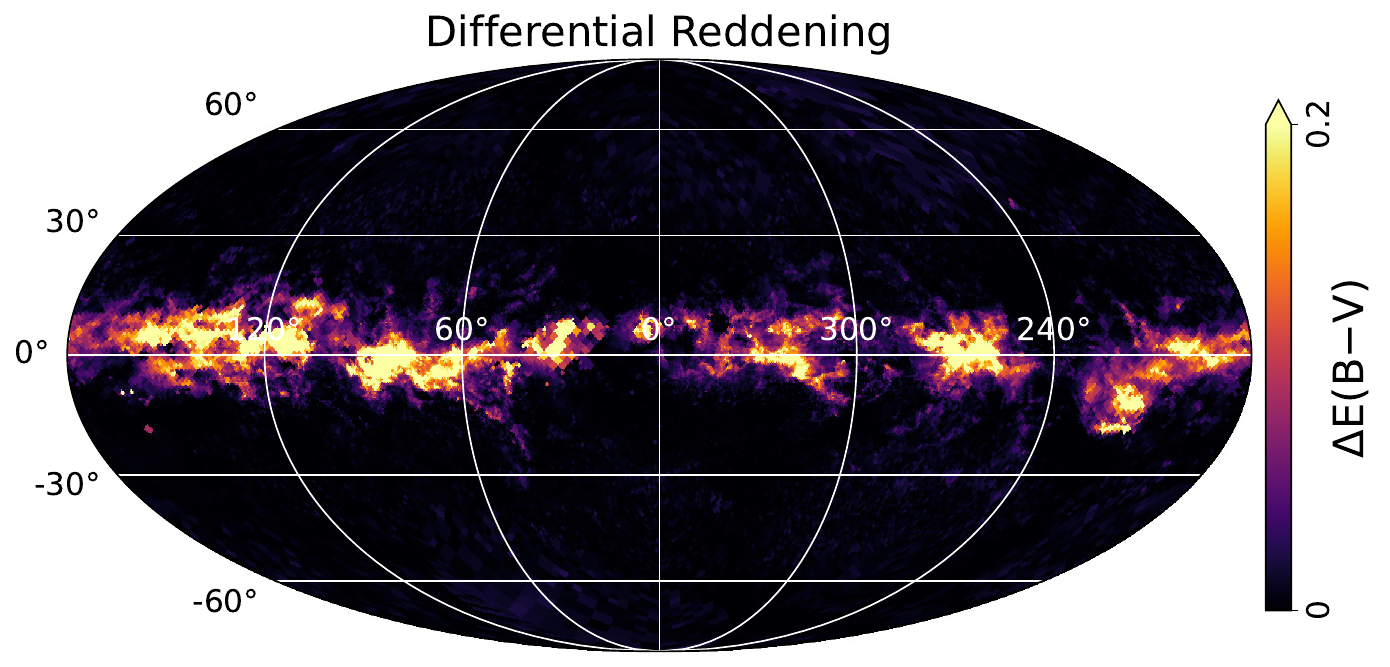}{0.42\textwidth}{\textbf{(d) $630\leq d \leq1000$~pc }}}
  \gridline{\fig{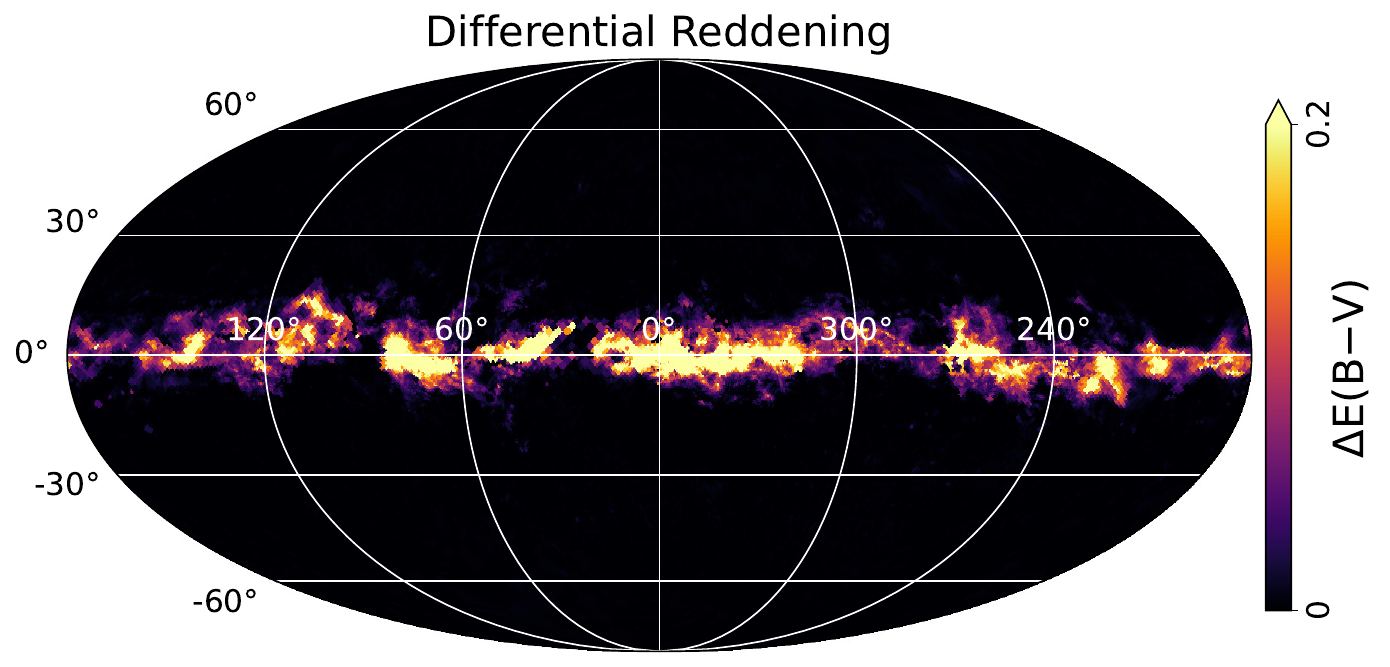}{0.42\textwidth}{\textbf{(e) $1000\leq d \leq1584$~pc }}
                \fig{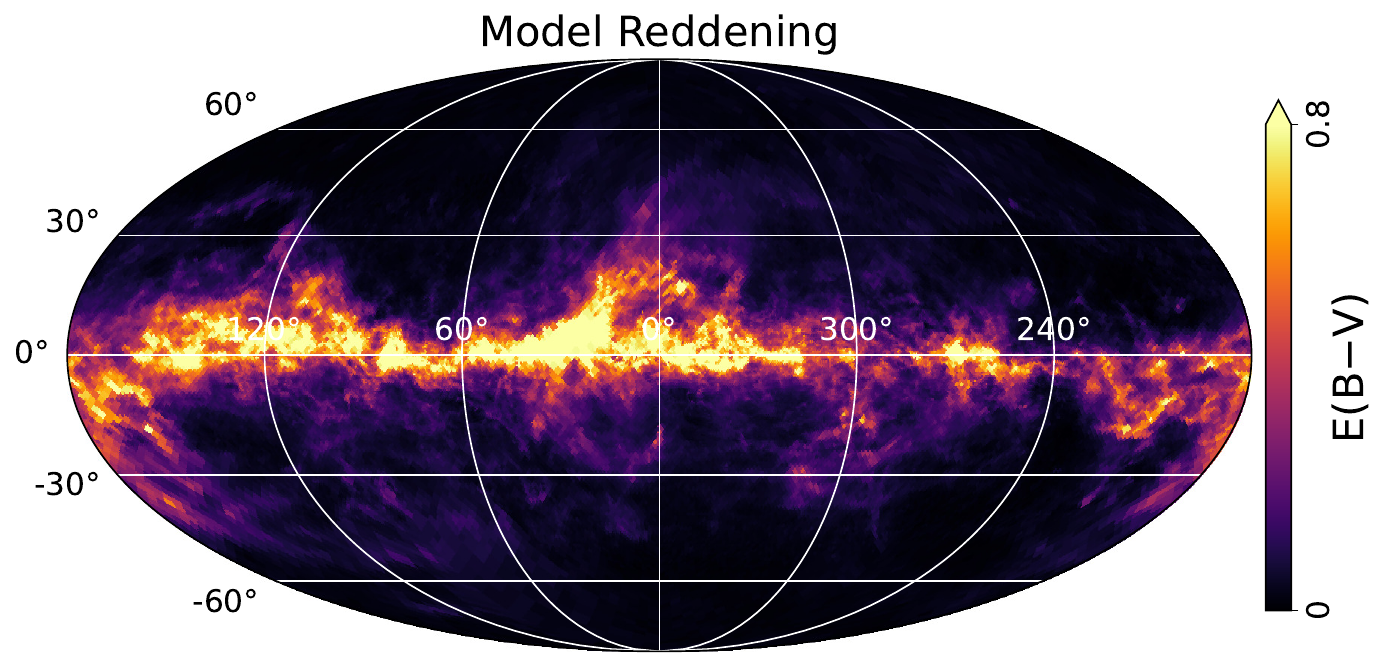}{0.42\textwidth}{\textbf{(f) $d \leq2,511$~pc }}}
\caption{Differential reddening of the XP MS sample based on Case~A in selected distance bins. Reddening maps are displayed in a Mollweide projection of the Galactic coordinate system using the multiresolution HEALPix.}
\label{fig:ebvmap}
\end{figure*}

Figure~\ref{fig:avgmap} displays the distribution of average $\ebv$ in the Galactic coordinate system for the four sets of parameter estimation utilized in our analysis (Table~\ref{tab:tab1}). In all panels, a prominent large-scale structure of clouds in the local volume is observed along the Galactic plane. However, finer structures become more evident when these maps are divided into narrower distance bins, as shown in Figure~\ref{fig:ebvmap}, on the same multiresolution grid as depicted in panel~(b) of Figure~\ref{fig:num}. It presents cross sections of our 3D extinction map from Case~A within selected distance intervals. The $\Delta \ebv$ indicates an increment of reddening induced by dust clouds in a given distance bin, calculated using our best-fitting models in equation~(\ref{eq:1}) along each line of sight.

The majority of dark clouds and cloud complexes identified in previous CO surveys \citep[e.g.,][]{dame:01} are discernible in Figure~\ref{fig:ebvmap}; see also \citet{schlafly:14} for distances to individual clouds. Focusing solely on the prominent structures among these, the extended molecular cloud complex in the constellation of Ophiuchus ($\langle d \rangle\sim100$~pc) becomes visible in the nearest distance bin (panel~(a)), situated toward the Galactic center at $(l, b) \sim (0^\circ, +20^\circ)$. This complex overlaps with the Aquila Rift ($\langle d \rangle\sim100$~pc), centered at $(l, b) \sim (30^\circ, +5^\circ)$, which gains prominence in the larger distance bin (panel~(b)). At $(l, b) \sim (300^\circ, -15^\circ)$, the Chamaeleon cloud complex is visible \citep[][$\langle d \rangle\sim200$~pc]{voirin:18}. In the direction of the Galactic anticenter, the Taurus--Perseus--Auriga cloud complex ($\langle d \rangle\sim100$--$300$~pc) is also discernible within these volumes, located at $(l, b) \sim (170^\circ, -10^\circ)$. In panels~(b) and (c), the Orion Nebula ($\langle d \rangle\sim400$~pc) centered approximately at $(210^\circ, -20^\circ)$ and the Cepheus and Polaris Flares ($\langle d \rangle\sim400$~pc) at $(l, b) \sim (110^\circ, +20^\circ)$ come into play. Beyond this point, a predominantly flattened structure of dust prevails along the Galactic plane at greater distances.

\begin{figure*}
\centering
\gridline{\fig{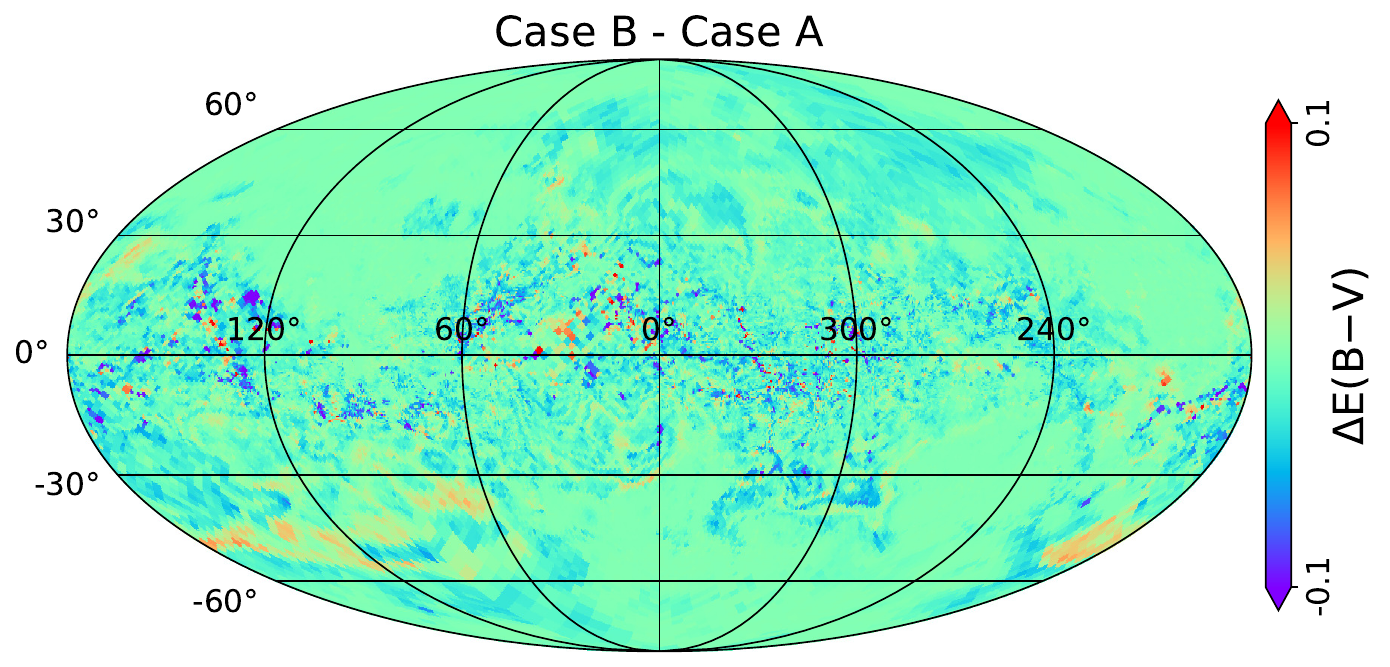}{0.32\textwidth}{\textbf{(a) $d=251$~pc }}
              \fig{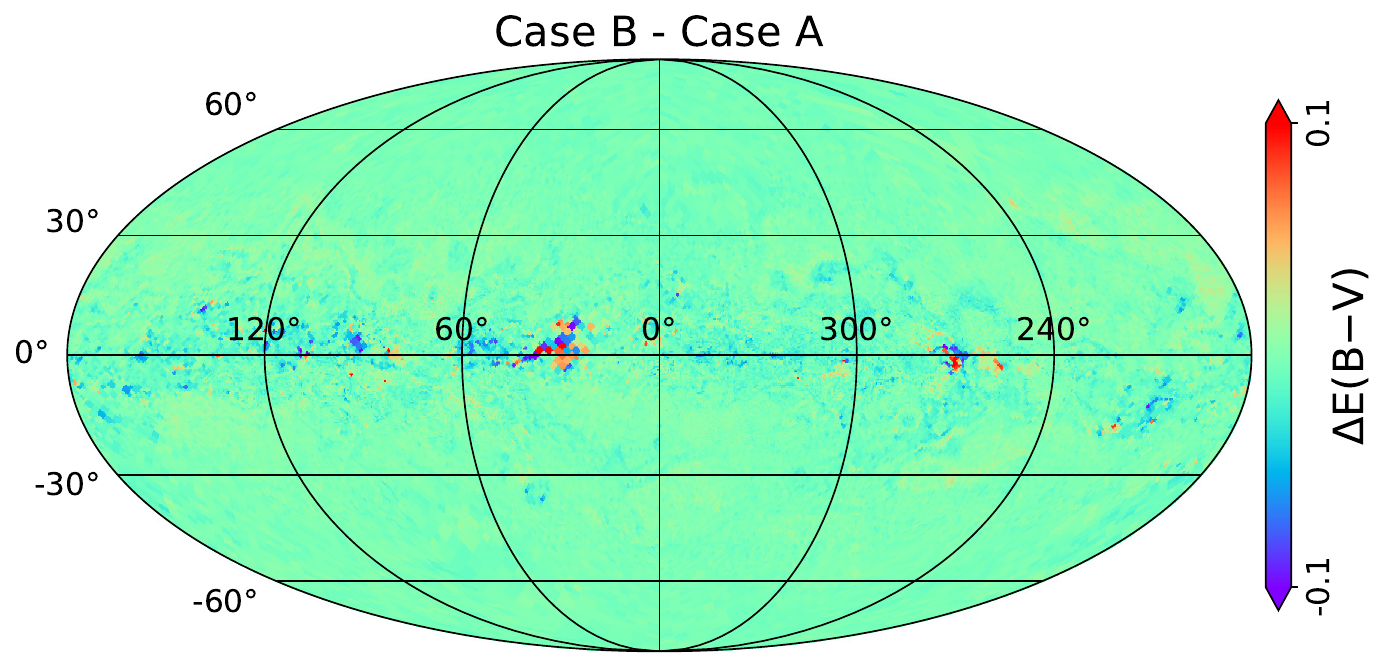}{0.32\textwidth}{\textbf{(b) $d=630$~pc }}
              \fig{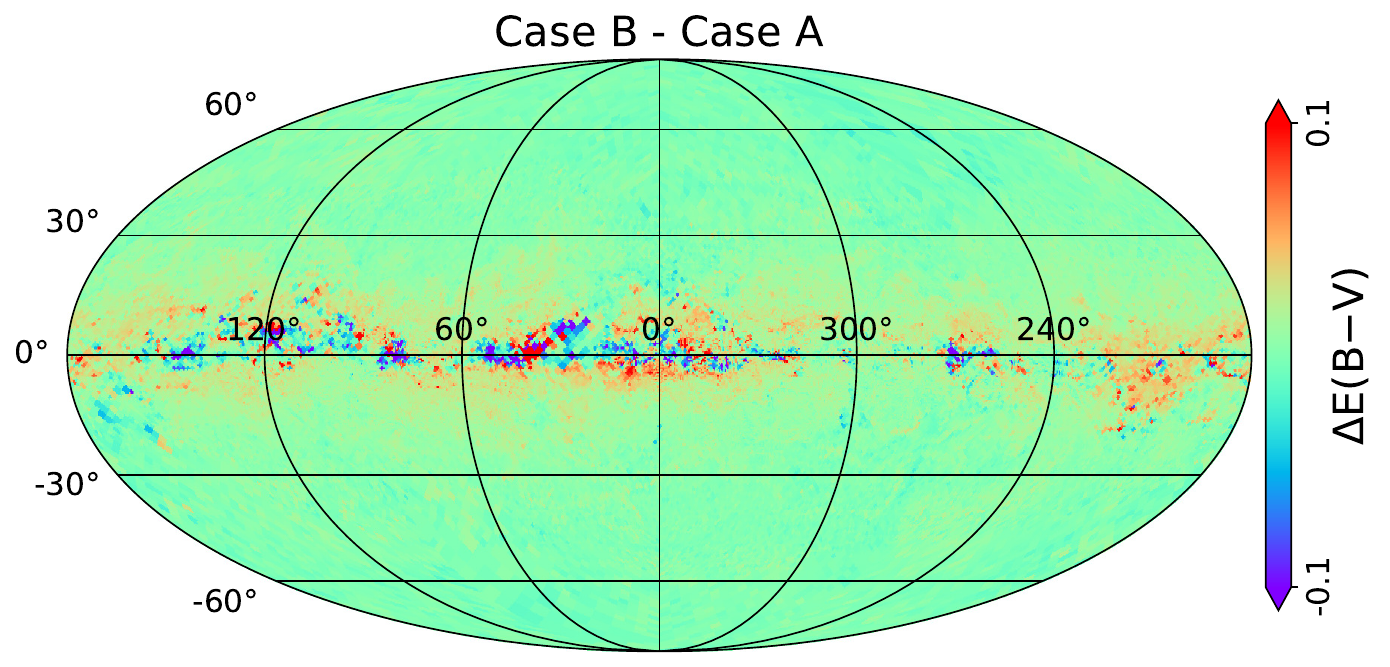}{0.32\textwidth}{\textbf{(c) $d=1584$~pc }}}
\gridline{\fig{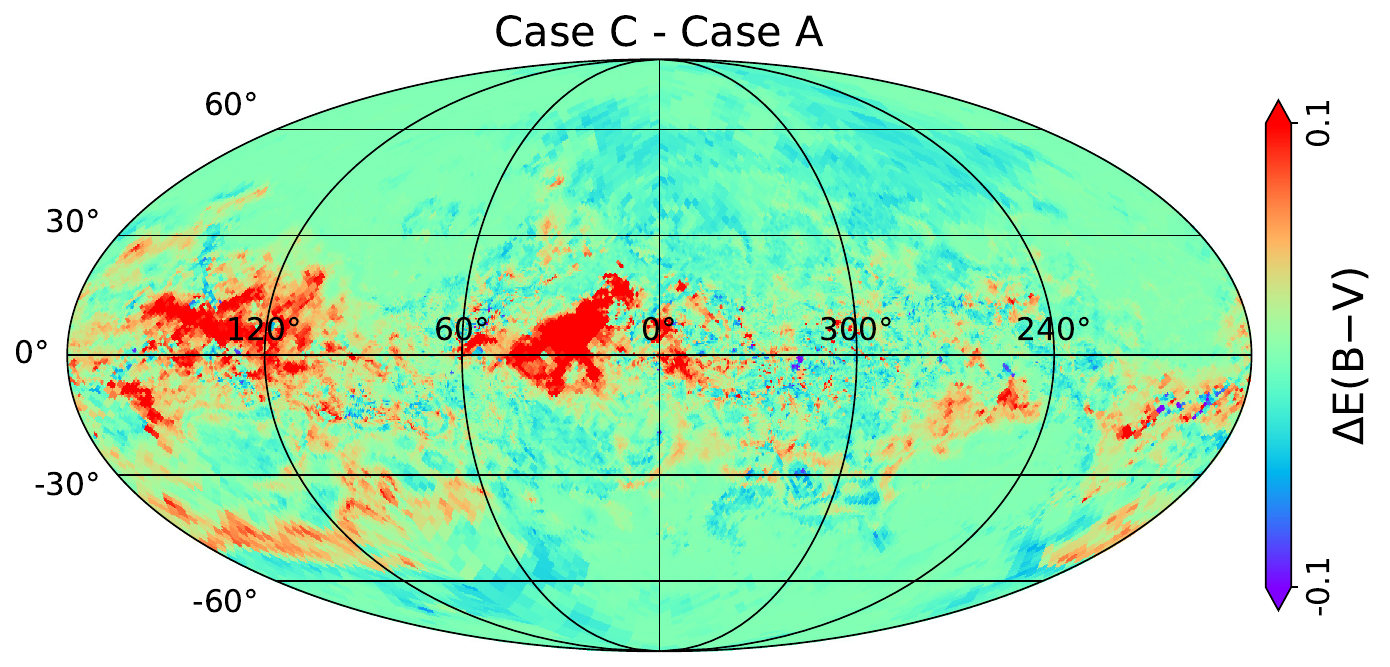}{0.32\textwidth}{\textbf{(d) $d=251$~pc }}
              \fig{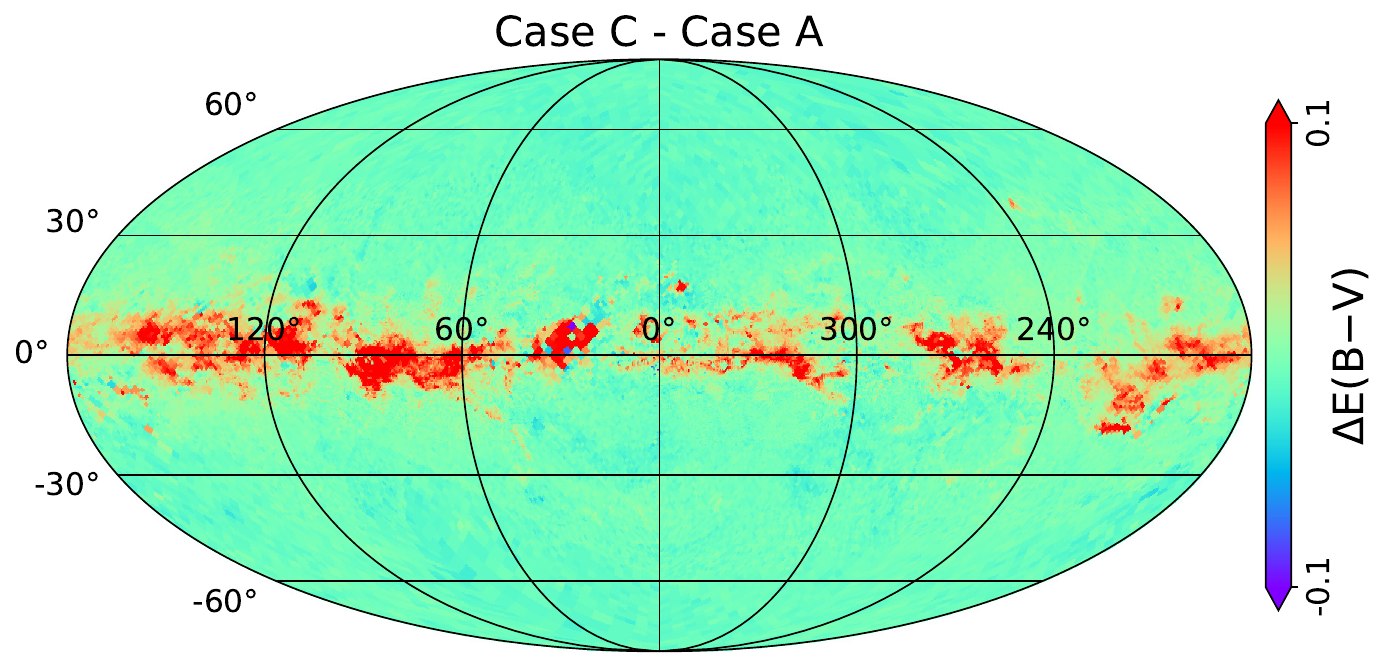}{0.32\textwidth}{\textbf{(e) $d=630$~pc }}
              \fig{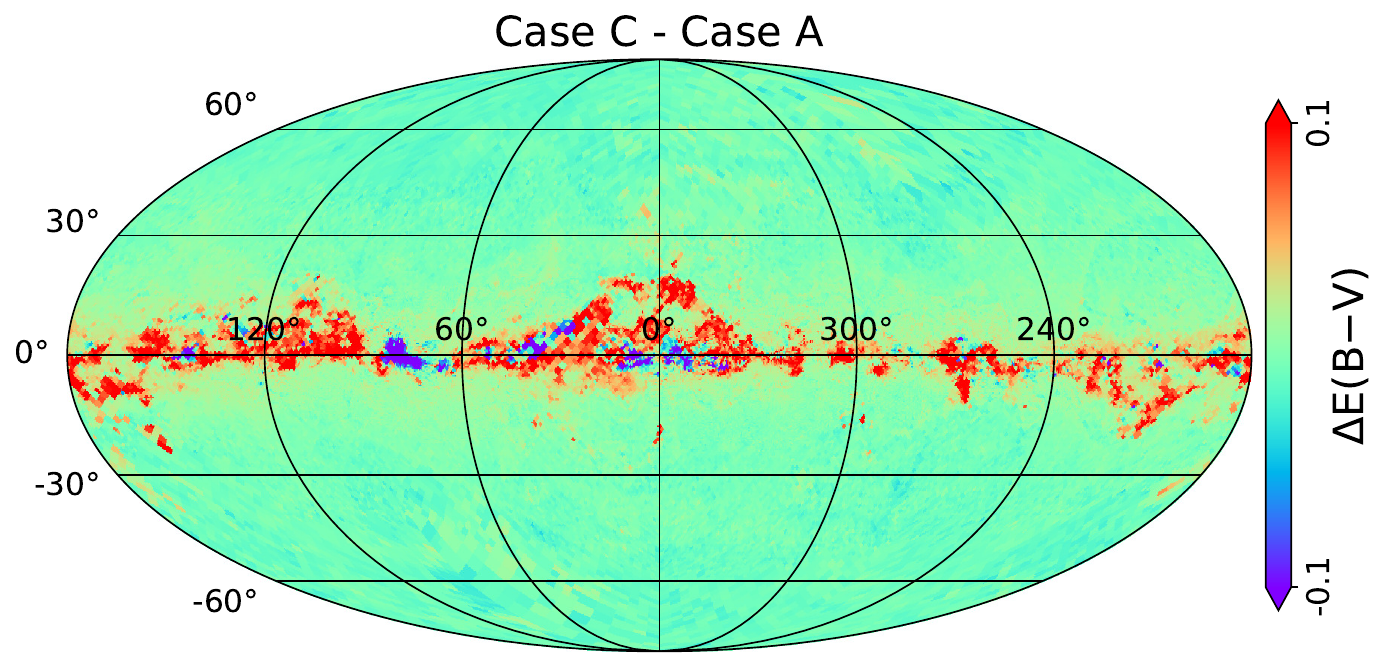}{0.32\textwidth}{\textbf{(f) $d=1584$~pc }}}
\gridline{\fig{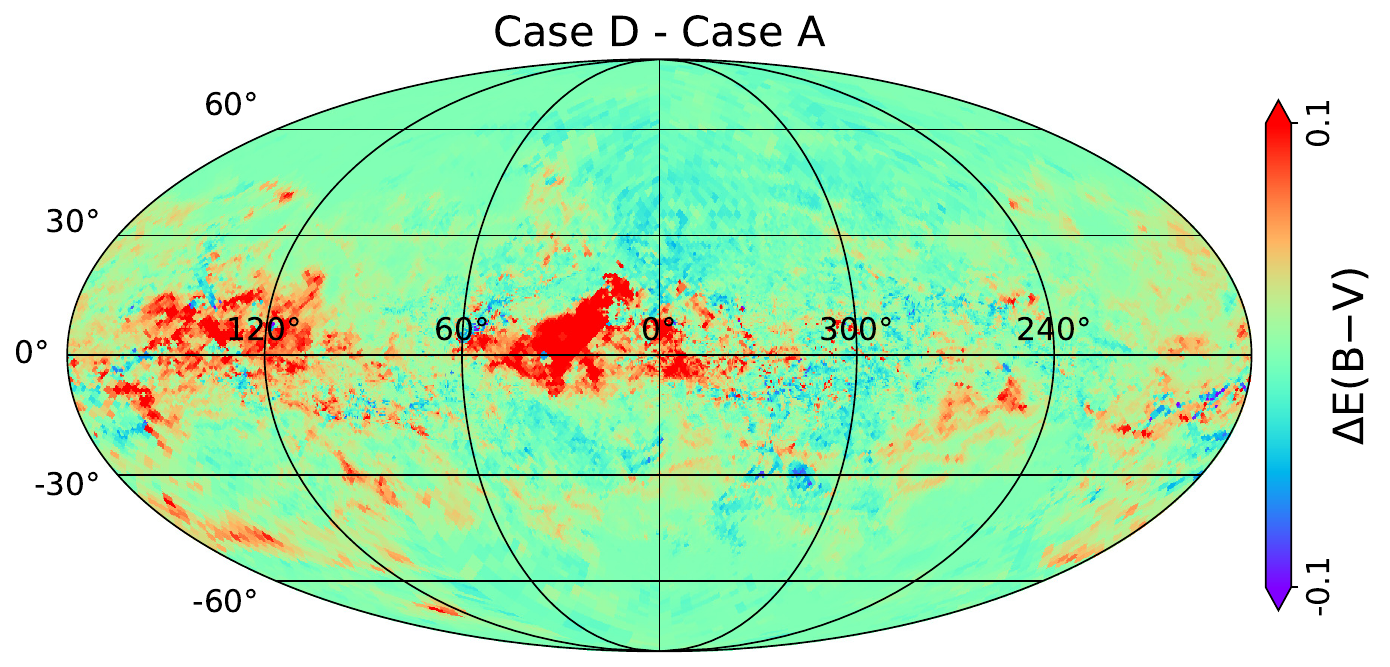}{0.32\textwidth}{\textbf{(g) $d=251$~pc}}
              \fig{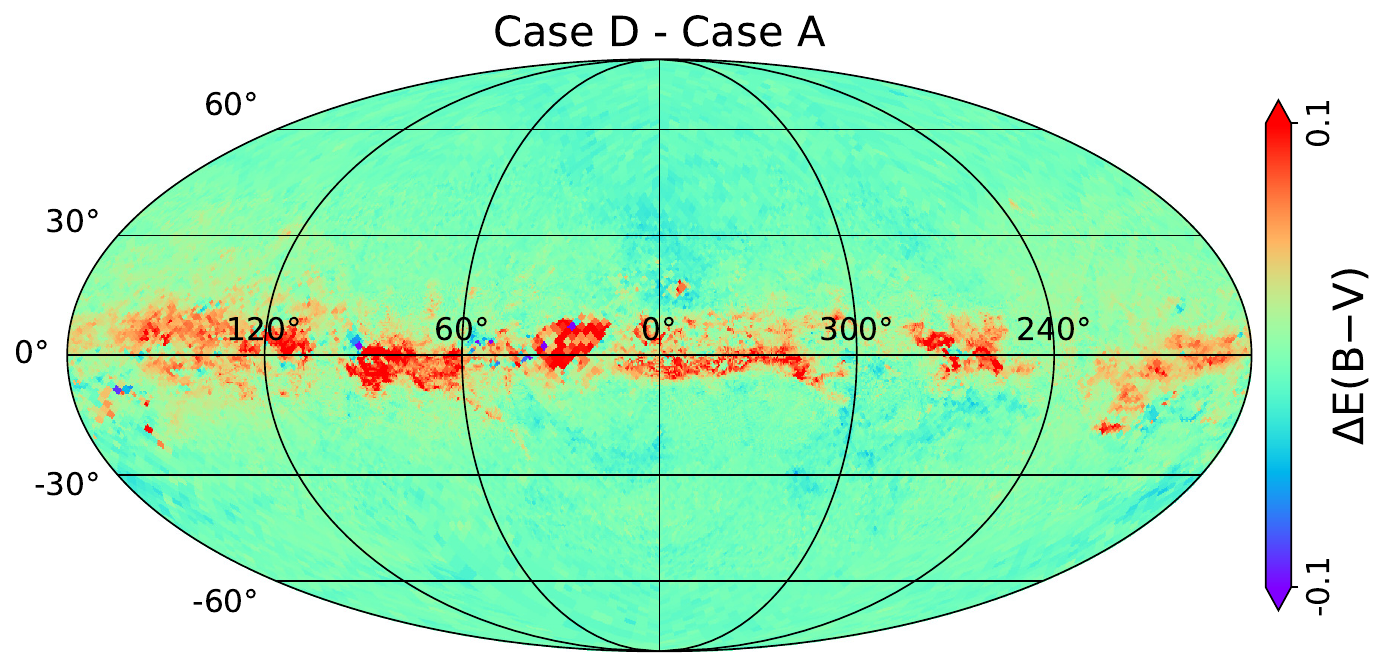}{0.32\textwidth}{\textbf{(h) $d=630$~pc }}
              \fig{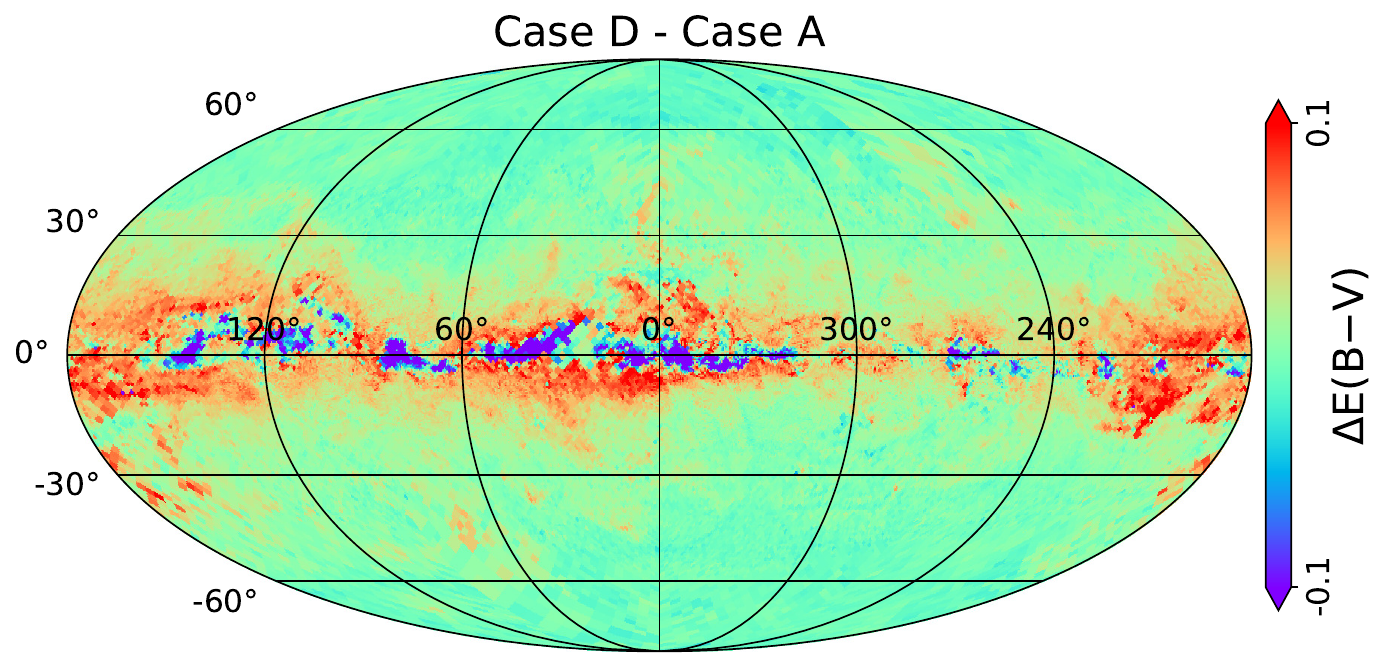}{0.32\textwidth}{\textbf{(i) $d=1584$~pc }}}
\caption{Difference in modeled $\ebv$ between different cases of parameter estimates (Table~\ref{tab:tab1}) in selected distance bins.}
\label{fig:compmap}
\end{figure*}

To gain insight into the systematic variations among different $\ebv$ maps, we compared different solutions with each other in Figure~\ref{fig:compmap} at three representative distance bins ($251$, $630$, and $1584$~pc). As a base case, we chose Case~A, and subtracted cumulative (rather than differential) $\ebv$ of the other maps from it. Overall, the differences are minimal when comparing Case~A with Case~B, suggesting that incorporating Gaia parallaxes as a constraint in our parameter estimates leads to internal consistency, irrespective of the choice of calibration. On the contrary, the solutions derived from the pure spectrophotometric approach (Case~C and Case~D) display substantial deviations from Case~A. Specifically, Figure~\ref{fig:compmap} reveals noticeable discrepancies in the direction of the Aquila Rift and the Cepheus Flare at a distance of $251$~pc. These discrepancies can be attributed to the large uncertainties in $\ebv$ and $\dmn$ for Case~C and Case~D, leading to a smearing of the $\ebv$ vs.\ $\dmn$ relation (see Fig.~\ref{fig:raw}).

In addition, large deviations are evident at $630$ and $1584$~pc along the Galactic plane, where there is substantial foreground extinction. The discrepancy in $\ebv$ is most evident with Case~D, indicating that this difference arises from a redder MS turnoff and the stronger correlation in our parameter estimates when the Gaia's prior on distance is absent. This may also indicate that our modeling encounters challenges when attempting to accurately reproduce the observed relationship between $\ebv$ and $\dmn$ at low Galactic latitudes. Because our modeling approach involves deriving structural parameters using stars in the adjacent HEALPix cells (\S~\ref{sec:model}), the problem may be attributed to the presence of compact, dense molecular clouds, causing $\ebv$ values to change rapidly over relatively short angular and heliocentric distance intervals.

Despite the substantial differences illustrated in Figure~\ref{fig:compmap}, we note that the models from Case~A or B are favored over those of Case~C and Case~D, primarily due to more precise measurements of $\ebv$ and $\dmn$. Hence, while the observed differences among different solutions serve as {\it upper} limits for the uncertainties in our estimates, the systematic difference between Case~A and Case~B may represent a more realistic measure of the true internal accuracy of our $\ebv$ estimates. Within the range of $|b|\la10\arcdeg$, the $\ebv$ difference between Case~A and Case~B tends to increase with distance. At a distance of $1584$~pc, approximately $1\%$ of the area exhibits a deviation greater than $\Delta\ebv=0.1$~mag. Above this latitude threshold, however, the median discrepancy between the two cases remains minimal at all distances: after pixelating the maps into HEALPix cells with a base resolution of $N_{\rm side}=128$, effective $2\sigma$ uncertainties, calculated from the $15.9$th and $84.1$st quartiles, amount to $0.022$, $0.010$, and $0.012$~mag at distances of $251$, $630$, and $1584$~pc, respectively. Taking these factors into consideration, it is evident that the internal accuracy of our $\ebv$ models is approximately on the order of $\sim0.005$~mag at $|b| \ga 10\arcdeg$.

\begin{figure*}
\epsscale{0.65}
  \gridline{\fig{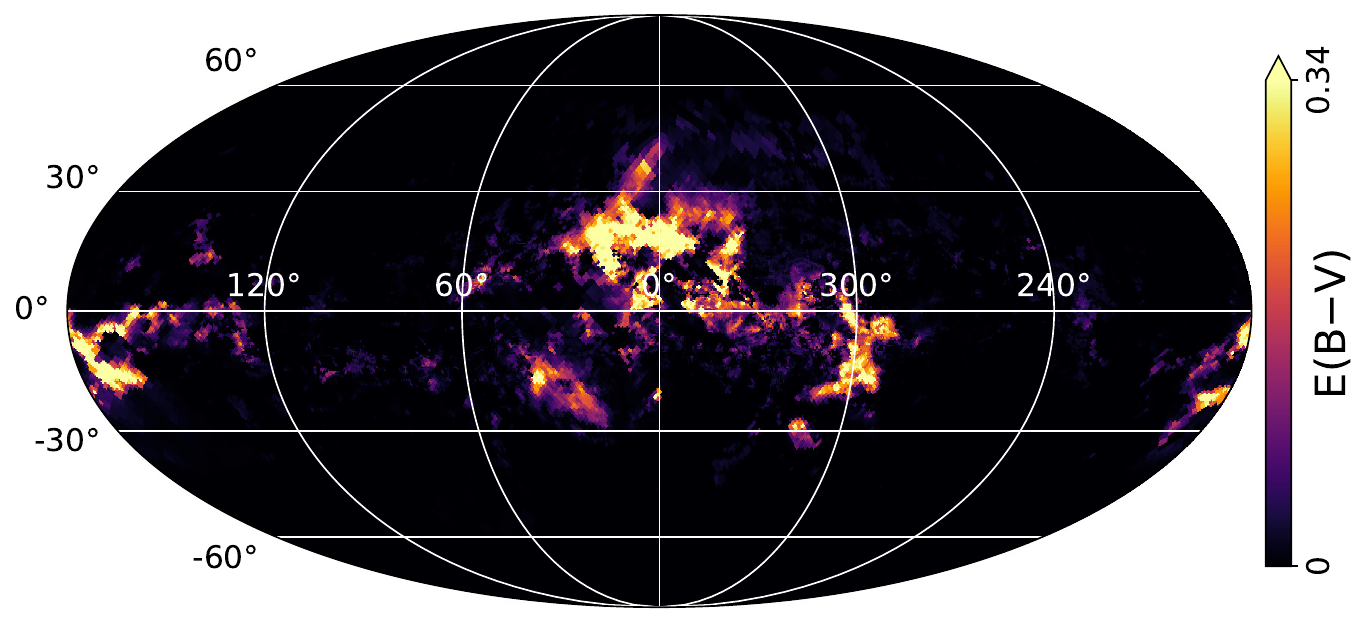}{0.4\textwidth}{\textbf{(a) $d=100$~pc }}
                \fig{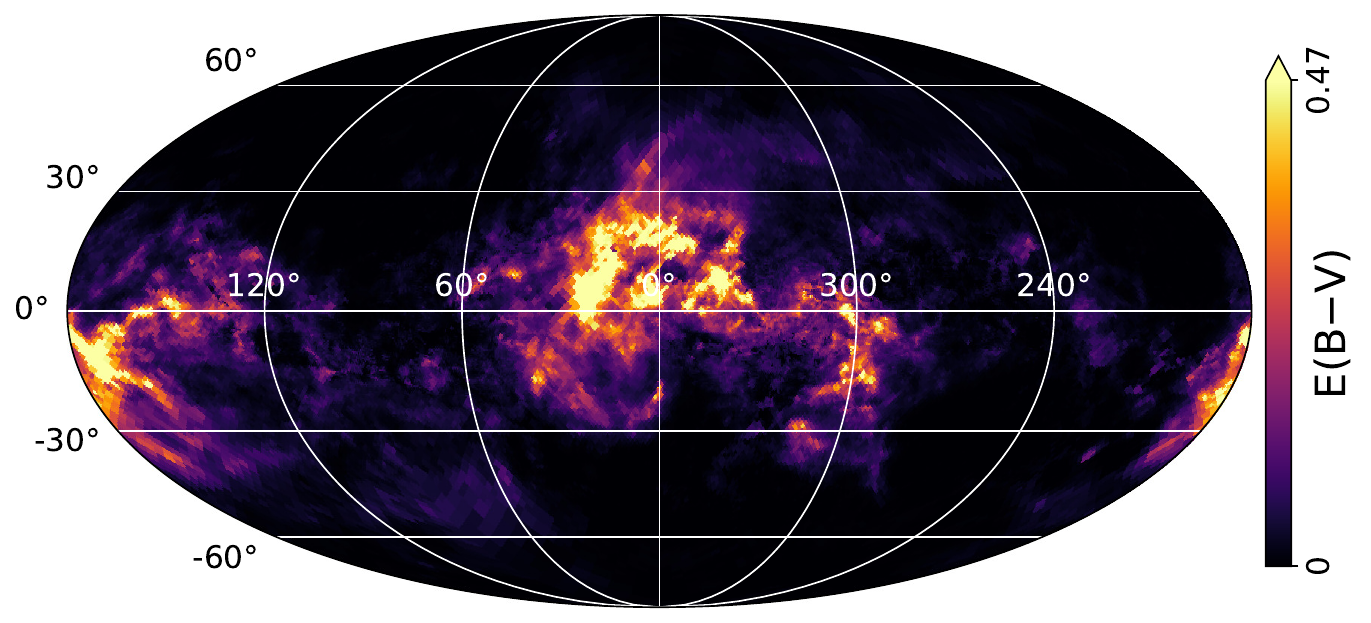}{0.4\textwidth}{\textbf{(b) $d=251$~pc }}}
  \gridline{\fig{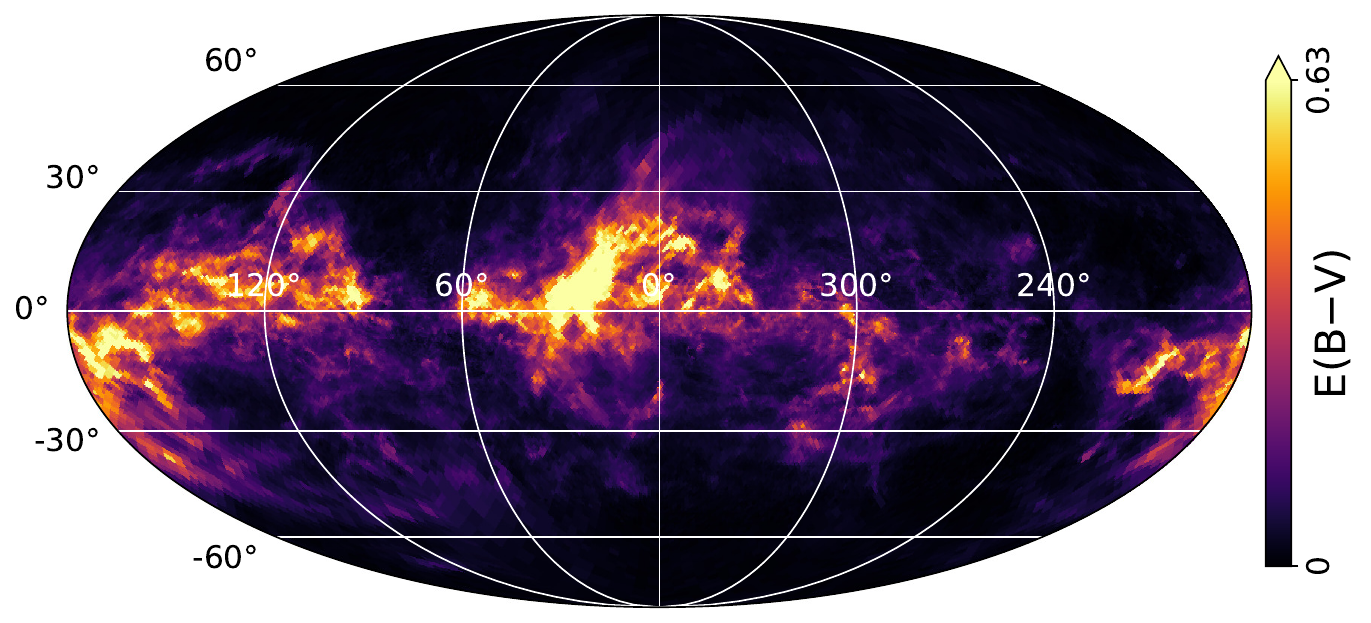}{0.4\textwidth}{\textbf{(c) $d=630$~pc }}
                \fig{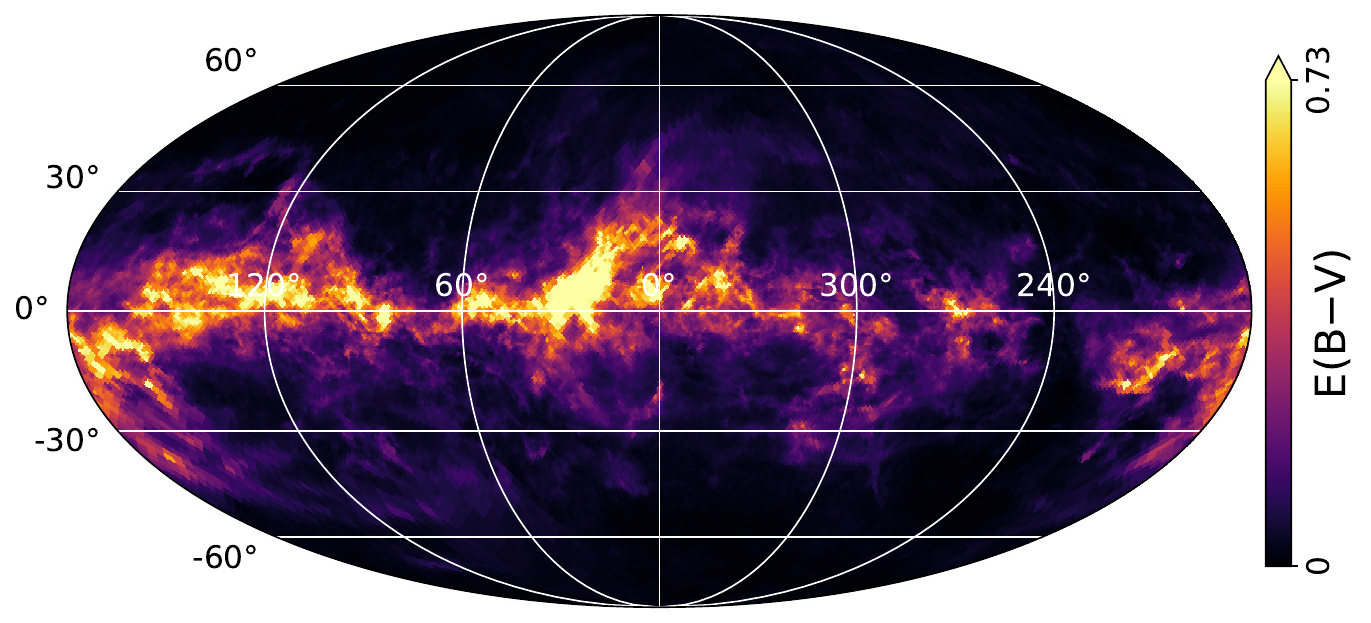}{0.4\textwidth}{\textbf{(d) $d=1000$~pc }}}
  \gridline{\fig{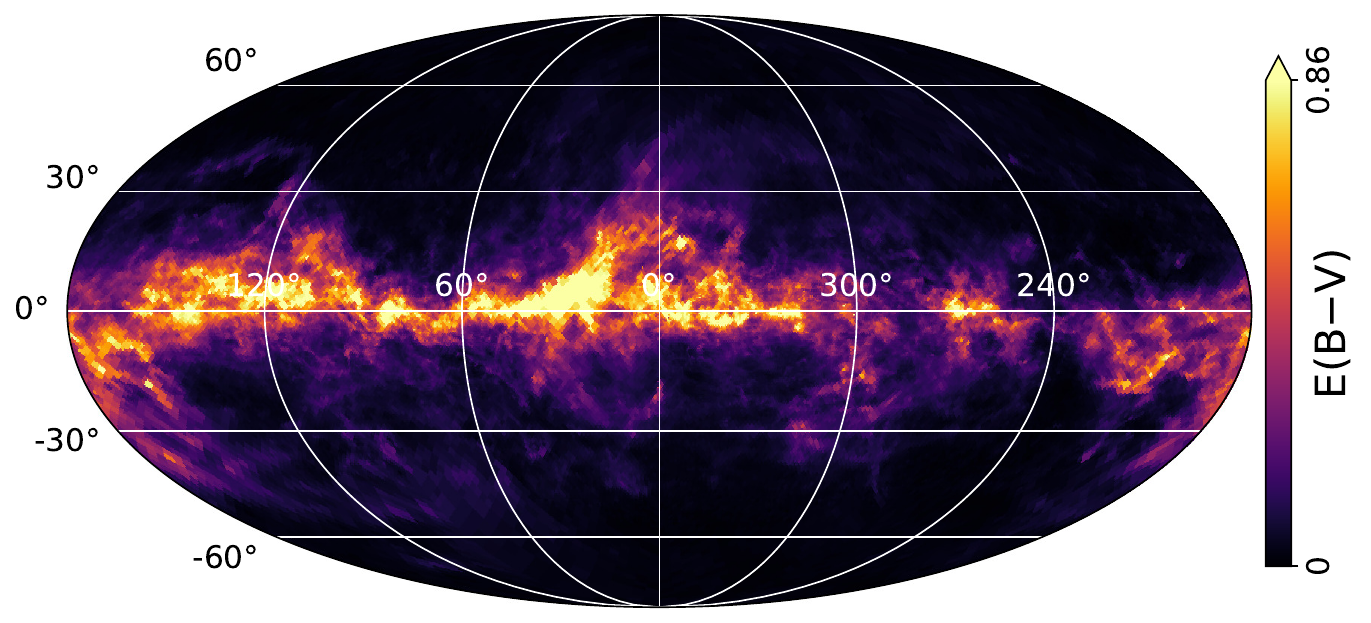}{0.72\textwidth}{\textbf{(e) $d=1584$~pc }}}
\caption{Average $\ebv$ from Case~A and Case~B (Table~\ref{tab:tab1}) in the Galactic coordinate system as a function of heliocentric distance.}
\label{fig:ebvmap2}
\end{figure*}

Both Case~A and Case~B rely on parameter estimates with constraints from Gaia parallaxes but diverge in terms of their model calibration. Nonetheless, their extinction models are equally valid, as we do not favor any specific calibration method for our isochrones (also see discussions in Paper~IV). Hence, we merged the extinction models from both cases by computing their averages in bins of $\Delta\dmn=0.02$ mag, spanning from $\dmn=5$~mag to $\dmn=12.4$ mag. The resulting composite 3D reddening map is displayed in Figure~\ref{fig:ebvmap2} in selected distance bins. We incorporated the case at a distance of $100$~pc in panel~(a) to ensure completeness, although our models exhibit reduced accuracy for distances less than $\sim250$~pc owing to the limited number of nearby stars in the XP MS sample.

\section{Comparisons with Previous Extinction Maps}\label{sec:comp}

\subsection{\citet{schlegel:98}}

\begin{figure}
\centering
\gridline{\fig{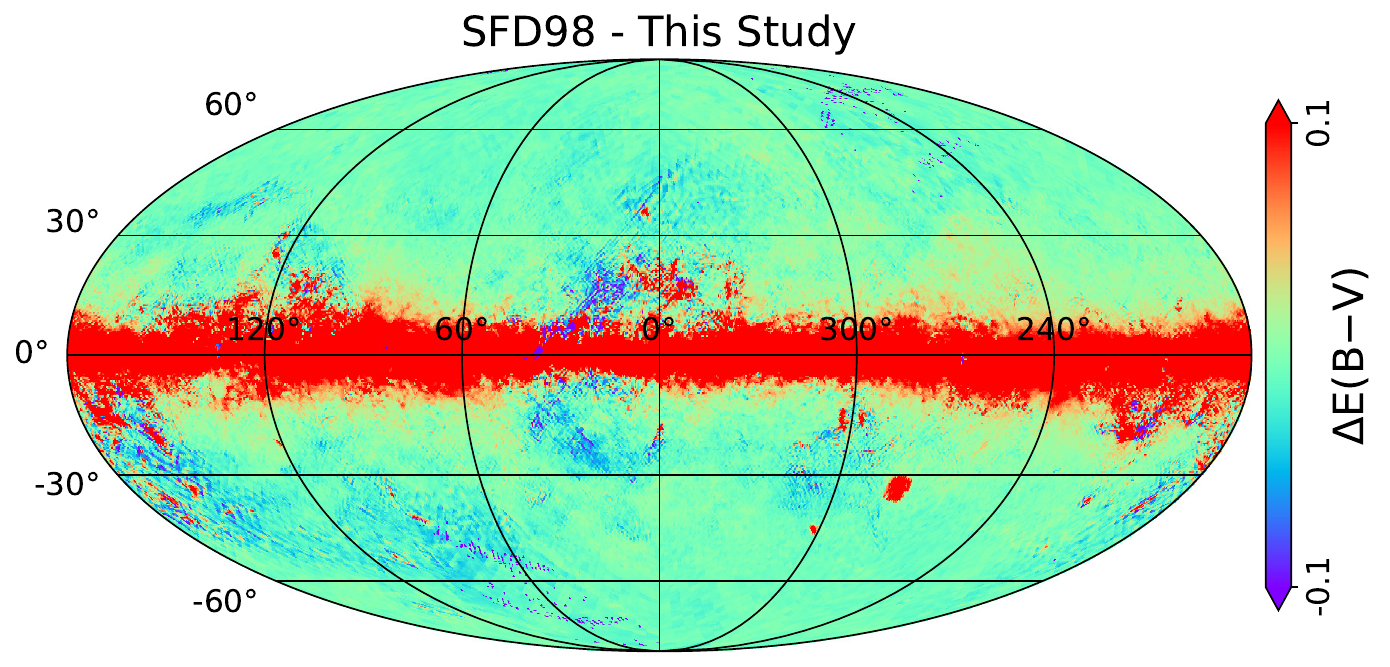}{0.42\textwidth}{\textbf{ }}}
\gridline{\fig{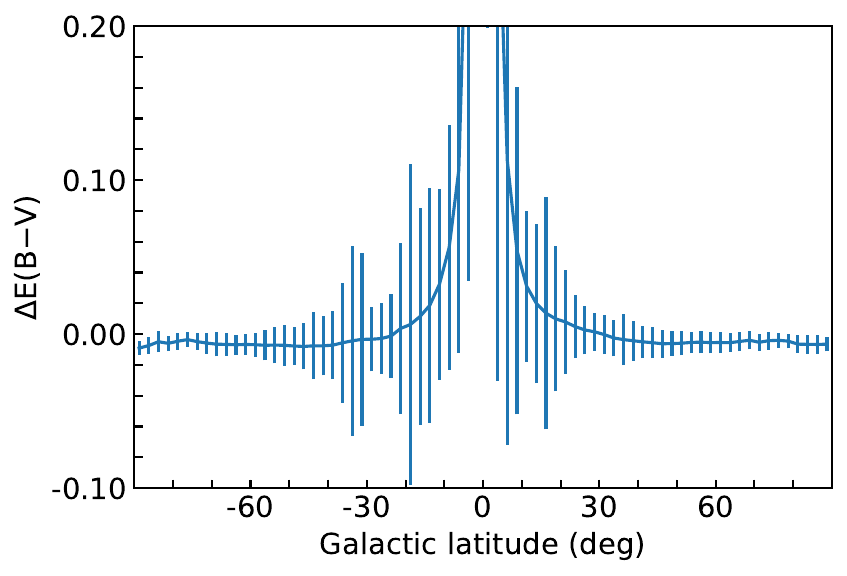}{0.38\textwidth}{\textbf{ }}}
\caption{Comparison to the $\ebv$ map of \citet{schlegel:98}. Top panel: differences from the averaged $\ebv$ map at $d=1584$~pc (panel~(e) in Fig.~\ref{fig:ebvmap2}), in the sense of their $\ebv$ minus our values. Bottom panel: the solid line represents the median difference in $\ebv$, with error bars indicating the standard deviation.}
\label{fig:sfd}
\end{figure}

In Figure~\ref{fig:sfd}, we present a comparison between our averaged reddening map, developed in the preceding section (Fig.~\ref{fig:ebvmap2}), and the widely used extinction map from \citet[][hereafter SFD98]{schlegel:98}. The latter map offers a cumulative measure of the dust column along each line of sight, utilizing an all-sky far-infrared dust emission map. In this comparison, we selected our map at a distance of $1584$~pc, or $\dmn=11$~mag, to ensure an adequate assessment, leveraging reddening measurements from stars located at sufficiently large distances. We applied a scaling factor of $0.86$ to the SFD98 map to correct for the systematic difference from stellar locus measurements \citep{schlafly:11}. Given that the SFD98 dataset has a finer spatial resolution ($\sim6\arcmin$) in comparison to our baseline map with $N_{\rm side}=128$ (having a mean spacing of $30\arcmin$), we computed the median $\ebv$ from SFD98 for our XP MS sample within each HEALPix cell.

The result, presented in Figure~\ref{fig:sfd}, reveals an excess in the SFD98 dust map near the Galactic plane. This surplus is attributed to a significant volume of dust located beyond the distance covered by our reddening map. This variation is also clearly evident in the bottom panel, where the median difference in $\ebv$ is illustrated against Galactic latitude, organized into bins of $\Delta b = 2.5\arcdeg$; it exceeds $\Delta \ebv = 0.1$~mag for $|b| < 10\arcdeg$. A similar surplus in $\ebv$ is also evident in the direction of the Magellanic Clouds, at $(l, b)=(280\arcdeg, -33\arcdeg)$ and $(303\arcdeg, -44\arcdeg)$. While SFD98 appropriately accounts for the dust present within these galaxies, our localized dust map does not encompass them. Despite the inherent limitations in this comparison, however, Figure~\ref{fig:sfd} demonstrates a rapid convergence of both maps for $|b| > 10\arcdeg$. Specifically, at $|b| = 30\arcdeg$, the difference in $\ebv$ is merely $0.003$~mag, and it remains within $0.01$~mag at larger $|b|$.

While this level of agreement already lends substantial support to the accuracy of the zero-point in our $\ebv$ measurements, there is a tantalizing possibility that the small residual discrepancy observed at high Galactic latitudes ($\sim0.005$~mag) can be attributed to a zero-point offset within SFD98 itself. Based on more accurate measurements of dust temperature and column density using Planck far-infrared observations, \citet{planck:14} revealed that the original $\ebv$ estimates in SFD98 are smaller by approximately $0.003$--$0.006$~mag in the diffuse interstellar medium ($\ebv < 0.04$~mag). This discrepancy is of the same order as the difference observed at high Galactic latitudes in Figure~\ref{fig:sfd}.

\subsection{\citet{green:19}}

\begin{figure*}
\centering
  \gridline{\fig{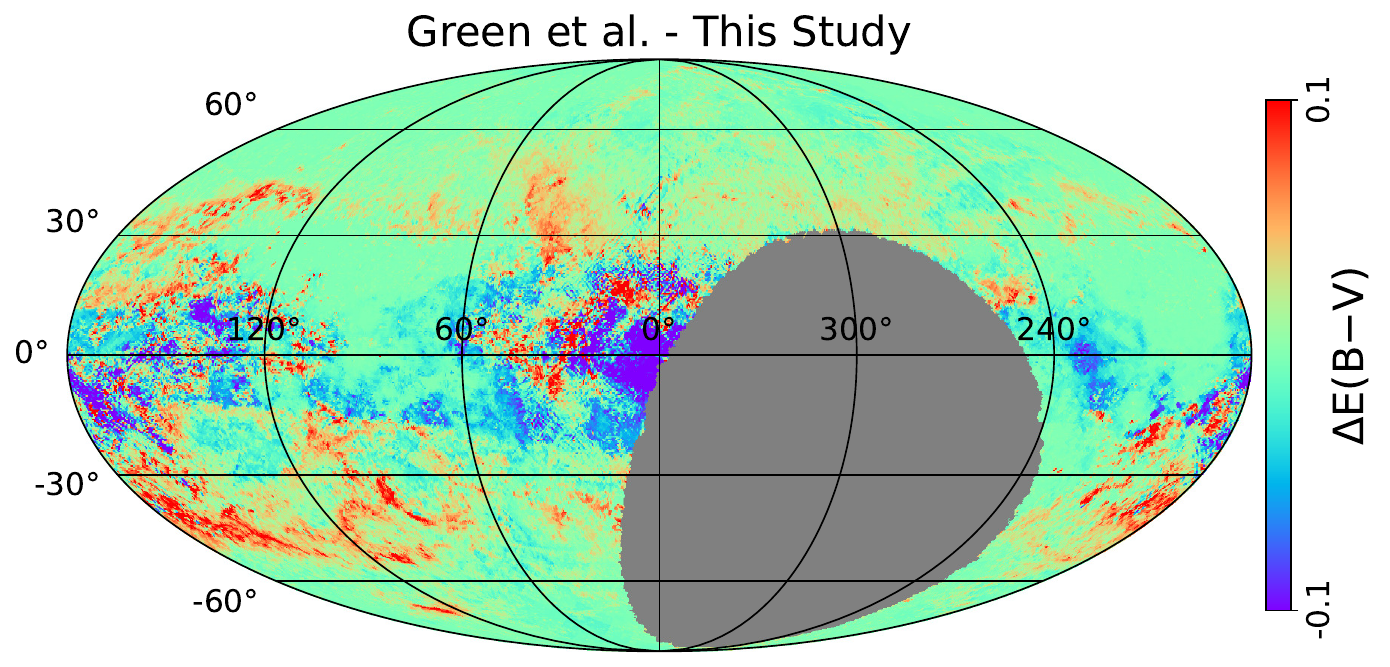}{0.3\textwidth}{\textbf{(a) $d=251$~pc }}
                \fig{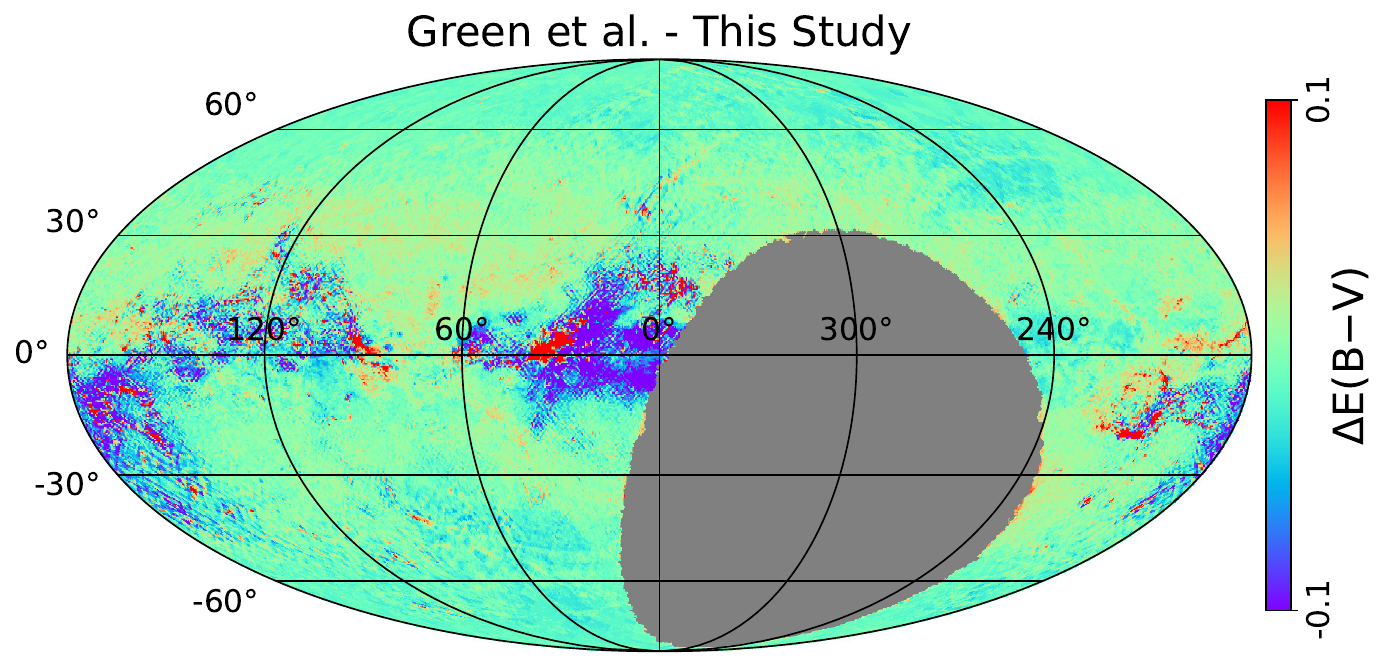}{0.3\textwidth}{\textbf{(b) $d=630$~pc }}
                \fig{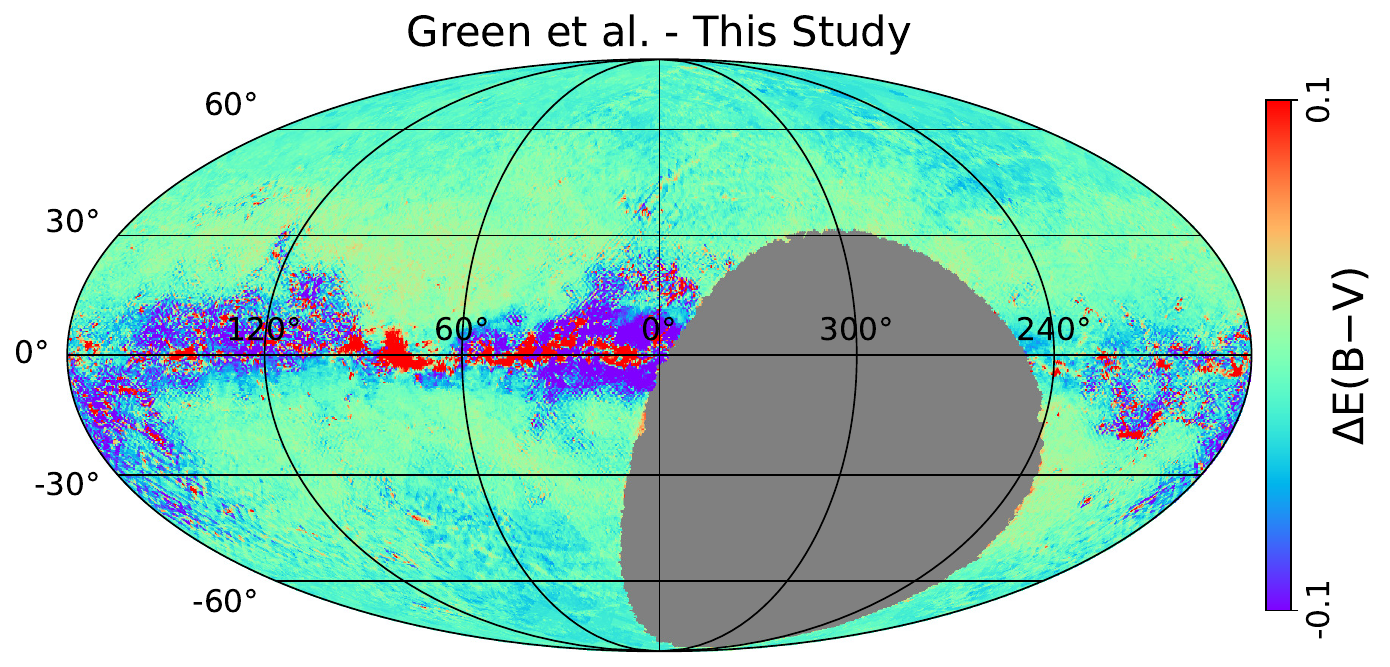}{0.3\textwidth}{\textbf{(c) $d=1584$~pc }}}
  \gridline{\fig{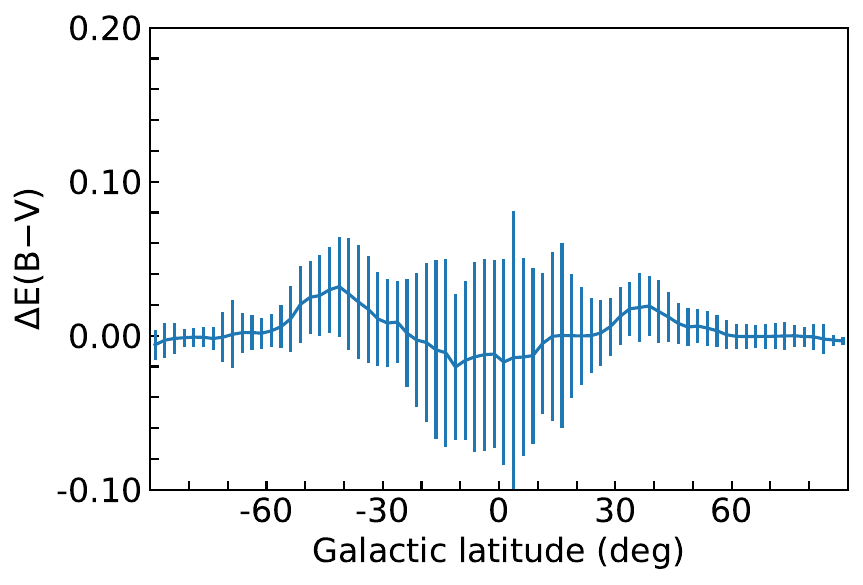}{0.26\textwidth}{\textbf{(d) $d=251$~pc }}
                \fig{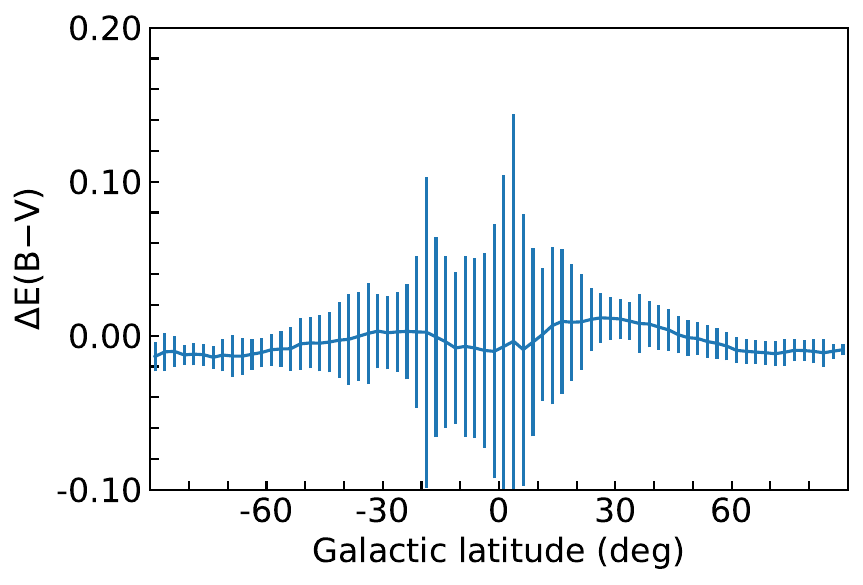}{0.26\textwidth}{\textbf{(e) $d=630$~pc }}
                \fig{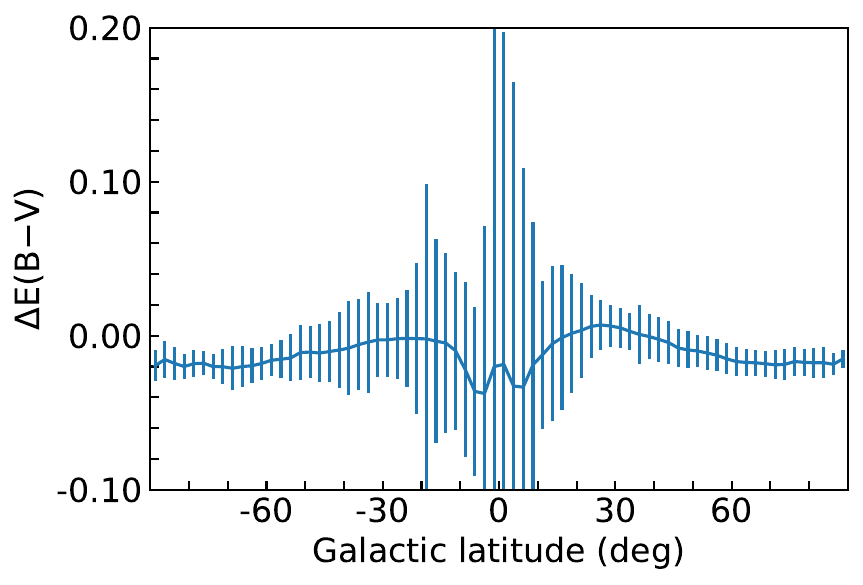}{0.26\textwidth}{\textbf{(f) $d=1584$~pc }}}
\caption{Comparison to the $\ebv$ map of \citet{green:19}. Top panels: comparisons are shown with respect to our averaged $\ebv$ maps at selected distance bins ($d=251$, $630$, and $1584$~pc), in the sense of their $\ebv$ minus our values. Bottom panels: median deviations as a function of Galactic latitude in intervals of $2.5\arcdeg$. Error bars denote their standard deviations.}
\label{fig:green}
\end{figure*}

A more rigorous comparison of our extinction map can be facilitated by utilizing the 3D extinction map presented in \citet{green:19}. They employed data from Pan-STARRS~1 (PS1) and 2MASS photometry in conjunction with Gaia DR2 parallaxes and constructed a high-resolution extinction map by subdividing HEALPix cells down to a level of $N_{\rm side}=1024$. The top panels of Figure~\ref{fig:green} illustrate a comparison between our averaged extinction map and their `SFD-like' $\ebv$ values in the Galactic coordinate system at selected distance bins, $\dmn=7$, $9$, and $11$~mag. For consistency, we scaled their reddening values by a factor of $0.86$, similar to our treatment of the SFD98 map. To ensure a consistent spatial resolution, we degraded their high-resolution map to match our base HEALPix resolution of $N_{\rm side}=128$.

As illustrated in Figure~\ref{fig:green}, the extinction map provided by \citet{green:19} is not available for regions with decl.\ less than $-30\arcdeg$ owing to limitations in PS1 photometry. Nevertheless, it is evident that regions with noticeable discrepancies are more prominent in areas characterized by relatively high values of $\ebv$, although the discrepancy does not follow a consistent pattern. In the bottom panels, the median deviations with respect to Galactic latitude are shown as solid lines in bins of $2.5\arcdeg$. To provide a sense of the dispersion, the standard deviations of these differences are represented by error bars. At a distance of $251$~pc, the median deviation reaches up to $\Delta\ebv=0.03$~mag, while it stays below $0.015$~mag at $630$~pc. At a distance of $1584$~pc, a more pronounced deviation is evident near the Galactic plane, but the median difference remains modest at $|b|>10\arcdeg$, rarely exceeding $\Delta\ebv=0.02$~mag. The observed discrepancies may arise from differences in the adopted distances (Gaia DR2 vs.\ DR3), extinction laws, fundamental stellar color relationships, and the input data. Moreover, in regions characterized by high extinction, the lower angular resolution of our map could also influence the observed differences.

\subsection{\citet{zhang:23b}}

\begin{figure*}
\centering
  \gridline{\fig{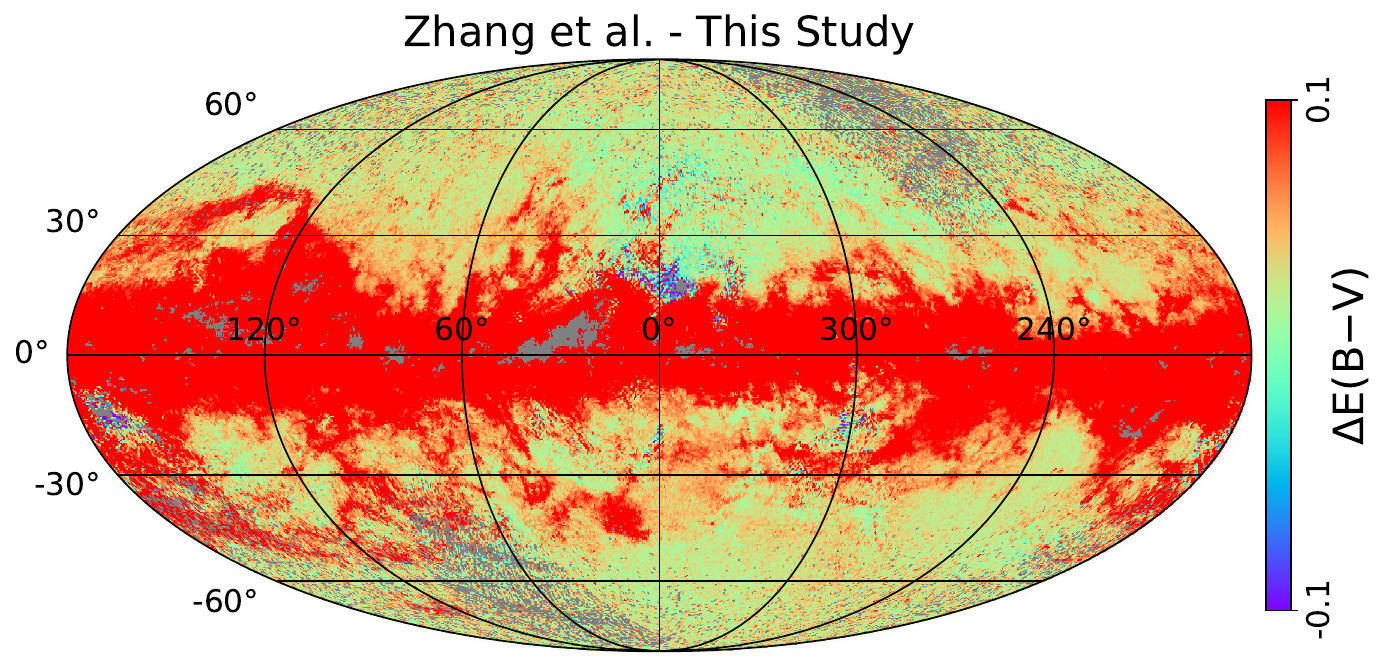}{0.3\textwidth}{\textbf{(a) $d=251$~pc }}
                \fig{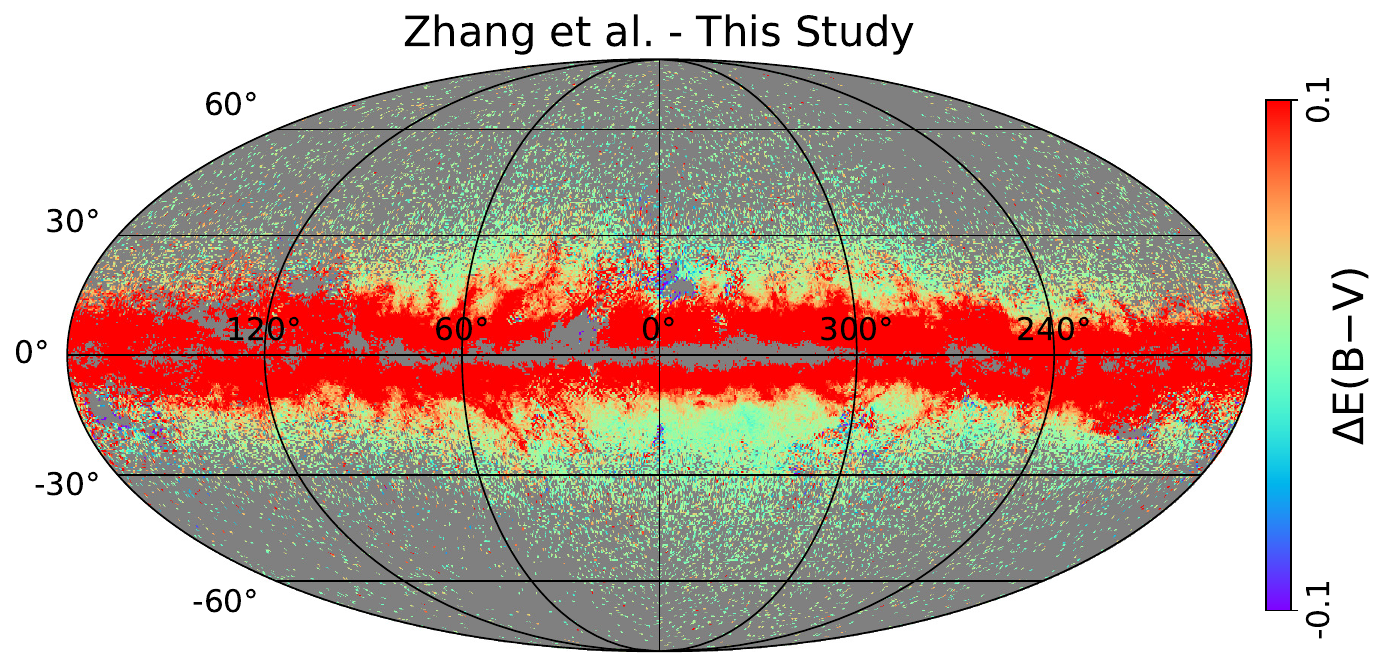}{0.3\textwidth}{\textbf{(b) $d=630$~pc }}
                \fig{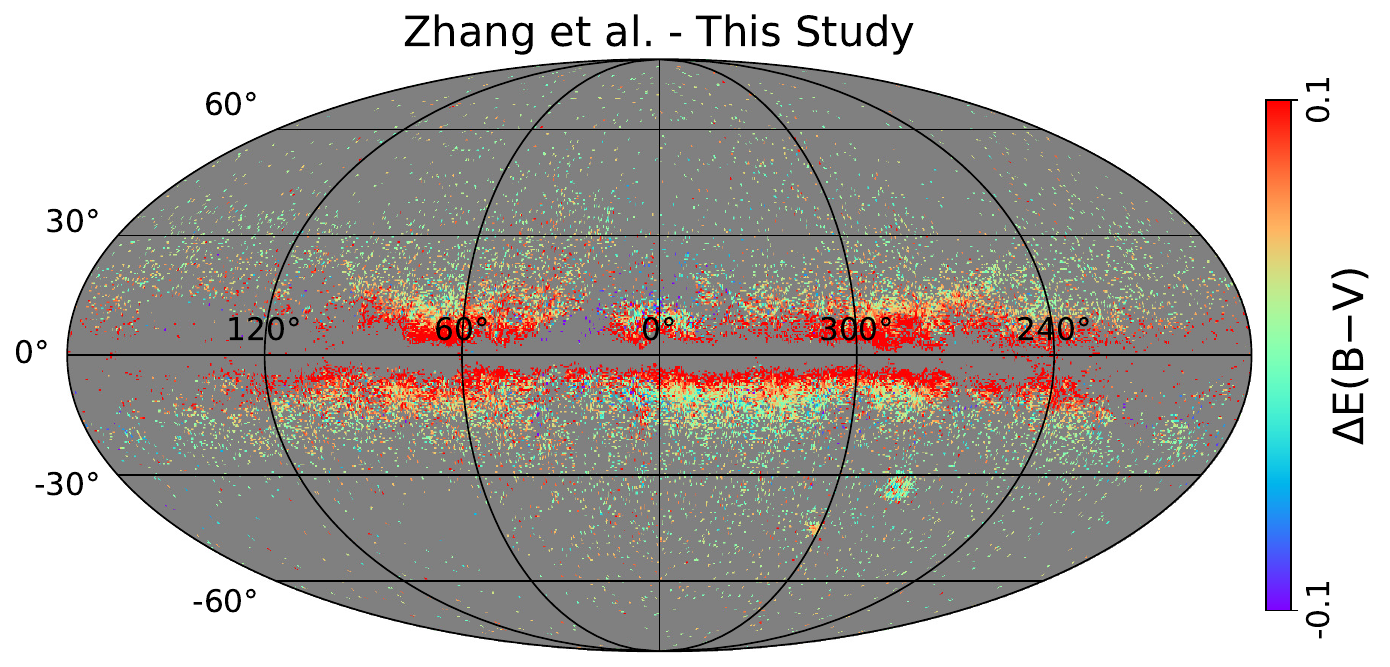}{0.3\textwidth}{\textbf{(c) $d=1584$~pc }}}
  \gridline{\fig{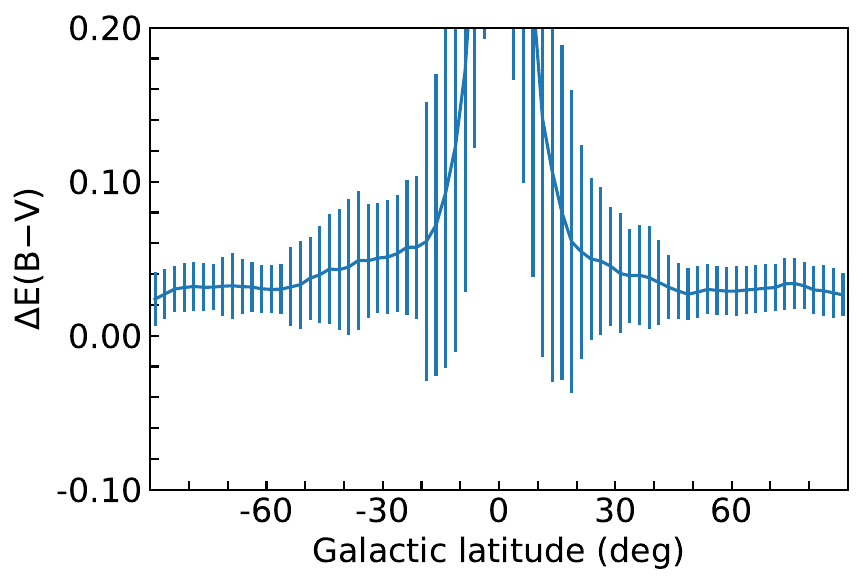}{0.26\textwidth}{\textbf{(d) $d=251$~pc }}
                \fig{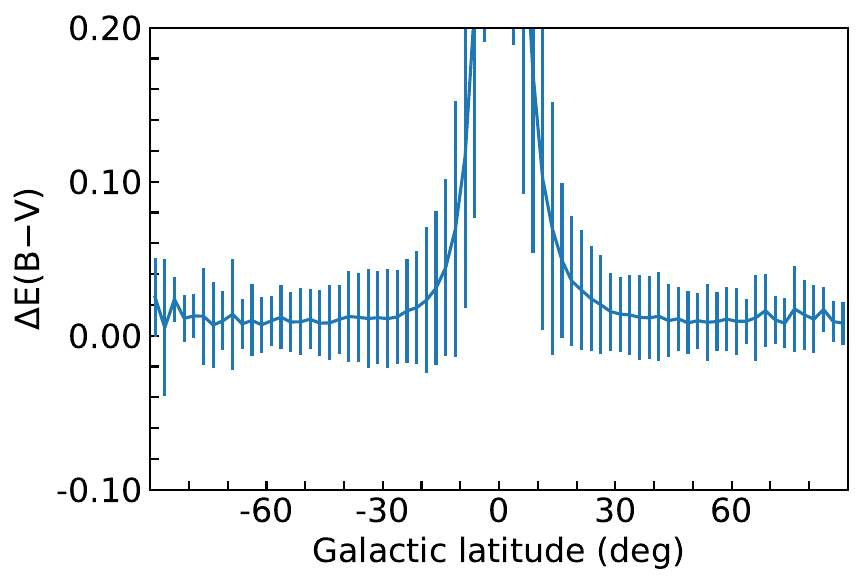}{0.26\textwidth}{\textbf{(e) $d=630$~pc }}
                \fig{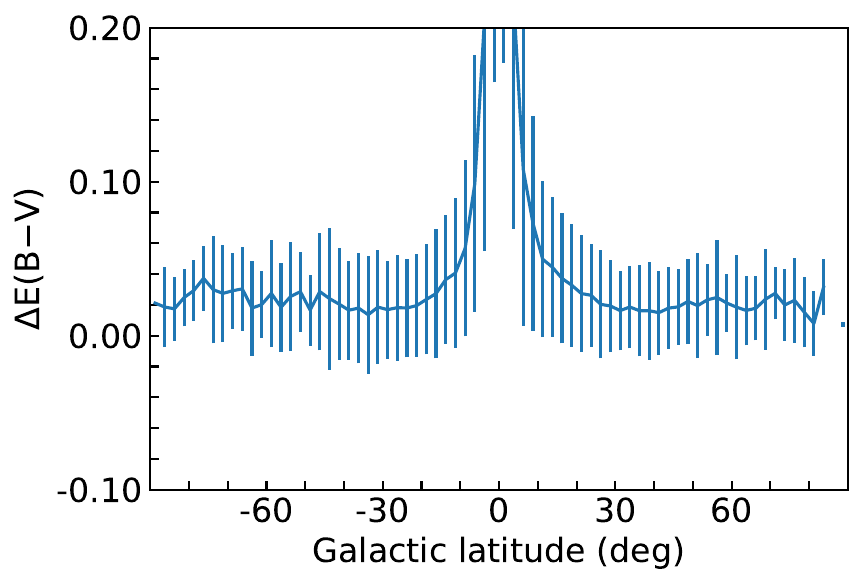}{0.26\textwidth}{\textbf{(f) $d=1584$~pc }}}
\caption{Same as in Figure~\ref{fig:green}, but displaying a comparison with \citet{zhang:23b}.}
\label{fig:zhang}
\end{figure*}

Figure~\ref{fig:zhang} provides a comparison to more recent, all-sky extinction measurements presented in \citet{zhang:23b}. Similar to ours, they utilized XP spectra to derive line-of-sight extinction. Following their recommendations, we applied a set of quality flags, including {\tt quality\_flags} $<8$, {\tt teff\_confidence} $>0.5$, {\tt feh\_confidence} $>0.5$, $\sigma(\teff)<500$~K, $\sigma({\rm [Fe/H]})<0.5$~dex, and a maximum $20\%$ uncertainty in distance. In each of the HEALPix cells with $N_{\rm side}=128$, we computed a weighted mean of their reddening estimates (denoted as `$E$'), which closely aligns with the $\ebv$ scale used in our work, differing by $1\%$--$3\%$. The comparisons depicted in Figure~\ref{fig:zhang} focus on three specific distance bins, where stars from the \citeauthor{zhang:23b} catalog are situated within $|\Delta\dmn|<0.1$~mag of the given distance modulus. There are numerous missing pixels in these maps (gray colors), which are primarily due to the lack of measurements that pass the aforementioned quality flags in \citeauthor{zhang:23b} within this narrow distance interval.\footnote{The situation can potentially be improved using extinction models, but \citet{zhang:23b} only made their estimates of individual stars' reddening public. There is a recent study \citep{edenhofer:23} that utilized \citeauthor{zhang:23b} estimates to construct a 3D extinction model, but it is limited to $1.25$~kpc.}

Unlike the earlier comparison with \citet{green:19}, their reddening estimates reveal substantial discrepancies from ours throughout all distance bins. At high Galactic latitudes ($|b| > 30\arcdeg$), the difference in reddening amounts to up to $\Delta\ebv\approx0.05$~mag. Given the agreement of our reddening measurements at these latitudes with both SFD98 and \citet{green:19}, the observed deviation is plausibly attributed to an inherent offset in \citeauthor{zhang:23b}. Further concern arises from the presence of positive residuals originating from the Large and Small Magellanic Clouds, even within these local volumes (panel~(e)): $\Delta\ebv=0.019\pm0.006$~mag and $0.044\pm0.009$~mag, respectively. This suggests a potential misclassification of bright stars belonging to the Magellanic Clouds as local Milky Way stars in their study.

\begin{figure*}
\centering
  \gridline{\fig{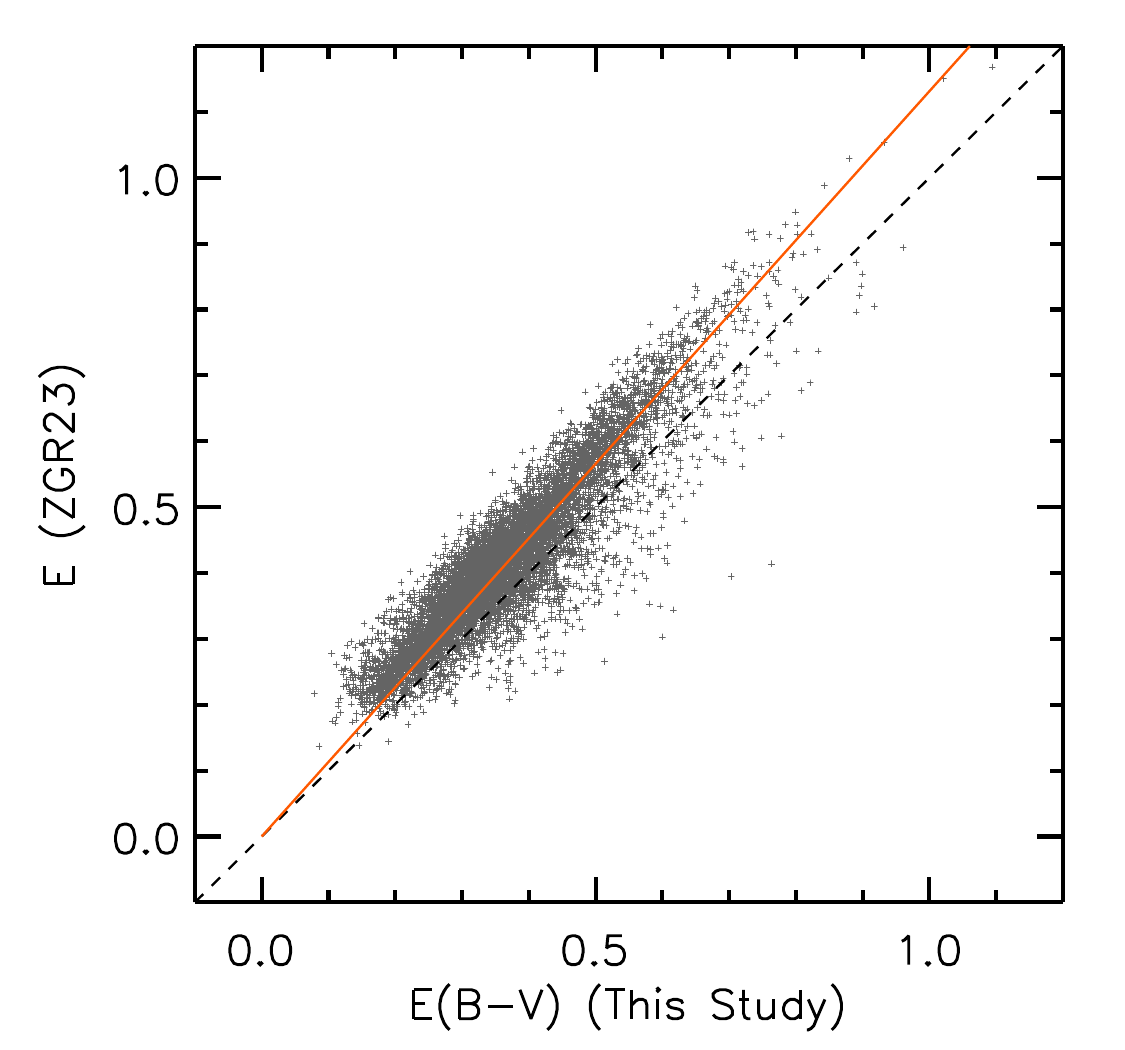}{0.34\textwidth}{\textbf{ }}
                \fig{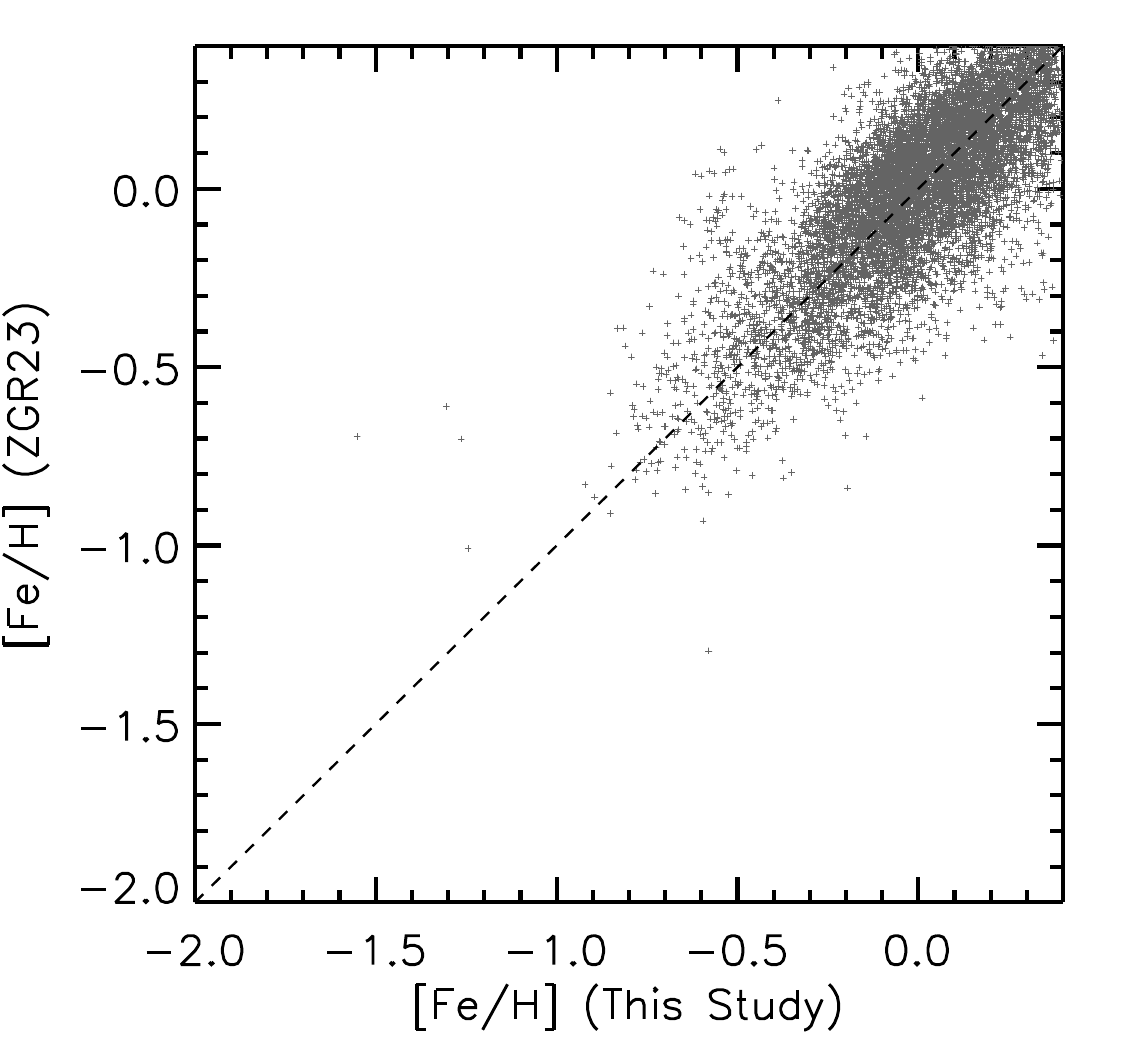}{0.34\textwidth}{\textbf{ }}}
  \gridline{\fig{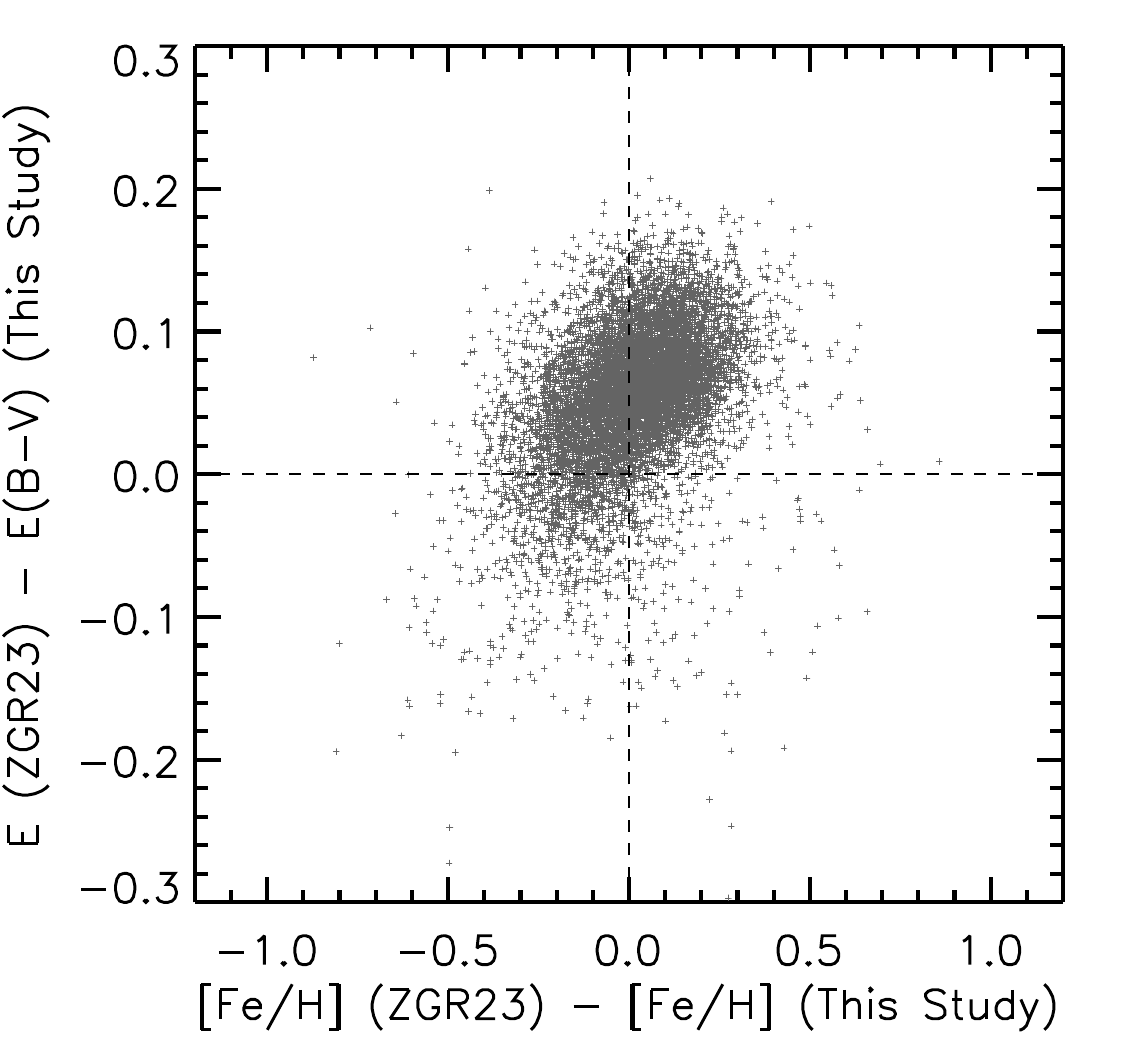}{0.34\textwidth}{\textbf{ }}
                \fig{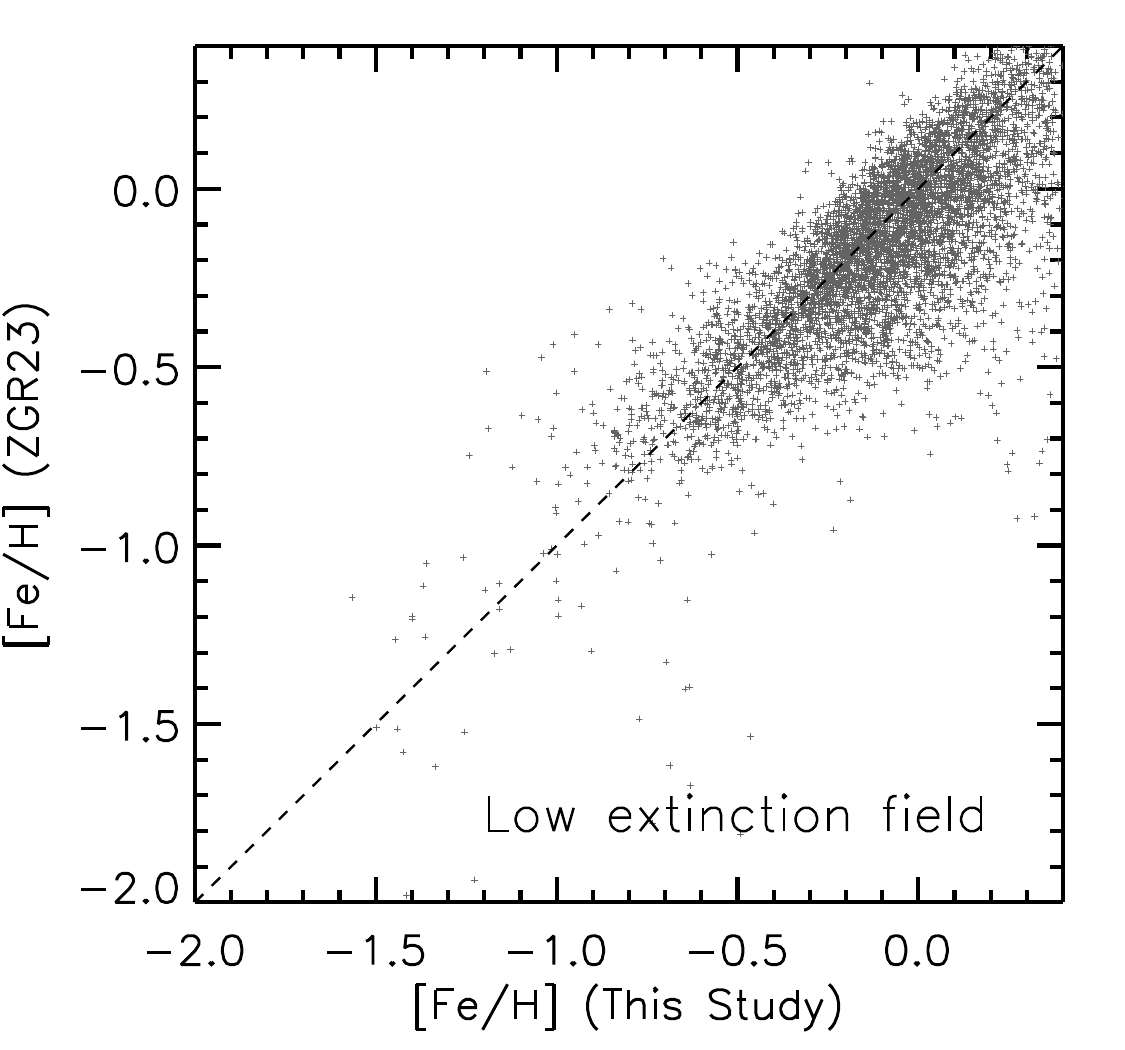}{0.34\textwidth}{\textbf{ }}}
\caption{Comparison of reddening and metallicity estimates between \citet{zhang:23b} and Case~A for the identical XP sample from a region with a $3\arcdeg$ radius, centered at $(l,b)=(20\arcdeg,20\arcdeg)$. An additional [Fe/H] comparison is presented in the bottom right panel along a line-of-sight with low extinction at $(l,b)=(300\arcdeg,45\arcdeg)$. Unity lines are represented by dashed lines, while null differences are displayed in the bottom left panel.}
\label{fig:zhang2}
\end{figure*}

Star-by-star comparisons of reddening and [Fe/H] are illustrated in Figure~\ref{fig:zhang2}. To facilitate this comparison, we restricted our analysis to stars within a $3\arcdeg$ radius centered at $(l,b)=(20\arcdeg,20\arcdeg)$, utilizing the same quality flags for data selection. The linear fit to the data indicates that their $E$ estimates are $13\%$ larger than our values (orange line in the top left panel), which is still significantly larger than expected from the different definitions of reddening parameters. While there is a significant correlation between reddening and [Fe/H] in the parameter estimation (bottom left panel), the discrepancy cannot be attributed to a systematic deviation in their metallicity scale, as the two independent methods yield essentially consistent metallicity estimates (top right panel). This is further supported by the comparison in the bottom right panel, showing [Fe/H] along a line of sight at $(300^\circ, 45^\circ)$ with small foreground extinction ($E(B-V) < 0.1$). In Appendix~\ref{sec:comp2}, we further validate our parameter estimation by comparing it with metallicity estimates from GALAH and APOGEE.

\section{Systematic Changes in the Extinction Laws}\label{sec:rv}

\begin{figure}
\centering
\epsscale{0.65}
  \gridline{\fig{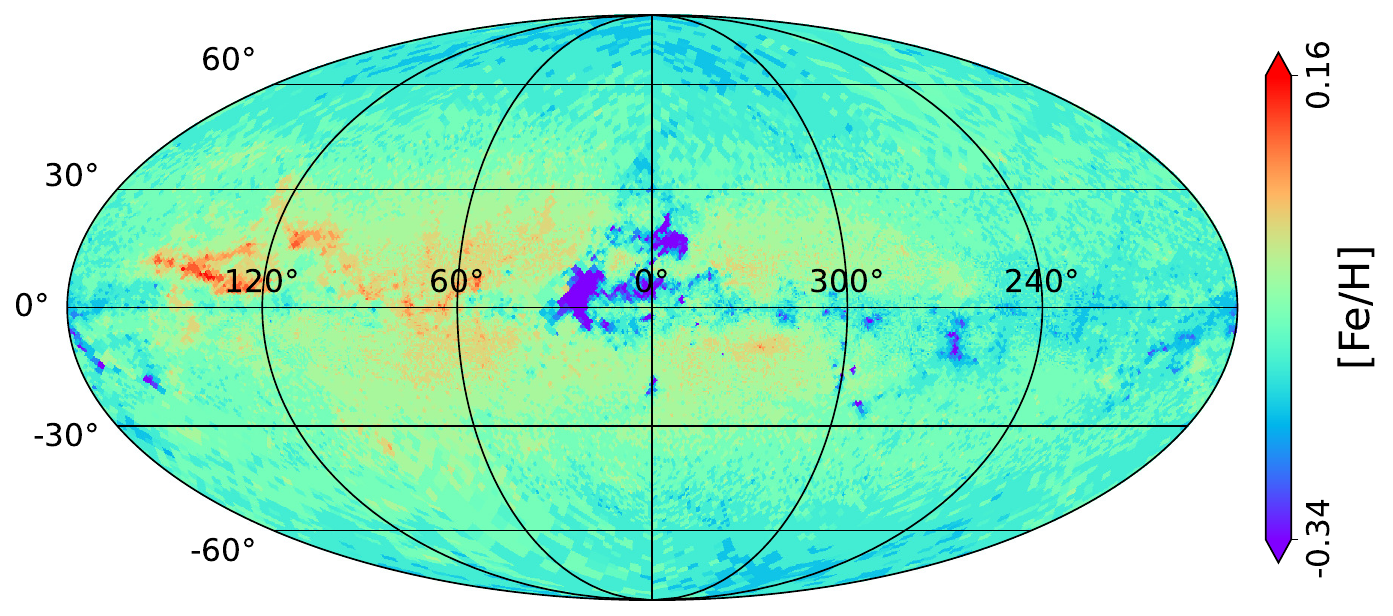}{0.48\textwidth}{\textbf{(a) Mean [Fe/H] from Case~A }}}
  \gridline{\fig{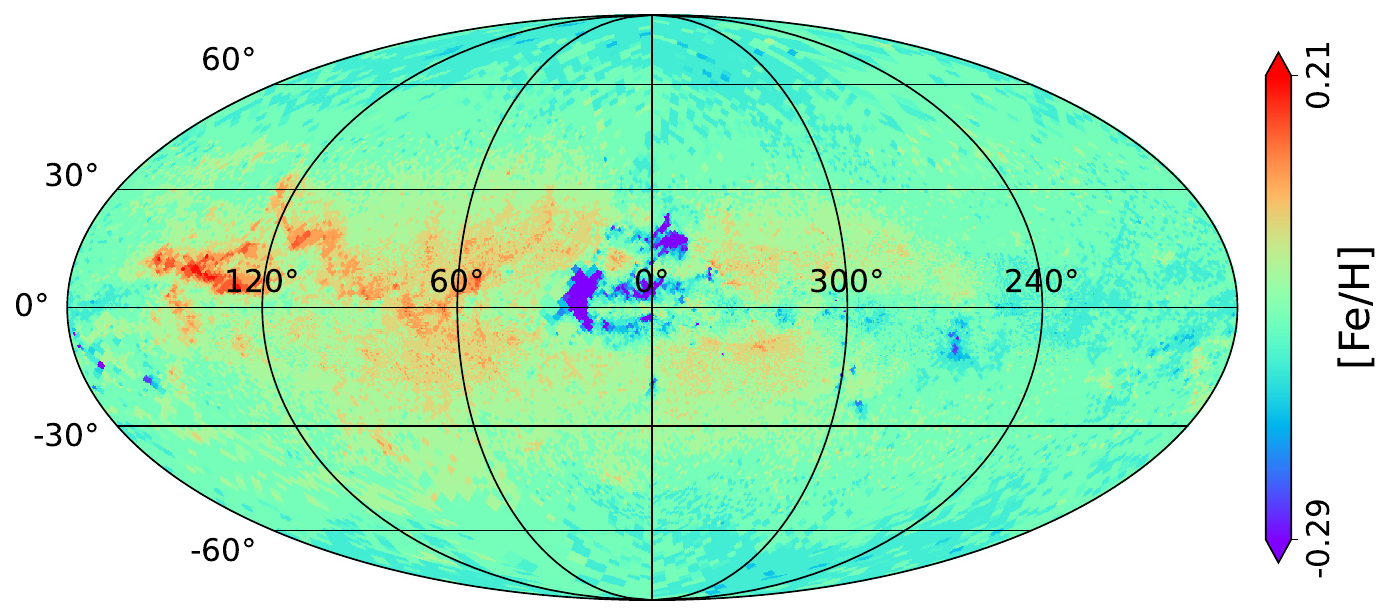}{0.48\textwidth}{\textbf{(b) Mean [Fe/H] froms Case~B }}}
\caption{Mean metallicity estimates on the sky in the Galactic coordinate system assuming $R_V=3.1$, for (a) Case~A and (b) Case~B.}
\label{fig:feh}
\end{figure}

Figure~\ref{fig:feh} illustrates the mean metallicity distribution of the XP MS sample from Case~A (panel~(a)) and Case~B (panel~(b)). Apart from the scale difference in metallicity between these maps ($\Delta {\rm [Fe/H]}\approx0.05$~dex), there are small-scale systematic variations in [Fe/H]. For instance, the region near $l\sim130\arcdeg$ and $b\sim5\arcdeg$ exhibits relatively higher [Fe/H] compared to the surrounding areas, while an opposite trend is observed near $l\approx355\arcdeg$ and $b\approx15\arcdeg$. These variations appear to be more pronounced in regions with higher $\ebv$, but the spatial trend does not appear to be entirely correlated with it (see Fig.~\ref{fig:avgmap}). In addition to these small-scale angular variations, stars in the first and second Galactic quadrants generally exhibit higher mean metallicity compared to the opposite side of the Galaxy. On the contrary, all-sky metallicity estimates obtained using the General Stellar Parametriser from Spectroscopy \citep[GSP-Spec;][]{recioblanco:23} reveal a more symmetric distribution of stars observed with the {\it Gaia} RVS \citep[][see their Fig.~4]{gaia:recioblanco:23}. Within a distance of $500$~pc, their metallicity distribution appears nearly uniform, while exhibiting a flattened structure in the $1$~kpc $< d < 2$~kpc bin.

The observed systematic trend can be interpreted as indicative of systematic changes in the extinction curve along various lines of sight within the Milky Way. In our previous analysis, we held $R_V$ fixed at $3.1$ \citep{fitzpatrick:99} during the parameter estimation. However, as initially studied by \citet{cardelli:89} and as corroborated by numerous subsequent studies, including \citet{fitzpatrick:99}, it is well established that extinction curves can vary substantially, ranging from $R_V\sim2$ to $6$. This variability has been attributed to differences in the distributions of dust grain sizes within the clouds along different lines of sight.

Indeed, the lines of sight displaying systematically lower [Fe/H] in Figure~\ref{fig:feh} are associated with dense molecular clouds, such as Ophiuchus, located near $(l,b)\approx(355\arcdeg,15\arcdeg)$, Taurus $(170\arcdeg,-15\arcdeg)$, Orion $(210\arcdeg,-20\arcdeg)$, and Chamaeleon $(300\arcdeg,-15\arcdeg)$. Previous studies \citep[e.g.,][]{vrba:84,cardelli:89,whittet:01} found that the denser parts of these regions exhibit higher $R_V$ ($3.5$--$6$) than in the diffuse interstellar medium, which have been attributed to the increase in grain size. On the contrary, stars within the Cassiopeia and Cepheus clouds, encompassing the region $110\arcdeg<l<140\arcdeg$ and $0\arcdeg<b<20\arcdeg$, consistently exhibit higher [Fe/H] values. These stars, located on the periphery of the Cepheus flare shell, an ancient supernova remnant, are associated with a cloud complex that interacts with supernova shocks \citep{grenier:89}. Earlier estimates of $R_V$ in the adjacent region suggest a value of $R_V=2.9$, which is slightly lower than the canonical value \citep{zdanavicius:02}.

To account for the influence of varying extinction laws along different lines of sight within the local volume, we employed the $R_V$ dependence formula outlined in \citet{fitzpatrick:19}. However, the use of their relation resulted in a mild reduction in the sample size with valid parameter estimates that passed our quality criteria. For sight lines exhibiting moderate to heavy foreground extinction ($\langle\ebv\rangle\sim0.1$--$1$), the number of stars remains within $5\%$--$10\%$ for both Case~A and Case~B. However, the magnitude of the reduction reaches $\sim5\%$--$30\%$ for Case~C and Case~D owing to some stars falling outside the model grid. This is related to the fact that the extinction curve at the standard $R_V=3.1$ in \citet{fitzpatrick:19} differs by up to $10\%$ on the short-wavelength side compared to the original curve in \citet{fitzpatrick:99}, which was utilized in the above analysis. Hence, the derived extinction also increases by this amount when adopting the standard $R_V=3.1$ curve from \citet{fitzpatrick:19}. Despite this, the decrease in the number of stars indicates that the extinction curve may not properly leverage the overall spectral energy distribution of stars. Consequently, we concluded that the original extinction curve presented by \citet{fitzpatrick:99} provides a more internally consistent result for our set of calibrated models.

Therefore, we utilized the extinction law from \citet{fitzpatrick:99} at $R_V=3.1$, while incorporating the $R_V$ dependence described in \citet{fitzpatrick:19}. This was accomplished by comparing $A_\lambda/\ebv$ at $R_V=3.1$ in both cases and then scaling the $A_\lambda/\ebv$ relations at arbitrary $R_V$. We conducted a repeat analysis of Case~A and Case~B by restricting our parameter search to $1 \leq R_V \leq 6$.\footnote{While \citet{fitzpatrick:19} originally formulated the $R_V$ dependence over the range $2.5 \leq R_V \leq 6$, we extended the lower limit in our modeling to account for uncertainties in our parameter estimates. Nonetheless, in our mapping, most lines of sight exhibit $R_V$ values greater than 2.5.} We found $\epsilon_0=\{0.015$, $0.029\}$~mag for Case~A and Case~B, respectively, for the zero-point in equation~(\ref{eq:1}).

\begin{figure*}
\centering
  \gridline{\fig{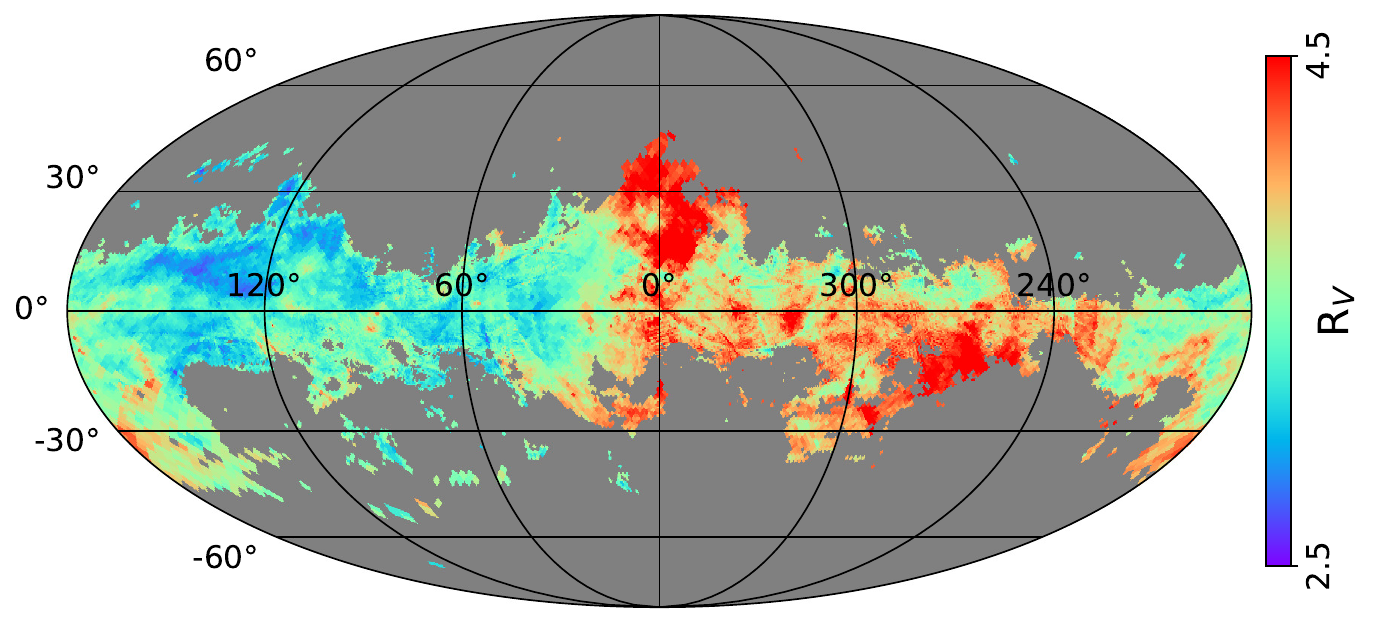}{0.42\textwidth}{\textbf{(a) Mean $R_V$ from Case~A and B }}
                \fig{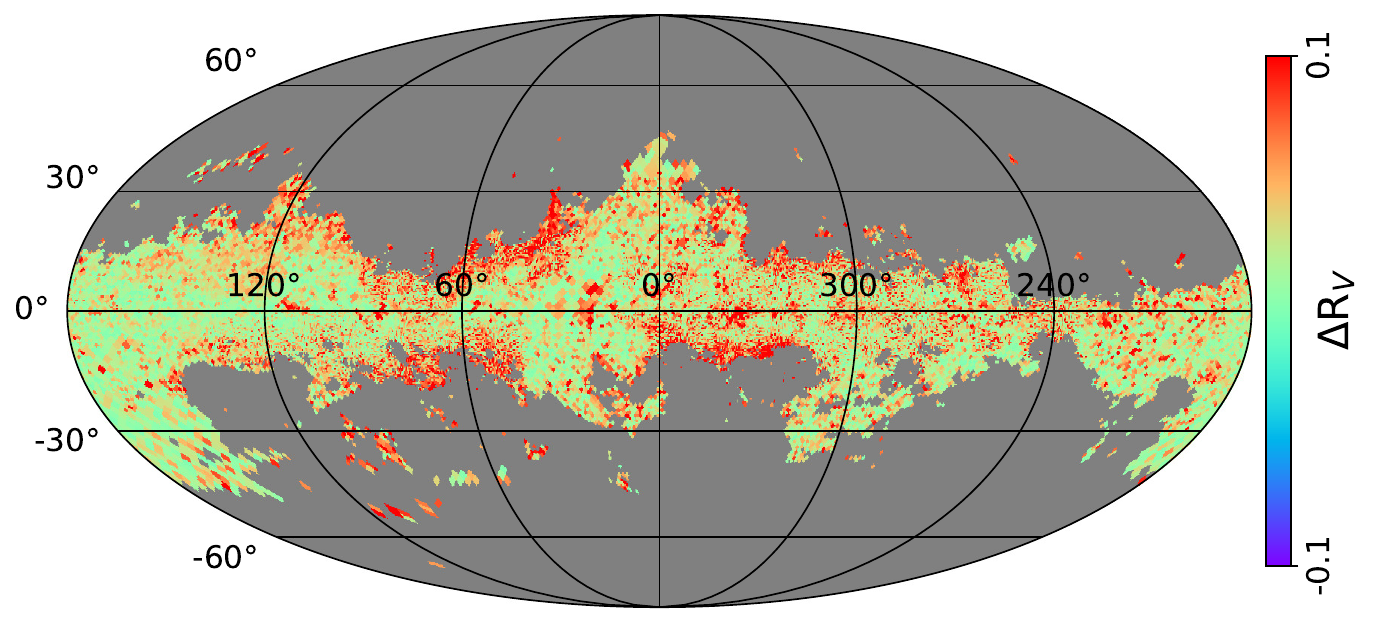}{0.42\textwidth}{\textbf{(b) Difference between Case~A and B }}}
  \gridline{\fig{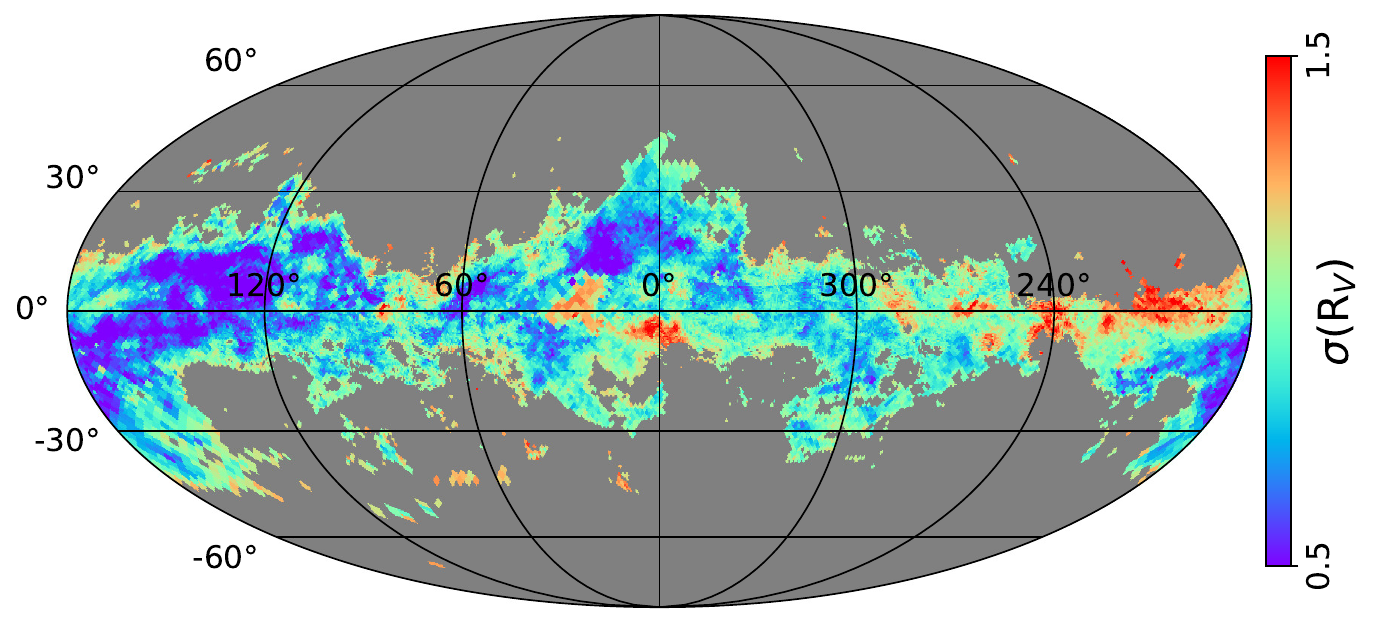}{0.42\textwidth}{\textbf{(c) Scatter in $R_V$ from Case~A }}
                \fig{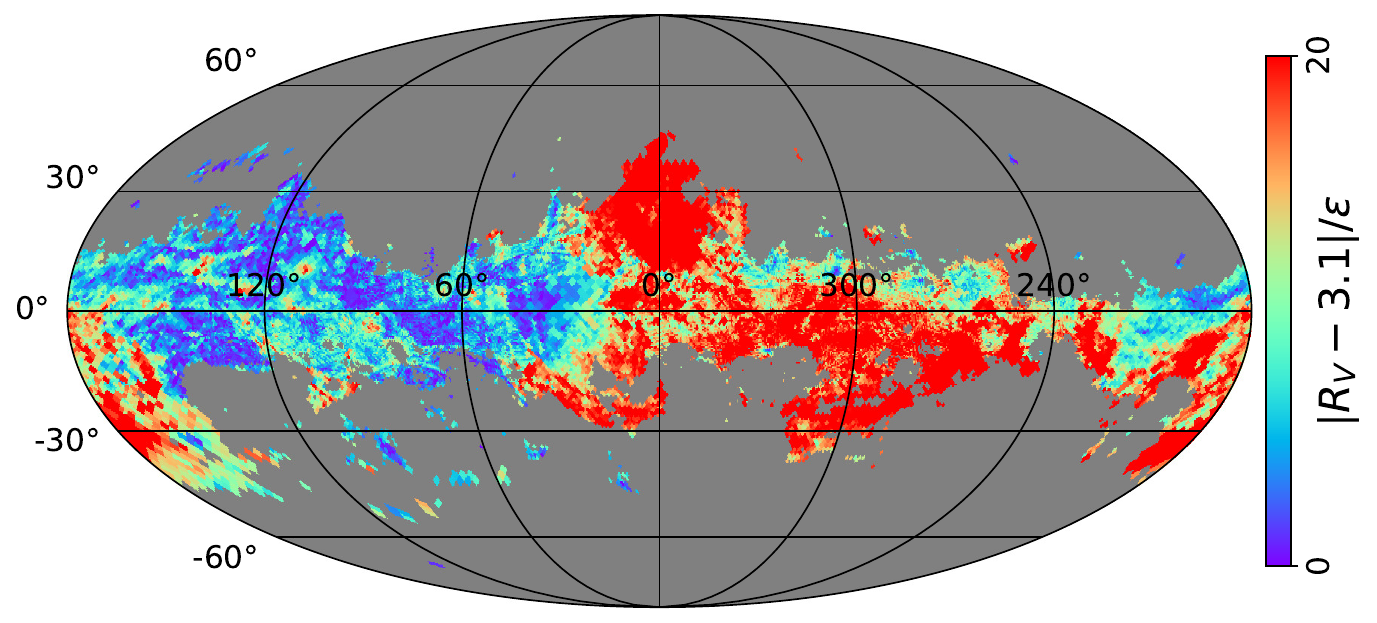}{0.42\textwidth}{\textbf{(d) Significance of $R_V$ deviation }}}
\caption{Mean and statistical properties of $R_V$ estimates from stars within $\dmn \leq 12$. Only regions with significant reddening ($\ebv\geq0.1$ at $\dmn=12$) are included. (a) Mean $R_V$ estimates from Case~A and Case~B. (b) Differences in $R_V$ between Case~A and Case~B. (c) Half of the difference between the $15.9$th and $84.1$st percentiles of $R_V$ from Case~A ($\sigma(R_V)$). (d) Significance of the deviation of $R_V$ from the fiducial case ($R_V=3.1$) in relation to the error in the mean $R_V$ ($\epsilon\equiv\sigma(R_V)/\sqrt{N}$).}
\label{fig:rv}
\end{figure*}

Panel~(a) of Figure~\ref{fig:rv} shows the distribution of mean $R_V$ values estimated using stars within $\dmn \leq 12$ from both Case~A and Case~B. Here we included only regions with significant reddening ($\ebv\geq0.1$ at $\dmn=12$), as $R_V$ can be reliably obtained in such areas. As shown in panel~(b), the median $R_V$ estimates between the two cases differ by $\Delta R_V\la0.1$ within $|b|\la30\arcdeg$, beyond which our estimates become less accurate owing to the small amount of foreground extinction. On the contrary, the variation in $R_V$ observed along each line of sight is significant, as demonstrated in panel~(c), where we present half of the difference between the $15.9$th and $84.1$st percentiles of $R_V$ from Case~A. In the vicinity of the Galactic plane, this effective $1\sigma$ deviation, $\sigma(R_V)$, typically falls within a range of $0.5$--$1.0$ in each multiresolution HEALPix cell. Nonetheless, the error in the mean $R_V$, computed as $\epsilon\equiv\sigma(R_V)/\sqrt{N}$, where $N$ indicates the number of stars in each cell, is still about $3$ times larger than the difference between Case~A and Case~B (panel~(b)). This suggests that a significant dispersion in $R_V$ can be attributed, at least in part, to actual star-by-star variations.

The mean $R_V$ values shown in panel~(a) in Figure~\ref{fig:rv} range from $\approx3$ to $4.5$, with a bulk average of $3.7$, which is larger than the canonical $R_V=3.1$ \citep[e.g.,][]{cardelli:89,fitzpatrick:19}. If one takes $\epsilon$ as an effective $1\sigma$ uncertainty, our estimates reveal a significant change in $R_V$ across the sky (panel~(d)). The systematic trend between metallicity in Figure~\ref{fig:feh} (evaluated at $R_V=3.1$) and $R_V$ in Figure~\ref{fig:rv} is not entirely correlated with each other. Nonetheless, as noted above, areas where metallicities appear to be underestimated tend to have relatively high $R_V$ values (such as the Ophiuchi cloud), while the opposite trend is observed in the low-$R_V$ region (most strikingly in the Cassiopeia and Cepheus cloud complexes).

In earlier studies, \citet{schlafly:16} and \citet{zhang:23a} investigated large-scale variations in $R_V$ using spectroscopic data from APOGEE and LAMOST, respectively. Their 2D mapping is inherently limited by target availability, with the latter study covering a significantly larger area, yet restricted to the northern celestial hemisphere. Nevertheless, it is possible to make a qualitative comparison for the specific regions that are mentioned above. First, the regions with the highest $R_V$ in \citet{zhang:23a} exhibit $R_V\sim4$ and are observed near the Ophiuchi cloud, in agreement with our findings. Their map also reveals elevated $R_V$ values in patchy regions near the Orion and Taurus cloud complexes, consistent with our findings. Moreover, the Cassiopeia and Cepheus clouds in \citet{schlafly:16} exhibit low $R_V$ ($\sim3.0$), and there is also an indication of low $R_V$ in \citet{zhang:23a} near the border of the cloud complexes, a region not fully covered in their study. However, we note that their samples include giants and extend beyond our distance limit, reaching out to $4$--$5$~kpc. Consequently, their projected $R_V$ map may be influenced by distant stars that are not included in our sample.

\begin{figure}
\centering
\epsscale{0.65}
  \gridline{\fig{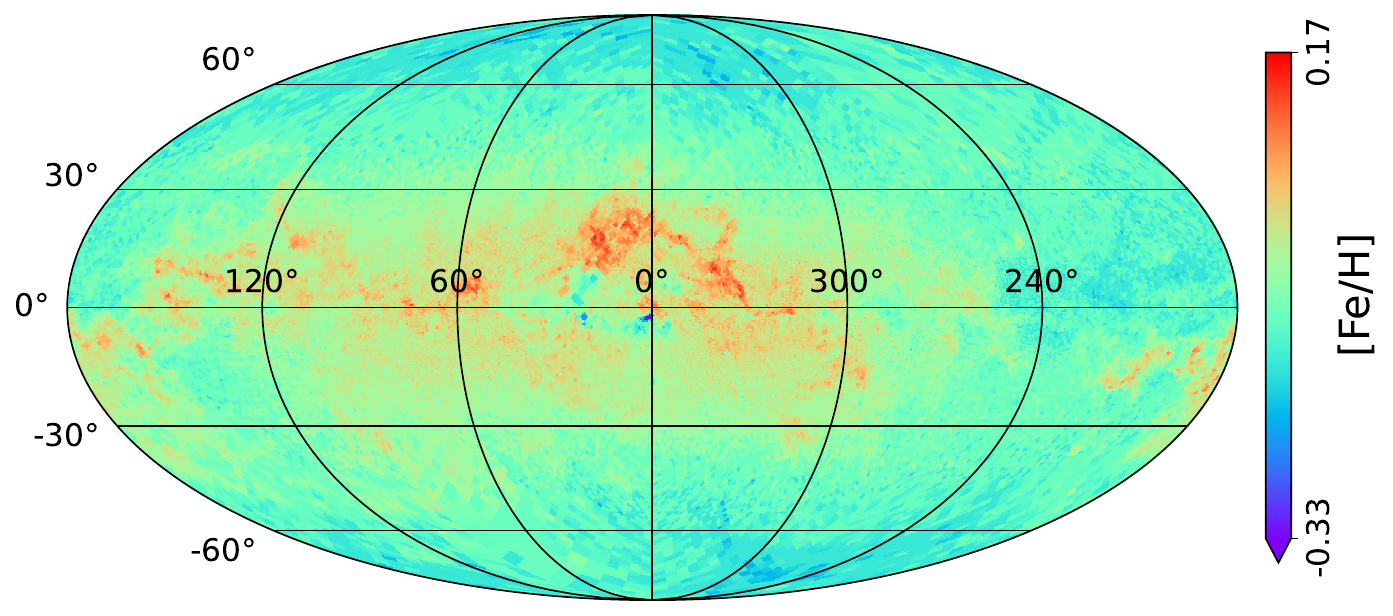}{0.48\textwidth}{\textbf{(a) Mean [Fe/H] from Case~A }}}
  \gridline{\fig{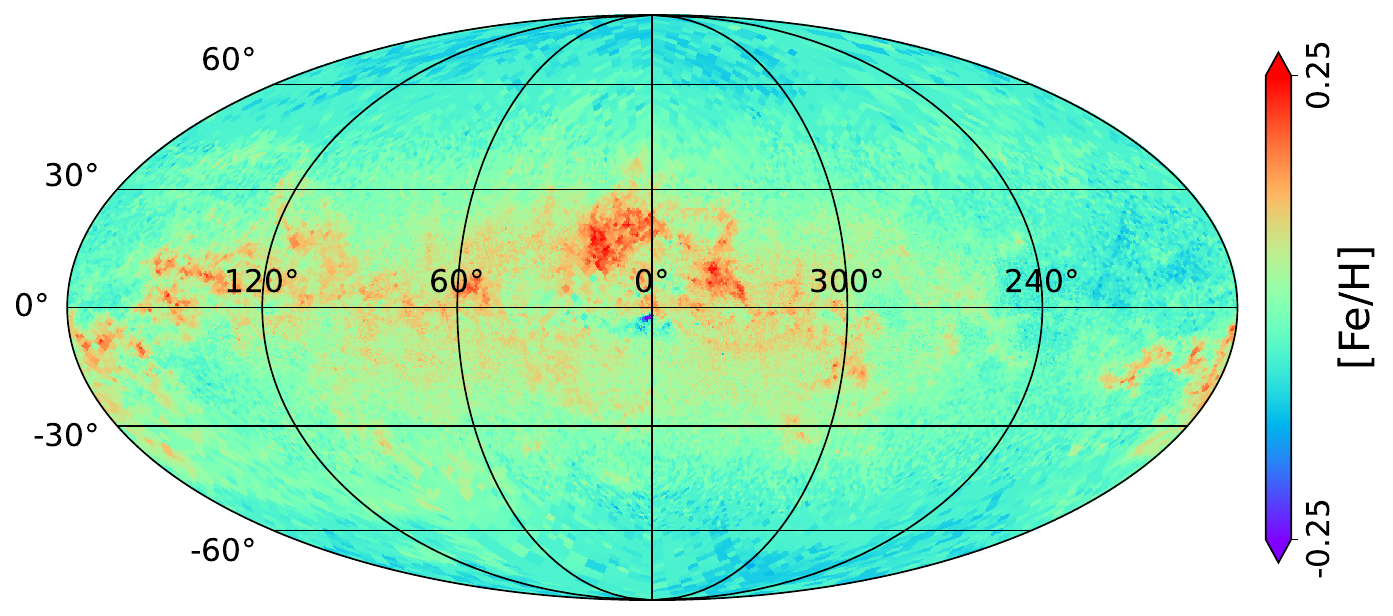}{0.48\textwidth}{\textbf{(b) Mean [Fe/H] froms Case~B }}}
\caption{Mean metallicity estimates from varying $R_V$ in the parameter search, for (a) Case~A and (b) Case~B.}
\label{fig:newfeh}
\end{figure}

The revised metallicity maps, which are based on varying $R_V$, are presented in Figure~\ref{fig:newfeh}. In contrast to Figure~\ref{fig:feh}, they reveal enhanced mean metallicities toward the Galactic center and are more symmetric with respect to both the Galactic plane and the Galactic prime meridian in both Case~A (panel~(a)) and Case~B (panel~(b)). In Figure~\ref{fig:newfeh}, the highest-metallicity regions are located near the Aquila Rift ($20\arcdeg$,$15\arcdeg$) and the Lupus dark clouds ($340\arcdeg$,$10\arcdeg$). Along with the high-metallicity regions near Taurus and Orion, they appear to follow the Gould Belt, a projected ring of a large number of bright stars, OB associations, and a significant amount of interstellar material, inclined by $\sim$20$^\circ$ with respect to the Galactic plane \citep[][and references therein]{zari:18}. Although this ring-like structure is not immediately evident in the RVS sample \citep{gaia:recioblanco:23}, the overall mean metallicity distribution in Figure~\ref{fig:newfeh}, characterized by higher metallicity along the Galactic plane, aligns with their findings, suggesting that the approach presented in this section produces more physically plausible results.

\begin{figure*}
\centering
\gridline{\fig{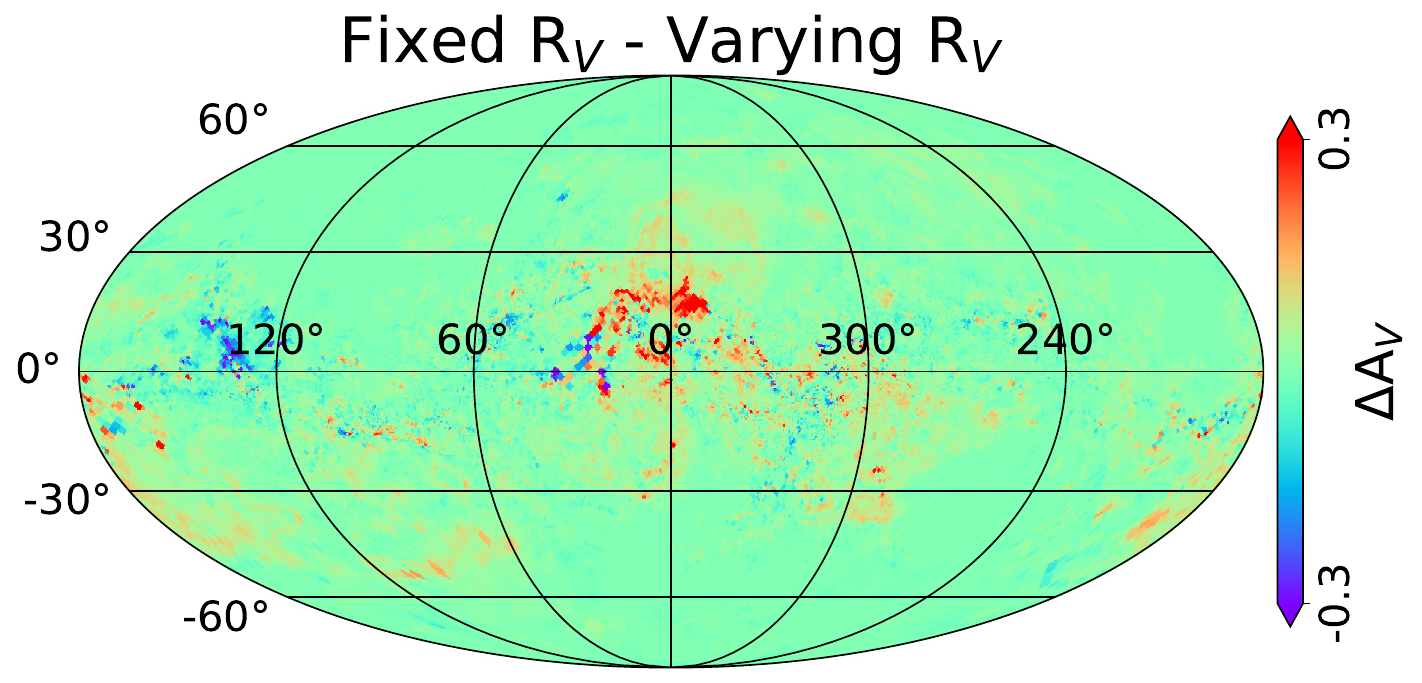}{0.32\textwidth}{\textbf{(a) $d=251$~pc }}
             \fig{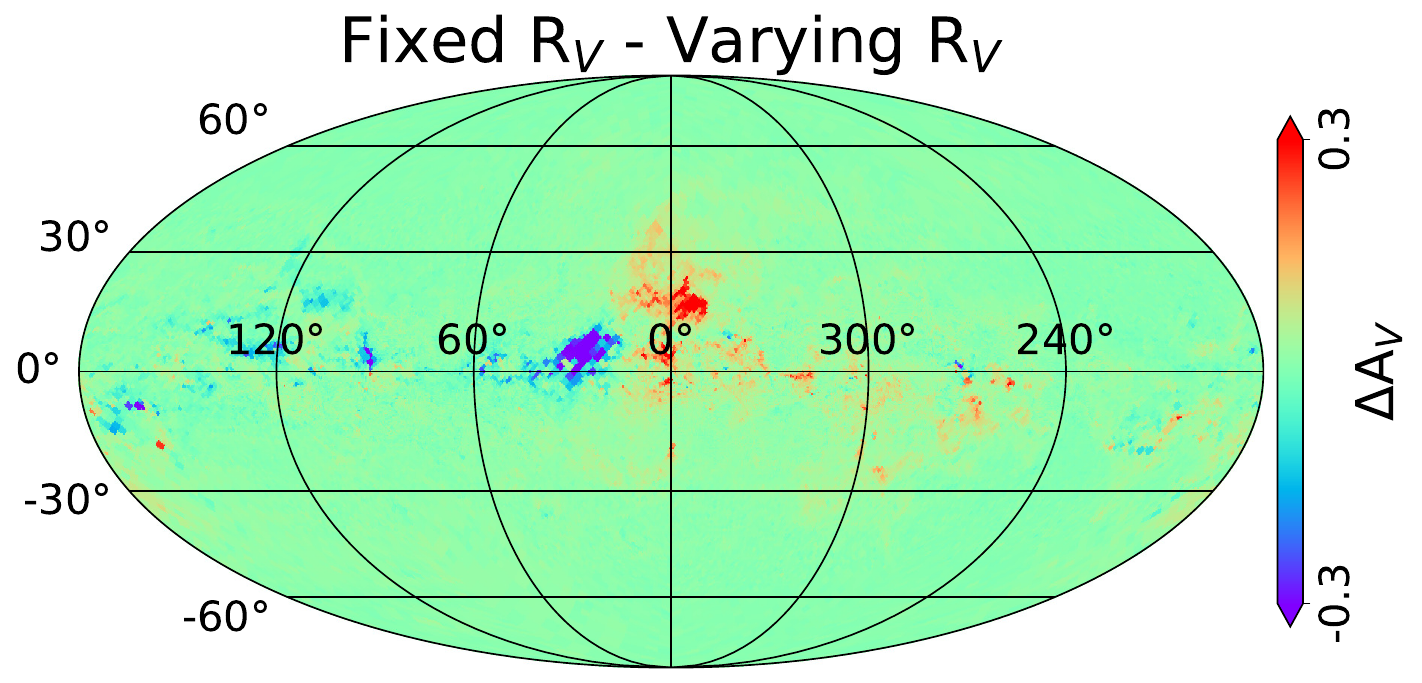}{0.32\textwidth}{\textbf{(b) $d=630$~pc }}
             \fig{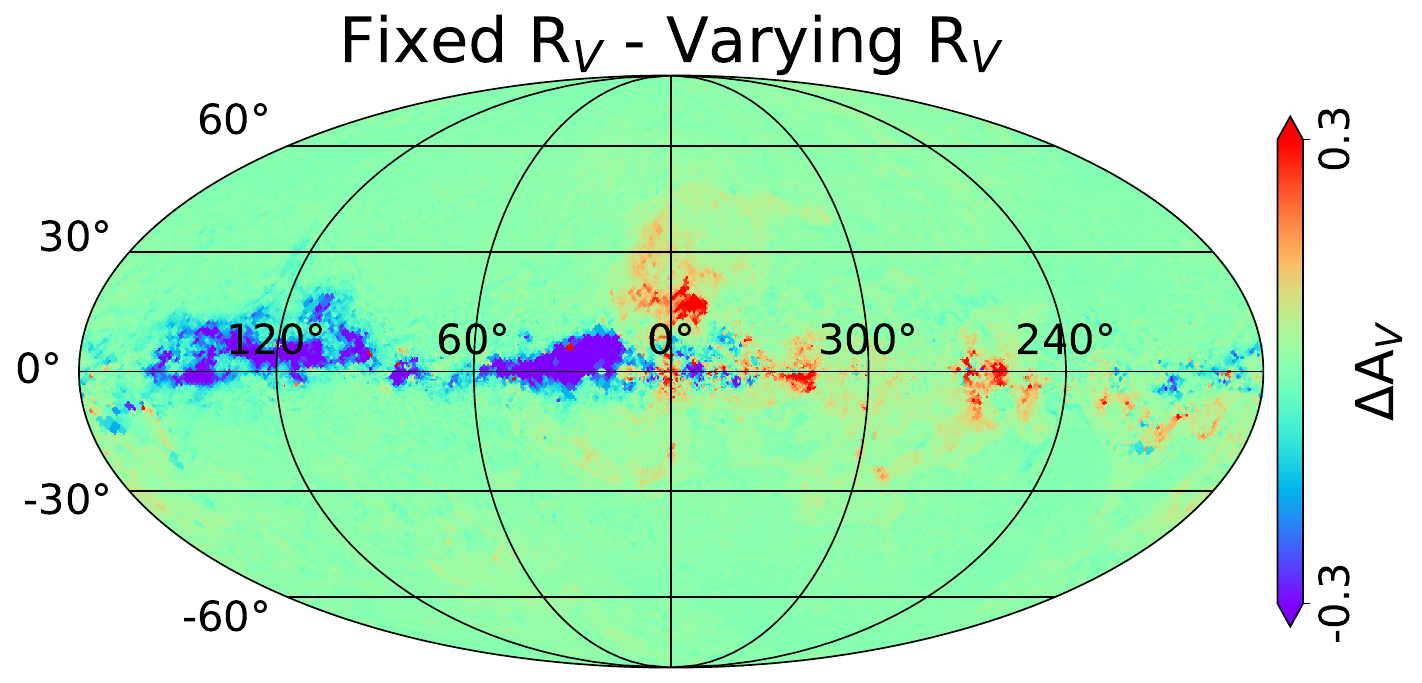}{0.32\textwidth}{\textbf{(c) $d=1584$~pc }}}
\caption{Differences in $A_V$ derived from Case~A when assuming a fixed value of $R_V=3.1$, compared to using a median $R_V$ (see text). The comparisons are illustrated at three representative distances.}
\label{fig:avdiff}
\end{figure*}

Despite this, the $\ebv$ data cube constructed using this approach does not necessarily surpass our base case with $R_V=3.1$ due to the limited precision of our $R_V$ measurements for individual stars, which would inevitably increase the uncertainty in the foreground extinction. Alternatively, one can compute a median value for $R_V$ along each line of sight and utilize it in estimating foreground extinction ($A_V$). The comparisons, presented in Figure~\ref{fig:avdiff} at three select distances, highlight differences in $A_V$ for two scenarios. The first is derived from Case~A, assuming a fixed value of $R_V=3.1$, while the second assumes a median $R_V$ computed from each cell. As demonstrated here, the most substantial discrepancies become evident in regions where $R_V$ notably deviates from the canonical value (as depicted in Figure~\ref{fig:rv}), effectively tracing regions with high dust columns. In regions with the highest $\ebv$ values, the difference in $A_V$ can reach a few tenths of a magnitude.

\section{Summary and Caveats}\label{sec:summary}

In this study, we utilized the extensive dataset of low-resolution spectra from Gaia DR3 to construct a 3D map of foreground dust reddening across the entire sky. This endeavor was made possible through the application of our newly developed, empirically calibrated stellar isochrones, as detailed in Paper~IV. Our approach involved modeling foreground extinction using a straightforward, two-step function along each line of sight, yielding results that exhibit a good agreement with widely used extinction maps by \citet{schlegel:98} and \citet{green:19}. In addition, our metallicity estimates show good agreement with high-resolution spectroscopic results from the GALAH survey and, to some extent, SDSS/APOGEE (see Appendix~\ref{sec:comp2}).

However, our map reveals a significant discrepancy, particularly at low Galactic latitudes, compared to the study by \citet{zhang:23b}, who provided reddening estimates based on the same XP spectra but through an empirical forward modeling technique. Both our method and that of Zhang et al.\ yield consistent metallicity estimates, suggesting that the reddening differences arise from systematic factors other than the metallicity scale.

While our mapping of foreground extinction provides comprehensive coverage in both the northern and southern hemispheres, there are certain limitations in our approach. First, our current results rely on calibrated isochrones for MS stars, which restricts this study to a relatively nearby volume compared to using giant stars. On average, our map extends up to $3$~kpc from the Sun, but the physical depth of the reddening cube varies across the sky owing to our sample being primarily constrained by the brightness limit of the XP spectra and the availability of accurate Gaia parallaxes. Consequently, some areas with nearby dust clouds have a limited distance range in our data cube.

Secondly, our reddening cube is constructed by modeling cumulative reddening as a function of distance using a simple two-step logistic function along each line of sight. This choice of model is primarily empirical, driven by the observed distribution of stars, rather than being motivated by the physical and structural properties of interstellar dust. Due to the trade-off between the number of stars within each HEALPix cell and the spatial resolution of the reddening map, our chosen mapping scheme did not offer HEALPix cell sizes large enough to generate a purely empirical representation of cumulative extinction.

Lastly, we presented two different cases of a reddening cube, one assuming a fixed value of $R_V$ and the other varying $R_V$ in our parameter search. Although the latter is more physically motivated, the limited precision of our measurements prevents us from accurately tracing fine-structure variations of $R_V$ along each line of sight. We have provided only mean $R_V$ estimates along each sight line, but adopting these mean $R_V$ values for subsequent analyses may potentially introduce bias into the results if the true $R_V$ of stars significantly deviates from its mean value for a given sight line.

The 3D reddening map presented in this paper holds significant value for our ongoing research within this series of papers. In our previous work, we utilized photometric data to infer the metallicities of stars, but the low-latitude regions near the Galactic plane posed challenges, as they were largely obscured by substantial extinction in optical imaging surveys. High-resolution spectroscopic surveys are particularly noteworthy in this regard, as they can offer the most detailed information on individual stars, such as their 3D kinematics and complete elemental abundances, even in the presence of large foreground extinction. Nevertheless, due to inherent limitations in spectroscopic surveys, such as the relatively stronger target selection bias and a smaller number of stars compared to photometry, it is crucial to establish a robust foundation for understanding the various Galactic stellar populations through the development of photometric metallicity mapping, as demonstrated in our series of works. Eventually, this will allow us to delve deeper into spatial distributions, dynamics, and chemical properties of the various (sub)structures that make up the Milky Way and gain a better understanding of its formation and evolution history.

\acknowledgements\

The authors gratefully acknowledge Young Sun Lee for his careful review of an early draft of the manuscript and for providing valuable feedback. D.A.\ also thanks Seo-Won Chang for meticulously examining the extinction cubes in the GitHub library. D.A.\ acknowledges support provided by the National Research Foundation of Korea grant funded by the Ministry of Science and ICT (No.\ 2021R1A2C1004117). T.C.B.\ acknowledges partial support from grant PHY~14-30152, Physics Frontier Center/JINA Center for the Evolution of the Elements (JINA-CEE), and from OISE-1927130: The International Research Network for Nuclear Astrophysics (IReNA), awarded by the US National Science Foundation. A.C.\ is supported by a Brinson Prize Fellowship at the University of Chicago/KICP.

This work presents results from the European Space Agency (ESA) space mission Gaia. Gaia data are being processed by the Gaia Data Processing and Analysis Consortium (DPAC). Funding for the DPAC is provided by national institutions, in particular the institutions participating in the Gaia MultiLateral Agreement (MLA). The Gaia mission website is {\tt https://www.cosmos.esa.int/gaia}. The Gaia archive website is {\tt https://archives.esac.esa.int/gaia}. This work has made use of the Python package GaiaXPy, developed and maintained by members of the Gaia Data Processing and Analysis Consortium (DPAC), and in particular, Coordination Unit 5 (CU5), and the Data Processing Centre located at the Institute of Astronomy, Cambridge, UK (DPCI).

{\it Software:} GNU parallel (Tange 2018).

\appendix

\section{Comparisons to High-resolution Spectroscopic Parameter Estimates}\label{sec:comp2}

\begin{figure*}
\centering
\epsscale{0.65}
  \gridline{\fig{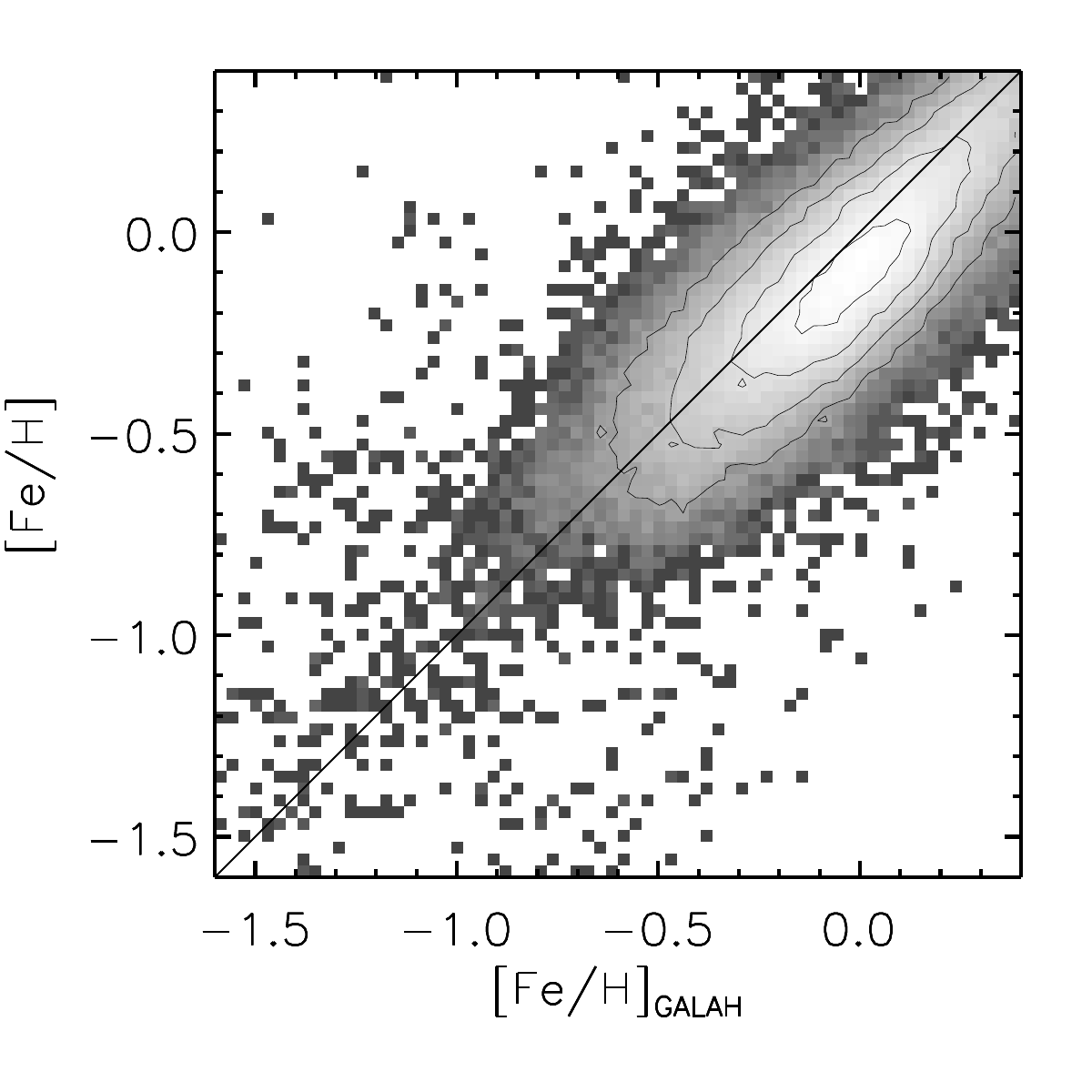}{0.32\textwidth}{\textbf{(a) Case~A vs.\ GALAH}}
                \fig{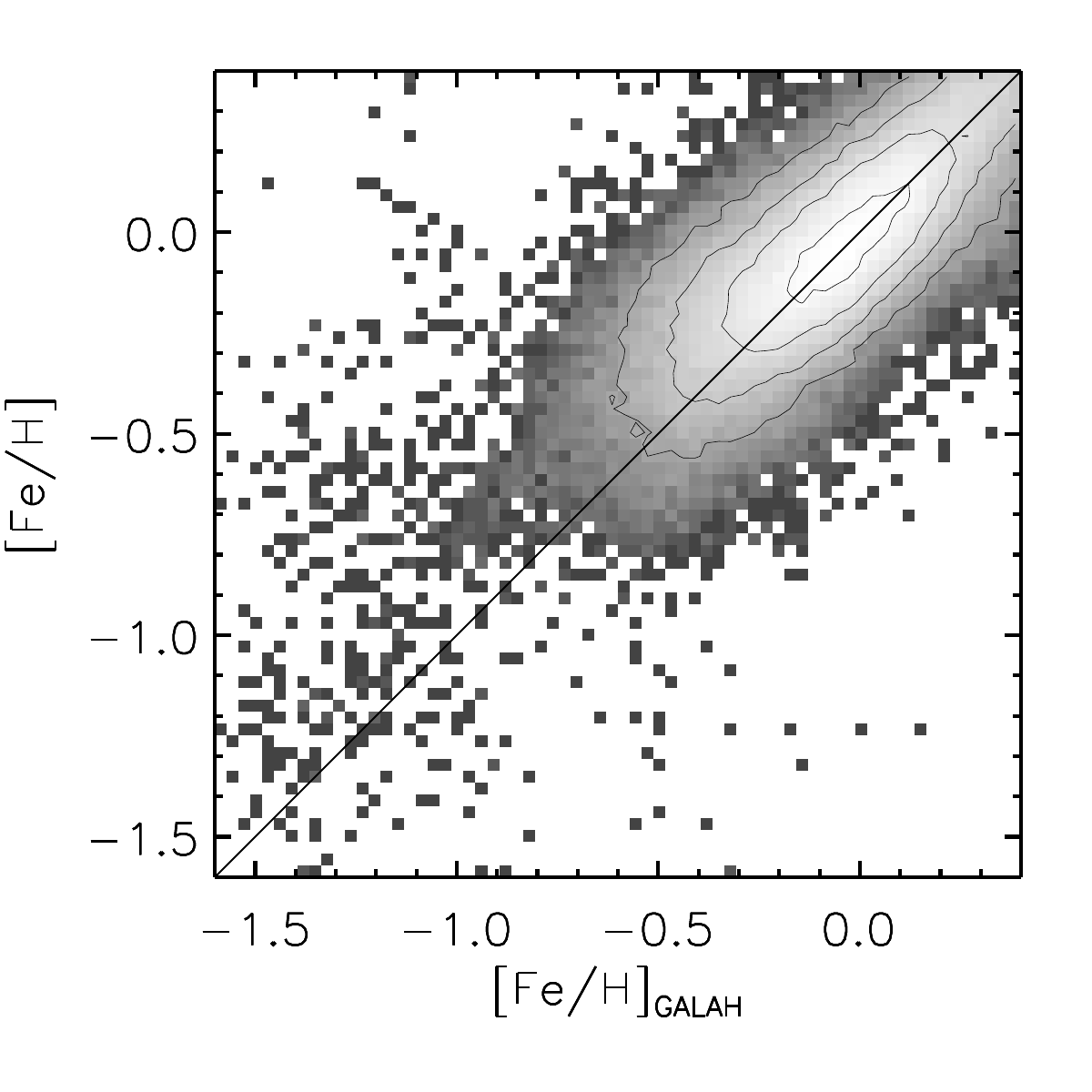}{0.32\textwidth}{\textbf{(b) Case~B vs.\ GALAH}}
                \fig{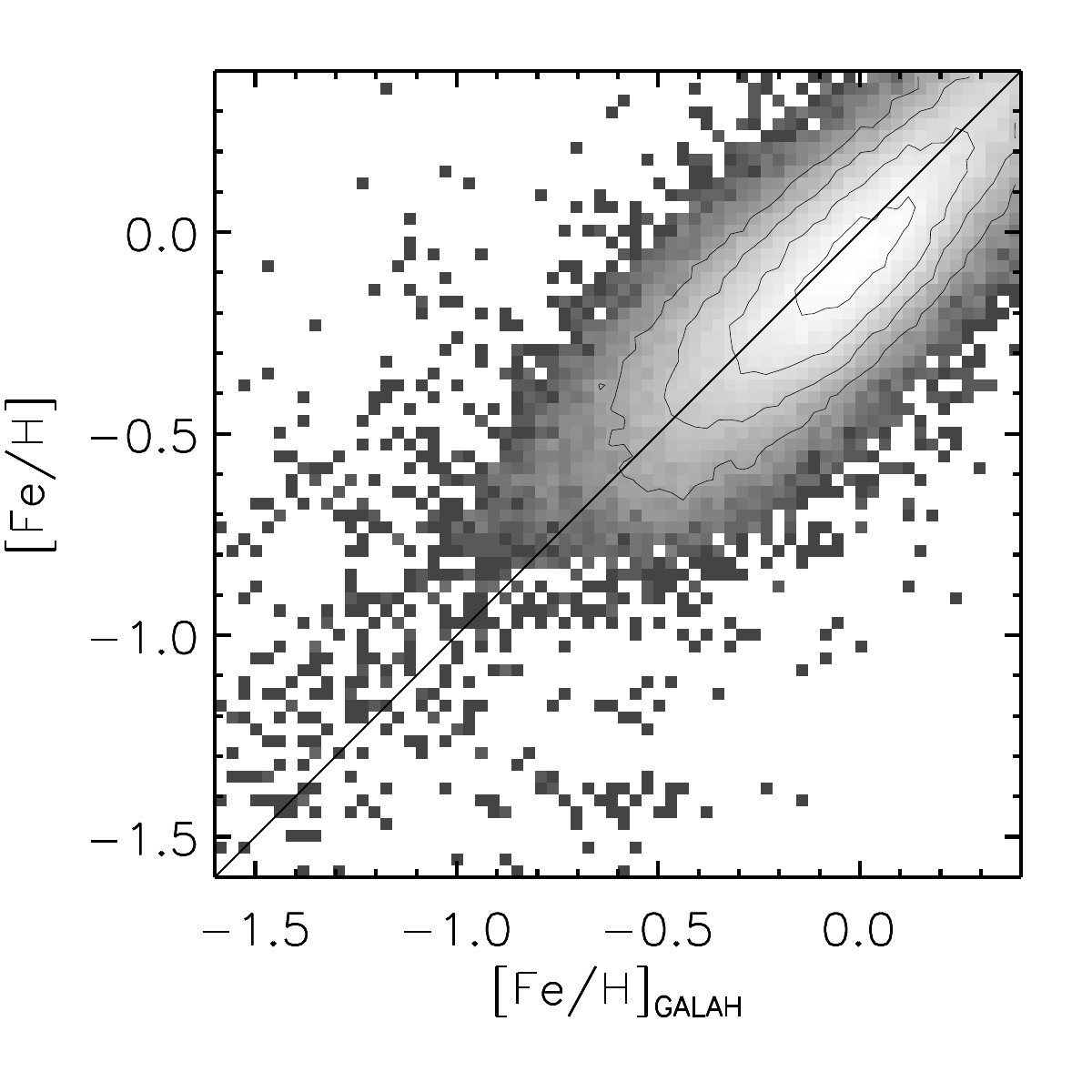}{0.32\textwidth}{\textbf{(c) Case~A (no UV) vs.\ GALAH }}}
  \gridline{\fig{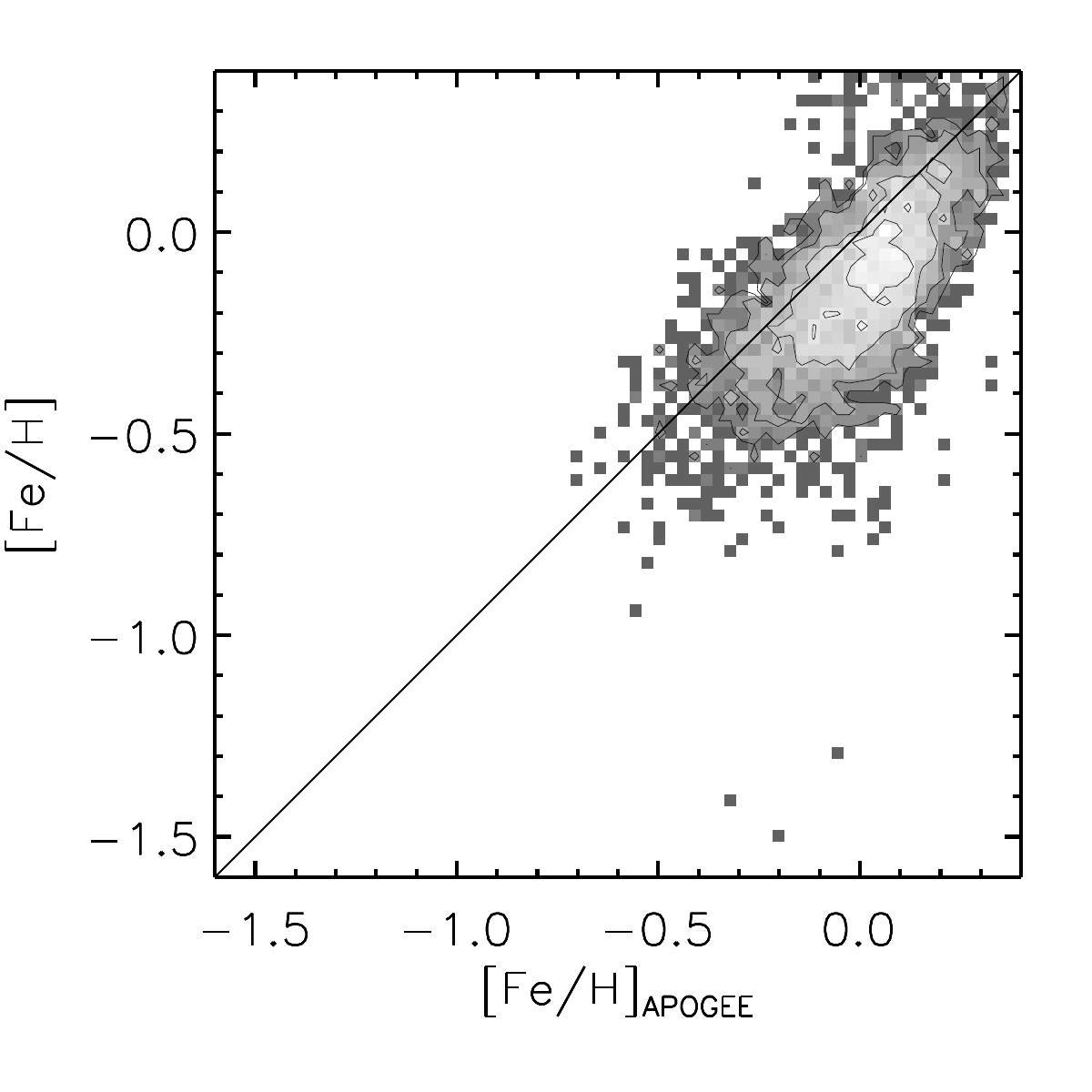}{0.32\textwidth}{\textbf{(d) Case~A vs.\ APOGEE }}
                \fig{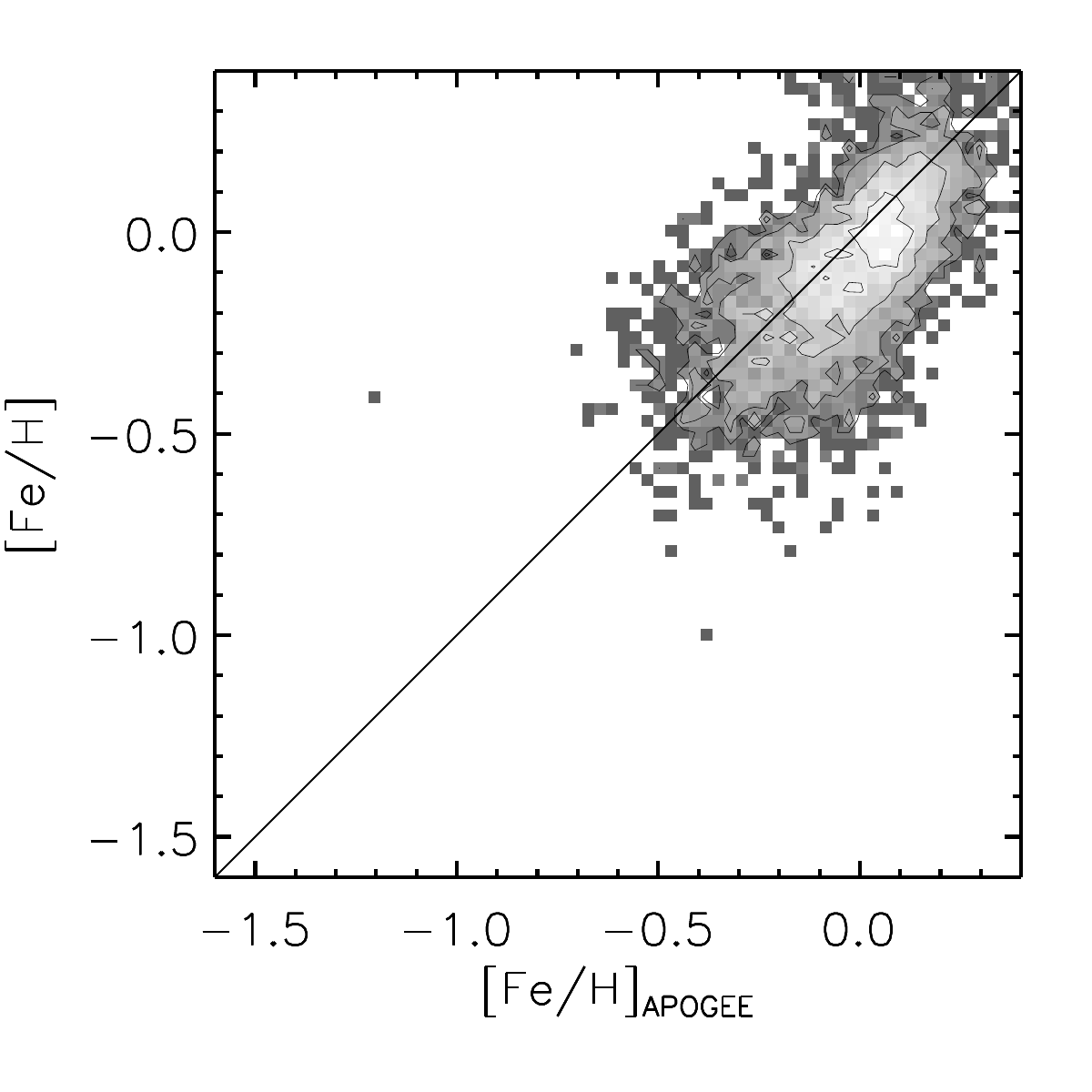}{0.32\textwidth}{\textbf{(e) Case~B vs.\ APOGEE }}
                \fig{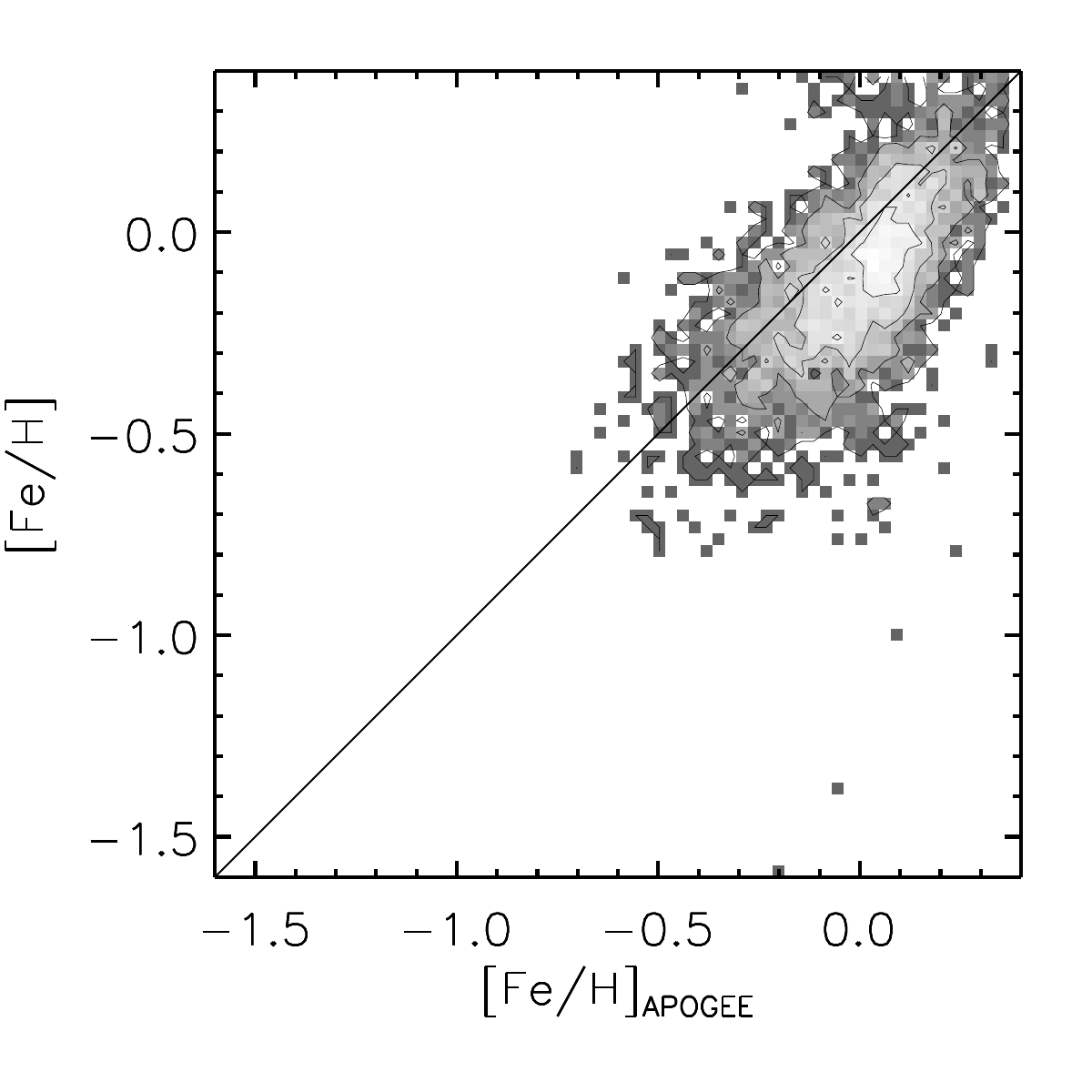}{0.32\textwidth}{\textbf{(f) Case~A (no UV) vs.\ APOGEE }}}
\caption{Comparisons of metallicity estimates in this study with high-resolution spectroscopic measurements from GALAH (top panels) and APOGEE (bottom panels). In the right panels, the same suite of comparisons is shown, but when our parameter estimates are restricted to $390$--$993$~nm, excluding UV portion of the spectra. The solid line represents a unity line.}
\label{fig:compspec}
\end{figure*}

In this appendix, we present a comparison between our metallicity estimates derived from XP spectra and spectroscopic measurements obtained through traditional line analysis methods. Figure~\ref{fig:compspec} illustrates these comparisons for two distinct datasets: GALAH DR3 \citep{buder:21} and APOGEE/SDSS DR17 \citep{apogee,sdss:dr17}, both based on high-resolution spectroscopy ($R\approx28,000$ and $R\approx22,500$, respectively). To conduct this comparison, we performed a cross-match between our catalogs and those of GALAH and APOGEE, using a search radius of $1\arcsec$. We specifically included our metallicity measurements with the following criteria: $\sigma(\ebv) < 0.3$, $3.5 < M_r < 7.5$, and $\sigma(\rm [Fe/H]) < 1.5$ as a minimum quality requirement, in alignment with the primary sample criteria for our work. In both datasets, we limited our analysis to stars with $\log{g}>3.5$ and $\ebv < 0.1$ in \citet{schlegel:98}. Additionally, we applied specific selection cuts for each spectroscopic dataset, ensuring that {\tt aspcapflag} is not set for APOGEE. For the GALAH sample, we required that the stellar parameter quality flag ({\tt flag\_sp}) and the overall iron abundance quality flag ({\tt flag\_fe\_h}) are not set.

In Figure~\ref{fig:compspec}, the first and second columns present data corresponding to Case~A and Case~B, respectively, in our study. These columns illustrate that the systematic differences between the two calibrations lead to an approximately $0.1$ dex difference in [Fe/H]. Despite these systematic distinctions between the two cases, these comparisons reveal that our metallicity estimates align well with those obtained from the GALAH survey, with differences within $\Delta{\rm[Fe/H]}\la0.1$ at ${\rm[Fe/H]}>-1$. While our estimates also exhibit reasonable agreement with the APOGEE dataset, it is worth noting that in Case~A there is a tendency for the underestimation of [Fe/H] by approximately $0.15$~dex. Such a scale difference in metallicity is not uncommon when comparing different spectroscopic abundance analyses. Considering the inherent systematics between these two spectroscopic samples, we interpret these results as satisfactory evidence that our metallicity scale remains reasonable and in line with existing spectroscopic measurements.

The final column in Figure~\ref{fig:compspec} illustrates the effect of incorporating UV flux in the XP spectra on our metallicity measurements, specifically for the comparisons with Case~A. In essence, there are negligible differences between the first and third columns, indicating that the exclusion of UV data has minimal impact on the parameter estimates. This outcome aligns with expectations from the fact that the uncertainty in flux measurements increases at wavelengths shorter than $400$~nm. While it is widely acknowledged that UV flux plays a crucial role in deriving stellar parameters from broadband photometry, our results indicate that even low-resolution spectra at $\lambda \ga 400$~nm can offer a wealth of information about photospheric metal abundance and foreground extinction.

\section{Comparisons to Cluster $\ebv$ Estimates in the Literature}\label{sec:clusters}

\begin{figure*}
\centering
  \gridline{\fig{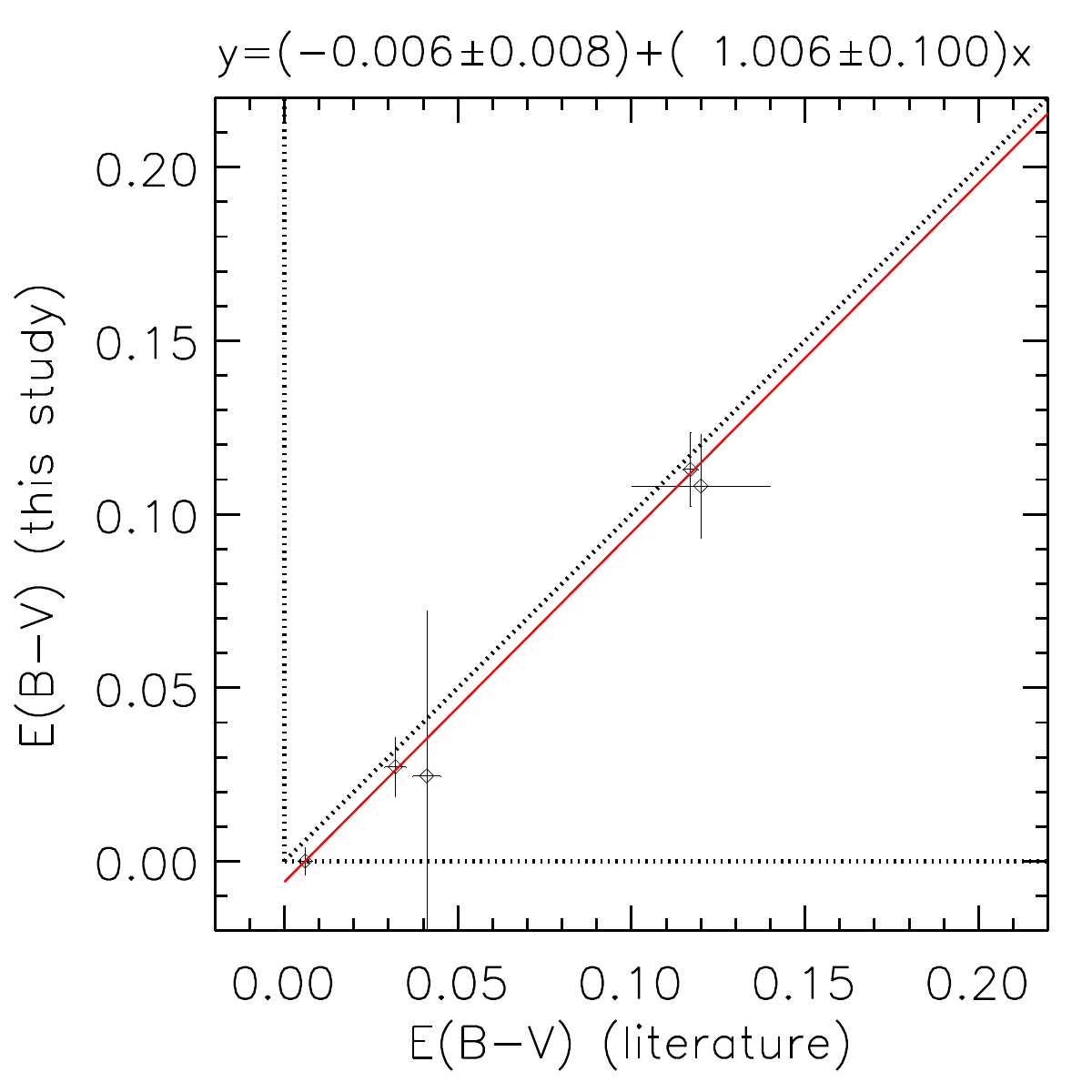}{0.32\textwidth}{\textbf{(a) \citet{an:07,an:19} }}
                \fig{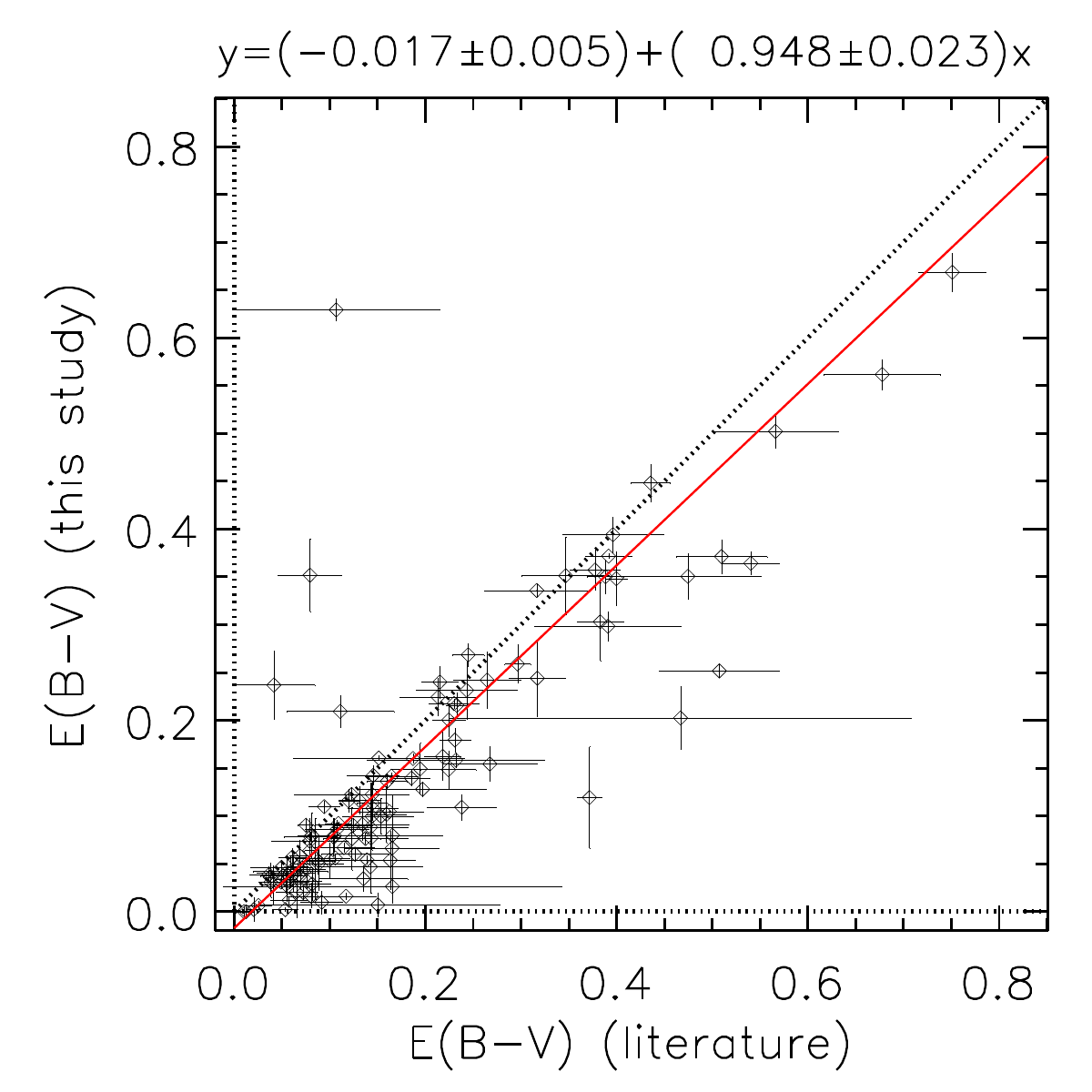}{0.32\textwidth}{\textbf{(b) \citet{dias:21} }}
                \fig{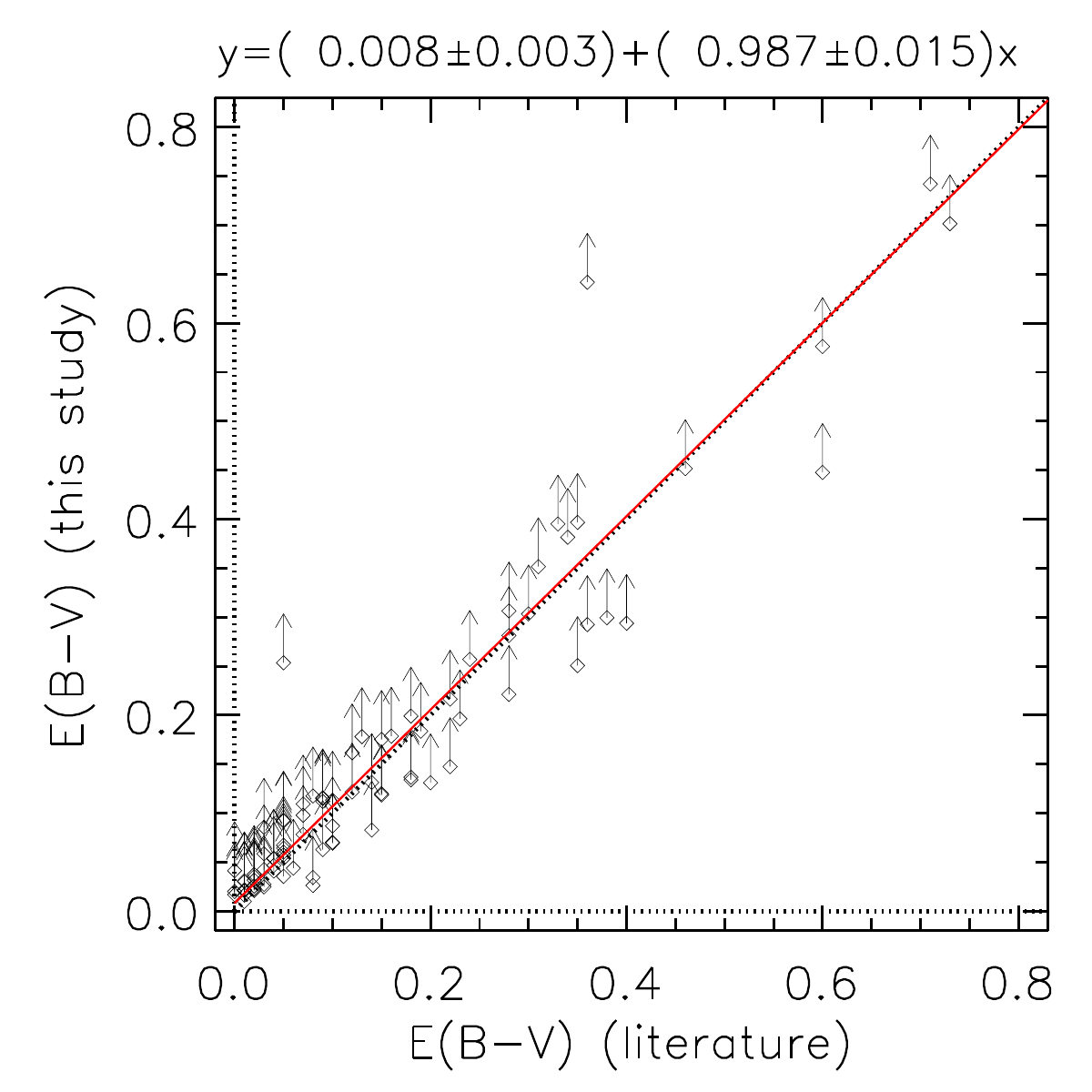}{0.32\textwidth}{\textbf{(c) \citet{harris:96} }}}
\caption{Comparisons with $\ebv$ estimates from the literature. (a) Open cluster samples with fundamental parameters from \citet{an:07,an:19}. (b) A homogeneous set of fundamental parameters for open cluster samples from \citet{dias:21}. (c) Compilation of cluster $\ebv$ for Galactic globular clusters as reported in \citet{harris:96}. In panel~(c), only the lower limit in our $\ebv$ estimates is displayed, as the cluster distances exceed the maximum threshold of our $\ebv$ cube. In all panels, the red solid line corresponds to the best-fitting regression to the data, with a zero-point and a slope displayed at the top, whereas dotted lines represent the unity and zero lines.}
\label{fig:clusters}
\end{figure*}

Figure~\ref{fig:clusters} provides a comparison of our $\ebv$ estimates, obtained from the averaged reddening cube at $R_V=3.1$, with those from the literature. In panel~(a), we display open cluster samples with precise fundamental parameters in \citet{an:07,an:19}. These clusters, namely Praesepe, the Pleiades, M67, NGC~6791, and NGC~2516 from left to right, have been essential resources in our prior tests for empirically calibrated isochrones in the Johnson--Cousins--2MASS photometric systems. Additionally, M67 and NGC~6791 have played a pivotal role in the empirical calibration of isochrones utilized in this work (Paper~III). Panel~(b) displays a comparison to a homogeneous set of fundamental parameter estimates for open cluster samples in \citet{dias:21}. Their estimates relied on fitting isochrones to the observed $G_{\rm BP}$ and $G_{\rm RP}$ photometric data in Gaia DR2. We assumed $R_V=3.1$ to convert their $A_V$ to $\ebv$, and we restricted their sample to those at $|b| > 10\arcdeg$.

In the right panel, we present a comparison to a compilation of cluster parameters for Galactic globular clusters from \citet{harris:96}. In this comparison, we display our $\ebv$ estimates as lower limits because the cluster distances extend beyond the maximum threshold of our $\ebv$ model. Nevertheless, our $\ebv$ lower limits exhibit a strong correlation with the values from \citet{harris:96}. This is because it is unlikely to encounter additional dust clouds beyond our distance limit at $|b| > 10\arcdeg$.

The error bars in both panels~(a) and (b) of Figure~\ref{fig:clusters} reflect the combined uncertainties, calculated as the quadrature sum of two components: half of the difference between Case~A and Case~B solutions and the propagated uncertainty from the adopted distance. The solid line in each comparison represents a linear fit to the data after applying a $3\sigma$ clipping, accounting for errors in both axes. In the right panel, we also made an attempt to fit the data using the maximum $\ebv$ value within our models along each line of sight (i.e., lower limits in $\ebv$ for distant clusters), as demonstrated by the solid line. The resulting best-fitting lines reveal a zero-point offset of $0.01$--$0.02$~mag. Our $\ebv$ estimates for open clusters are systematically smaller by $5\%$ compared to the values in \citet{dias:21}.

While \citet{dias:21} also adopted the extinction law by \citet{fitzpatrick:19}, as in this study, the systematic difference is further increased to approximately $16\%$ when comparing our estimates to the reddening cube with varying $R_V$. Therefore, other factors, such as the correlation between metallicity and reddening estimates, may be responsible for the observed systematic difference. Alternatively, a finite size of the angular resolution in our reddening map could be the cause if the open clusters are preferentially situated within extra dust clouds. In any case, a star-by-star comparison, such as shown in Figure~\ref{fig:zhang2}, would be invaluable in determining the origin of the systematic difference in $\ebv$.

\section{Accessing Extinction Data Cubes}\label{sec:access}

The primary data products resulting from this study can be accessed through the following link.\footnote{\tt https://github.com/deokkeunan/Galactic-extinction-map} Within this repository, we have made available two sets of data cubes. The first set, described in Section~\ref{sec:cube}, corresponds to $\ebv$ cubes derived for $R_V=3.1$. The second set, discussed in Section~\ref{sec:rv}, includes data cubes resulting from a parameter search involving varying values of $R_V$. Each of these data cubes contains the mean $\ebv$ values from Case~A and Case~B, using the multiresolution HEALPix scheme in the Galactic coordinate system. Additionally, we have included an example script to facilitate the utilization of these data sets.

{}

\end{document}